\documentclass[]{aastex631}

\usepackage{graphicx}
\usepackage{amsmath}
\usepackage{amssymb}
\usepackage{verbatim}
\usepackage{multirow}
\usepackage{bigdelim}
\usepackage{threeparttable}
\usepackage{float}
\usepackage{soul}

\newcommand{\usc}{Department of Physics \& Astronomy, University of South Carolina, Columbia, SC 29208, USA}
\newcommand{\geo}{Department of Biochemistry, Chemistry, \& Physics, Georgia Southern University, Statesboro, GA, 30458, USA}

\shorttitle{Dust in distant galaxies}
\shortauthors{Klimenko et al.}

\begin{document}
\title{Probing Dust Composition in Distant Galaxies with JWST Mid-IR Spectroscopy of Quasars with Foreground 2175 {\AA} Absorbers \textbf{II. Measurements of Grain Composition and Extinction Properties}}

\author[0000-0001-6730-2917]{Viacheslav V. Klimenko}
\affiliation{\usc}
\author{Varsha P. Kulkarni}
\affiliation{\usc}
\author{Monique C. Aller}
\affiliation{\geo}

\begin{abstract}

We present results from a mini-survey of dust spectral features arising in galaxies at redshifts $0.5 < z < 1.2$ in our James Webb Space Telescope (JWST) mid-infrared spectra of physically-unrelated background quasars. We analyze the JWST Mid-infrared Instrument (MIRI) Medium-Resolution  Spectrometer (MRS) spectra of five quasars presented in Klimenko, Kulkarni, \& Aller 2025a (Paper I) to determine the best-fit silicate mineralogies. Template profile fits to the 10 $\mu$m feature suggest the possible presence of crystalline silicates in three of the galaxies. This contrasts with the predominately amorphous silicate grains in the Milky Way diffuse interstellar medium (ISM).
We also measure the extinction curves using existing data from UV to mid-IR. Combining our results with past Spitzer IRS studies, we find that (i) the 10~$\mu$m silicate peak optical depth ($\tau_{10}$) is about three times stronger than expected for the local diffuse ISM over the range $A_V =0.1-2.0$, with $\tau_{10}/A_V$=$0.17\pm0.09$. (ii) The relative strength of the UV bump is similar to that in the local ISM. However, the ratio $\tau_{10}/A_{2175}$ is larger ($\sim0.1-1$), and appears to decrease with $A_V$, approaching the Galactic ISM value ($\sim 0.1$) at $A_V\sim1.5-2$. (iii) No significant correlation of $\tau_{10}/A_V$ with $R_V$. (iv) $\tau_{10}$ is strongly correlated with the gas-phase Mg~{\sc ii} absorption strength for the quasar sightlines. Possible interpretations include that some quasar sightlines probe dust in the circumgalactic medium (CGM), and that dust grains may have been significantly reprocessed in the ISM and CGM under conditions that may differ from those in the local ISM.
\end{abstract}



\section{Introduction}
\label{sec:intro}
Interstellar dust has been regarded as more of an  ``inconvenience" in many studies in extragalactic astronomy, and its effects are often ignored or inadequately accounted for. However, far from being an inconvenience,  interstellar dust is a vital ingredient of galaxies, since it has a profound effect both on the gas and star formation  properties of galaxies, and on their light output \citep[e.g.,][and references therein]{Mathis1990,Draine2003,Galliano2018}.

There is an increasing level of understanding of interstellar dust grain properties and their role in galaxy evolution, because of progress in both observational and theoretical studies over recent decades. The revolutions in ultraviolet (UV) and infrared (IR) space astronomy, including the Hubble Space Telescope (HST), the Spitzer Space Telescope (SST), the Herschel Space Observatory (HSO), and the James Webb Space Telescope (JWST), have resulted in a better understanding of the structure and properties of dusty regions within the Milky Way based on imaging and spectroscopic studies. The high sensitivity and (relatively) high spatial resolution of the Atacama Large Millimeter Array (ALMA) has also been allowing 
detection of emission from cold dust in wide-ranging environments. It is clear that dust grains have major impacts on the physical and chemical  properties of the interstellar medium (ISM), such as temperature, ionization, magnetic fields, and molecular content, and thereby on star formation. Furthermore, the bulk of stellar emission in galaxies below rest-frame wavelengths of 0.8 $\mu$m is obscured by dust. 
Galaxy evolution simulations, thus, increasingly include dust grain physics among their ingredients \citep[e.g.,][]{Narayanan2021, Narayanan2023}.

Recent observations have demonstrated that dust is important not only in the local Universe, but also at high redshifts. Detections of substructures have been made in 850 $\mu$m continuum observations of numerous dusty galaxies at redshifts $z > 2$ \citep[e.g.,][and references therein]{Hodge2020}. Highly dust-obscured, vigorous star formation has been detected by combining JWST and ALMA studies of sub-mm galaxies (SMGs) at $z \sim 1-6$ \citep[e.g.,][]{Hodge2025, McKinney2025}.  
Dust-obscured galaxies and active galactic nuclei (AGN) have been discovered at redshifts as high as 8 in many recent JWST studies \citep[e.g.,][]{Matsuoka2025, Finkelstein2025, Labbe2025}. Furthermore, combining JWST and ALMA studies reveals correlations between NIRCAM colors and ALMA dust continuum fluxes for galaxies \citep[e.g., the ALMA Large APEX BOlometer CAmera  Extended Chandra Deep Field South Submillimeter Survey (ALESS) of][]{Hodge2019},  illustrating that the reddening is caused by dust, as opposed to older stellar populations. 
Thus, dust grains were produced even at early cosmic epochs, and played a significant role even in early phases of the global star formation history. Given the ubiquity of dust, it is important to observationally determine whether dust grain properties, such as compositions and sizes, vary as a function of cosmic epoch or environment within a galaxy.

Extinction and attenuation curves provide one tool to investigate dust grain properties as a function of environment within a galaxy. Extinction (absorption and scattering) of (star) light by dust grains greatly affects the spectral energy distributions (SEDs) of galaxies \citep[e.g.,][]{Mathis1990, Calzetti2001}. The extinction laws in the Milky Way (MW) and in the Large and Small Magellanic Clouds (LMC, SMC) have been relatively well-determined, although substantial variations  exist among different sightlines [e.g., \cite{Gordon2009, Fitzpatrick2019, Gordon2021, Decleir2022, Gordon2023} (MW), \cite{GC1998, Gordon2003, Gordon2024} (SMC), \cite{MCG1999, Gordon2003} (LMC)]. Unfortunately, the dust extinction law is less well-determined in distant galaxies, so many extragalactic studies assume the extinction curves of those galaxies to be identical to those in the MW, LMC, or SMC, or to be like those in highly star-forming galaxies \citep[e.g.][]{Calzetti2000}. However, recent investigations with JWST of dust attenuation in z$\approx$2-12 galaxies suggests that the power-law slope of the attenuation curve and the 2175~\AA~bump strength decrease toward high redshift, possibly due to greater percentage of high redshift grains being larger and originating in supernovae ejecta \citep{Markov2025}. It should also be noted that attenuation curves are difficult to interpret, as they are complicated by the star-dust geometry in the galaxy. Deriving dust properties from attenuation curves is very hard, because it is necessary to disentangle the effect of dust properties from the geometry effects, which is only possible with radiative transfer modeling \citep[e.g.,][]{Salim2020}.

Dust grain chemical compositions, shapes, and structures can also be constrained using  absorption spectral features in UV  and Mid-IR (MIR)  imprinted by specific species of carbonaceous and silicate dust \citep[e.g.,][]{Draine2003}. Carbonaceous dust grains produce a strong UV feature at 2175~\AA~attributed to polycyclic aromatic hydrocarbons (PAHs), amorphous carbon, or graphitic carbon \citep[e.g.,][]{Draine1989,Sorrell1990,Duley1998,Papoular2009,Lin2023,Galliano2018}. The 2175~\AA~bump varies in strength, with no feature seen in some SMC sightlines \citep{Gordon2003, Gordon2024}, but it has been detected in galaxies out to z$\gtrsim$7 
\citep{Witstok2023,Markov2025}. In the MIR, a 3.4~$\mu$m aliphatic hydrocarbon feature is found in absorption in the diffuse Galactic ISM \citep{Decleir2025, Zeegers2025} and protoplanetary nebulae, as well as in some ultraluminous infrared galaxies (ULIRGs) and Seyfert nuclei \citep{Mason2004}. Aromatic hydrocarbons, such as the PAH features at 3.3, 6.2, 7.7, 8.6, 11.2, and 12.7~$\mu$m, are found in emission in a range of environments, such as photo-dissociation regions (PDRs), protoplanetary disks, planetary and reflection nebulae, the interstellar medium of Local Group galaxies, starburst galaxies, ULIRGs, and AGNs. The PAH line ratios reveal differences in properties such as grain charge, size and molecular structure, which are indicative of differences in local environment \citep[e.g.,][and references therein]{Tielens2008,Boulanger1996,Meeus2001,Smith2007,Chown2024,Schroetter2024}. Although they produce fewer spectral features, by mass, silicates are more abundant than carbon grains, comprising $\sim 2/3$ of the total grain mass \citep{Draine2007,Jones2017}. These grains primarily produce strong features near 10 and 18~$\mu$m associated with the Si-O stretching and O-Si-O bending modes respectively \citep{Draine2003, Galliano2018}.  However, the feature shapes and peak wavelengths can vary with grain chemistry, porosity, temperature, size and crystallinity \citep[e.g.,][]{Jager1998,Molster2003,Molster2005,Koike2003,Koike2006,Chiar2006,Dorschner1995}.

The mineralogy of the silicate grains in diverse environments can be investigated by comparison with laboratory spectra for different grains. In the MW diffuse ISM, the broad 10 and 18~$\mu$m features are consistent with amorphous silicates \citep[$\lesssim$1-5\% crystallinity,][]{Kemper2004,Li2007} of a predominately olivine (Mg$_{2x}$Fe$_{2-2x}$SiO$_4$) composition. By contrast, somewhat larger amounts of silicate crystallinity are found in environments ranging from comets to AGB outflows to AGN to protoplanetary disks to star-forming galaxies \citep[e.g.,][and references therein]{Molster2003,Molster2005,Watson2009,Duy2019,Tsuchikawa2022,Srinivasan2017,Spoon2006,Stierwalt2014}. The absence of such crystallinity in the MW diffuse ISM has been attributed to physical mechanisms such as heavy ion cosmic ray bombardment, which can amorphize crystalline silicates within $\sim$70~Myr \citep{Bringa2007}. In environments other than the MW diffuse ISM, such as stellar debris disks, more complex mineralogies such as phyllosilicates, as well as pyroxenes (Mg$_x$Fe$_{1-x}$SiO$_3$), are also sometimes found \citep{Currie2011,Mittal2015}. While silicate dust in the MW diffuse ISM is comparatively homogeneous, \citet{Decleir2025} still find significant variations in the peak wavelength and symmetry of the silicate absorption features along stellar sightlines, which can be attributed to mineralogical variations. Previous studies of quasar absorption systems (QASs), in both dense and diffuse sightlines, have found an even greater diversity of silicate absorption feature shapes, substructures, and peak wavelengths than seen in MW interstellar sightlines, which may be indicative of a larger diversity of grain mineralogies compared with the MW diffuse ISM \citep{Kulkarni2011,Aller2012,Aller2013,Aller2014,Kulkarni2016}. This potential dissimilarity with the local interstellar dust is a primary driver of this study, wherein we can now utilize higher-spectral resolution and signal-to-noise JWST Mid Infrared Instrument (MIRI) - medium resolution spectrometer (MRS) spectra to investigate whether silicate grains indeed differ in some higher-redshift systems, as assessed using the 10~$\mu$m silicate feature.

With the goal of improving direct, empirical knowledge about dust grains in distant galaxies, we have been studying dust spectral features in galaxies situated along the lines of sight to background quasars/AGNs with SST Infrared Spectrograph (IRS) and JWST MIRI MRS \citep[e.g.,][Klimenko, Kulkarni, \& Aller 2025a, submitted]{Kulkarni2007, Kulkarni2011, Aller2012, Aller2013, Aller2014}. Our previous studies based on SST IRS reported hints of differences between 
the silicate grains in distant galaxies and those in the MW, but they lacked the sensitivity or spectral resolution to give robust results on grain composition for many sources. Although it is unable to cover the 18~$\mu$m feature for our higher-redshift targets, the JWST MIRI MRS now offers a huge improvement in both sensitivity and spectral resolution compared to the SST IRS, promising to revolutionize studies of dust grain composition in distant galaxies, which are crucial for constraining models of grain evolution.

This paper follows upon Klimenko et al. 2025a (hereafter Paper I), wherein we presented our sample selection for the JWST observations, the analysis of the MIRI MRS spectra, and preliminary results on the silicate features in our sample of gas-rich absorber galaxies with strong 2175~{\AA} bumps. 
In this paper we present fits to the silicate features with various laboratory and astrophysical templates, and we compare the strength and shape of these features with extinction parameters and gas-phase metal abundances in samples of quasar sightlines and in the local ISM.
Our results, combined with those for local galaxies, offer important empirical input for models of dust grain processing and evolution. 
This paper is organized as follows: 
Section~\ref{sec:data} summarizes the data analyzed.  Section~\ref{sec:silicate_fits} presents the fitting to the 10~$\mu$m silicate feature. Section~\ref{sec:extinction} presents the analysis of extinction curves of foreground galaxies using archival and new observations. 
Section~\ref{sec:dicussion} describes our results for the 10~$\mu$m silicate feature and the extinction curve parameters, and compares them with results for the MW ISM. Section~\ref{sec:conclusions} summarizes our conclusions. 

\section{Data}
\label{sec:data}

\begin{figure*}
\begin{center}
        \includegraphics[width=1\textwidth]{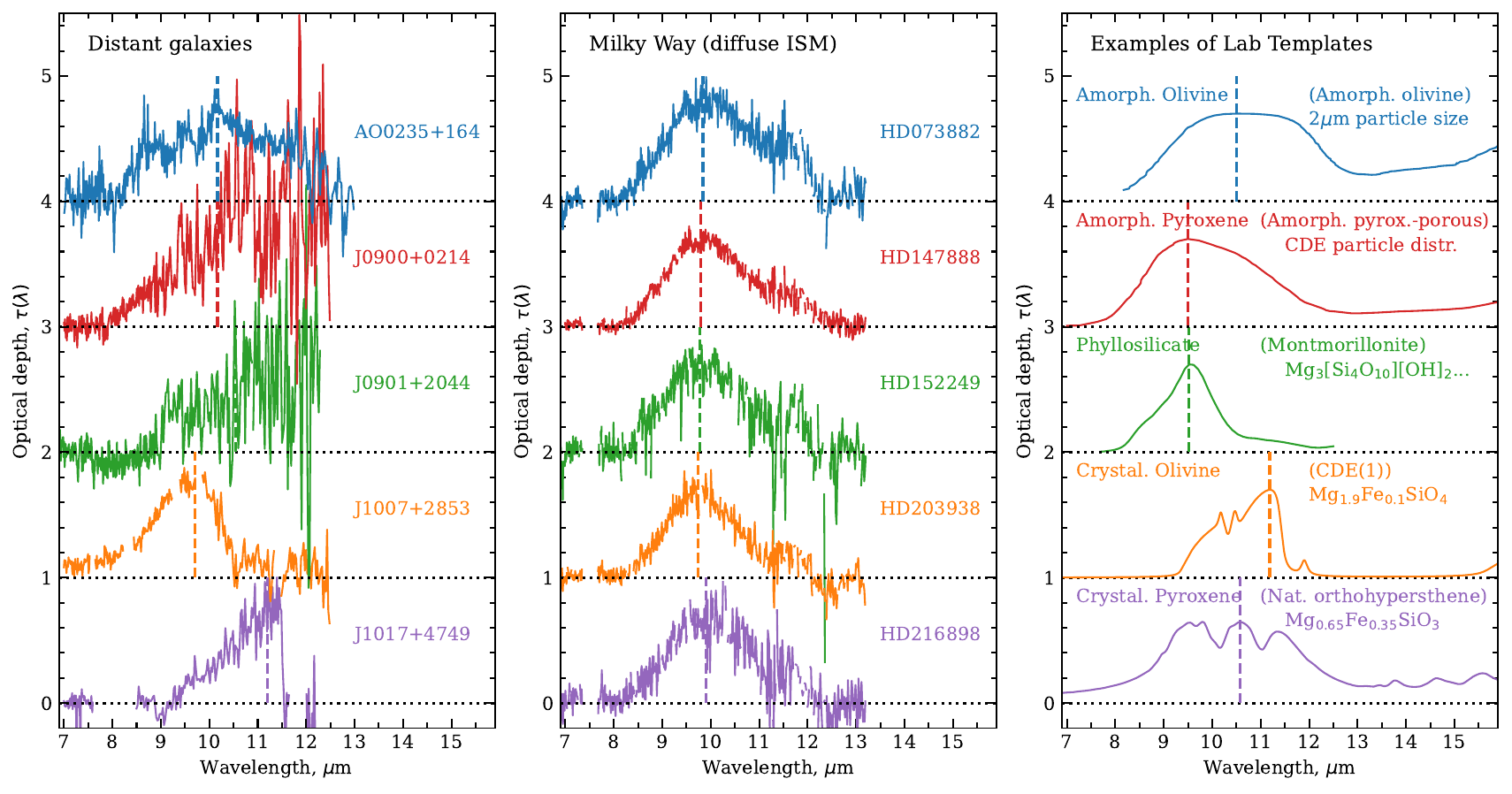}
        \caption{Comparison of 10~$\mu$m silicate optical depth profiles for our sample (left panels, MIRI MRS), a part of the MW sample (center panels, MIRI MRS, \citealt{Decleir2025}), and a subset of the laboratory-based templates (right panels,  \citealt{Aller2012}). Each profile is normalized to a peak optical depth of 0.7, and offset vertically by +1 units. The JWST data is binned by a factor of 10 for visual clarity. The vertical dashed line marks the peak wavelength position, as determined from the best-fit model. The legends in the right panels show the template names along with a brief description; see details in \cite{Aller2012}. All spectra are shown in the absorption rest frame.
        }
        \label{fig:tau-profiles-comparison}
\end{center}
\end{figure*}

The sample of quasars from our JWST program consists of 5 targets with intervening absorption systems at redshift $z=0.5-1.2$, presenting strong Mg~{\sc ii} absorptions and prominent 2175~\AA\ bump features. The program strategy, sample selection, data acquisition, and data reduction are presented in Paper~I. 
The quasar spectra were extracted from the MIRI MRS data cubes and normalized to the AGN continua, which have been individually constructed for each source. We detected the 10~$\mu$m silicate absorption feature in all five quasar sightlines at a significance level greater than $> 20\sigma$. The left panels of Figure~\ref{fig:tau-profiles-comparison} show the optical depth profiles of the 10~$\mu$m silicate absorption sources, calculated as $\tau(\lambda)=\ln\left(F_{\rm cont}(\lambda)/F_{\rm obs}(\lambda)\right)$. 
In Section~\ref{sec:silicate_fits}, we present the analysis of these 10~$\mu$m absorption profiles.

In Sections~\ref{sec:extinction} and \ref{sec:dicussion}, we extend our sample by including four quasar sightlines with strong 10~$\mu$m absorption features observed with the Spitzer IRS in our previous studies \citep{Kulkarni2007, Kulkarni2011, Aller2012, Aller2013, Aller2014}, allowing for a more comprehensive analysis of dust in distant galaxies.

\section{Analysis of Silicate Feature}
\label{sec:silicate_fits}

\subsection{Diversity of Overall Profile Shapes}

The 10~$\mu$m silicate profiles in our sample of quasar absorber sightlines show a wide variety of basic properties. Basic analysis of the optical depth profiles (see left panels in  Figure~\ref{fig:tau-profiles-comparison}) shows that (i) their peak wavelength ranges from 9.7~$\mu$m to 11.2~$\mu$m; (ii) their FWHM spans from 1.3~$\mu$m to 3~$\mu$m; and (iii) their degree of asymmetry varies, with some features showing a gentler decline on the short-wavelength side compared to the long-wavelength side. 

By contrast, the silicate features observed in the MW diffuse ISM with the SST IRS \citep{Shao2018, Gordon2021, Shao2024} and the JWST MIRI MRS \citep{Decleir2025} show more consistent profiles with diversity on a much smaller scale (see middle panels in Figure~\ref{fig:tau-profiles-comparison}). For instance, \cite{Shao2018}, \cite{Gordon2021} and \cite{Shao2024} found the peak wavelength of the 10~$\mu$m silicate feature to be in the range of 9.69 to 9.87~$\mu$m, 9.64 to 9.94 $\mu$m, and 9.49 to 9.96 $\mu$m, respectively, along the sight lines of samples of 5 OB stars, 15 OB stars and 49 stars of O, B, F, G, K, and M spectral classes. In a sample of 9 OB stars \cite{Decleir2025}, found peak wavelengths ranging from 9.47 to 9.90~$\mu$m, and a range of FWHMs from 1.73 to 1.94~$\mu$m. 

The right panels in Figure~\ref{fig:tau-profiles-comparison} illustrate laboratory-based templates for some of the minerals that could be observed in the ISM and other astrophysical environments \citep{Aller2012}. Crystalline silicates typically exhibit multiple sharp features, whereas amorphous silicates produce much smoother profiles that can still vary significantly in their asymmetry and peak position. It is clear that the laboratory templates show a wide variety of profiles, similar to the silicate profiles in quasar absorber sightlines. The variation in the silicate profiles in extragalactic sightlines versus those in the MW ISM sightlines may indicate differences in the chemical composition and morphology of dust grains in distant galaxies. 

\subsection{Different Approaches to Characterizing Silicate Strength}
Different studies have characterized the silicate strength in different ways. Our past studies \citep[e.g.,][]{Kulkarni2007,Kulkarni2011, Aller2012,Aller2014}, as well as several past studies by other groups \citep[e.g.,][]{Roche1984,Hanner1995,Spoon2006}, characterize the silicate strength in terms of the {\it peak optical depth}, $\tau_{10}$, derived from {\it profile fitting} of the 10 $\mu$m feature with observed/laboratory templates. Other studies 
express the strength in terms of $S_{sil}=\ln (f_{c}/f_{10})$, with $f_{c}$ and $f_{10}$ denoting the continuum flux and the  observed flux at the peak depth of the feature \cite[see, e.g.,][]{Hao2007}. A small difference between $\tau_{10}$ and $S_{sil}$ may arise if the best-fitting template does not perfectly match the observed profile (e.g., due to an instrumental artifact in the data). 

Another approach is based on deriving the peak optical depth from fitting with  theoretical profiles 
\citep[e.g.,][]{Gordon2021, Decleir2025}.
\cite{Gordon2021} fitted Spitzer photometry and spectroscopy of OB stars to determine MIR extinction curves consisting of a power law with the 10 and 18 $\mu$m silicate features approximated as modified Drude profiles. They characterized the silicate strength in terms of $S_{1}=2.5\log(e)\times\tau_{10}/A_V=1.086\times\tau_{10}/A_V$, the peak amplitude of the 10~$\mu$m silicate feature contribution to the extinction curve $A_{\lambda}/A_{V}$.  \cite{Decleir2025} fitted the 10~$\mu$m silicate feature in continuum-normalized  spectra of OB stars, and found better consistency with a skewed Gaussian profile than with a modified Drude profile.\footnote{As explained in \cite{Decleir2025}, this may be explained by the uncertainty in the continuum level near $\sim13~\mu$m, in the region where the continuum can be affected by the short-wavelength tail of the 18~$\mu$m silicate feature. However, because of increasing thermal noise in the longer-wavelength region of MIRI MRS spectra, 
the 18~$\mu$m silicate feature cannot be constrained directly from their spectra.}
We note that the Drude profile has an advantage over the Gaussian profile in its physical motivation for describing dust features, as it is directly related to a simple characterization of the optical properties of solids \citep[e.g.,][]{Massa1989}.

For the sake of consistency with our past work and several other literature studies, we characterize the silicate strength in terms of the {\it peak optical depth}  $\tau_{10}$. To account for possible systematic effects, due to the fitting method, we derive $\tau_{10}$ by fitting the the silicate profiles using the observed/laboratory templates, as well as by using theoretical profiles. However, the results for $\tau_{10}$ are found to be consistent across these different fitting methods.

\subsection{Laboratory/Astrophysical dust absorption templates}
\label{sec:silicate-templates}
We consider the observed/predicted optical depth profiles produced by a range of laboratory-based and modeled grain distributions and mineralogies, all of which we term {\it laboratory} templates in our analysis and discussion. 
We also compare our measured silicate feature absorption profiles with observed astrophysical optical depth profiles from a range of Solar System, Galactic, and extragalactic sources, which we term {\it astrophysical} templates. These laboratory and astrophysical templates are drawn from a sample of $>60$ such templates which were compiled for, and utilized in, a similar analysis
of the dusty absorber toward PKS~1830-211 by \citet{Aller2012}.

Laboratory templates span a range of grain chemical compositions, crystallinities, temperatures, and grain sizes. The majority of the laboratory templates are types of olivines, due to the fact that most previous analyses of extragalactic sources \citep[e.g.,][]{Kulkarni2010,Aller2012,Aller2014,Spoon2006} have identified
olivine compositions as best matching the observed data. Among the olivine grain varieties included in our template library are {\it 6 amorphous olivine} profiles, varying in particle shape and size, {\it 7 crystalline olivine} ($Mg_{2x}Fe_{2-2x}SiO_4$) species which vary in composition (i.e., $x$), {\it 4 crystalline olivine} ($Mg_{1.9}Fe_{0.1}SiO_4$) profiles which vary in particle shape, {\it 4 crystalline forsterite} ($Mg_2SiO_4$) species varying in temperature, and {\it 4 crystalline fayalite} ($Fe_2SiO_4$) species varying in temperature. We also include {\it 2 species of amorphous pyroxene}, an amorphous silica ($SiO_2$), {\it 5 species of  crystalline pyroxene} ($Mg_xFe_{1-x}SiO_3$) which vary in chemical composition (i.e., $x$), {\it 5 species of phyllosilicates}, two other blends of silicates, and four species of SiC. In total, 44 different templates are explored. We note that the largest differences between templates occur with variations in chemical composition and crystallinity, with more modest differences associated with differences in grain temperature. While the crystalline templates have significant substructure, as illustrated in \citet{Aller2012}, the amorphous templates are generally broader and more featureless, although the grain particle shape and size can affect the breadth of the feature and its peak absorption wavelength. Throughout our analysis, we utilize the same abbreviations to describe the profiles associated with these grain species as detailed in \citet{Aller2012} Table 10.
 
The astrophysical templates include {\it 2 Solar System comets} (Bradfield \& Levy),  1 Galactic template representing the {\it diffuse ISM} (WR118), a template for the Galactic Center GCS3, 4 sightlines probing {\it molecular clouds} or cloud complexes (Trap, Tau16, IC5146, $\rho$Oph),  2 Galactic templates from sightlines probing {\it T Tauri stars embedded in a molecular cloud} (emb-Tau, emb-$\rho$Oph),  2 templates  representing {\it stellar material} ($\mu$Cep, AGB), and { 3} templates from {\it sightlines to ULIRGs} that exhibit silicate grain crystallinity (ULIRG 00397-1312, ULIRG 06301-7934, and ULIRG 18443+7433). Additionally, we consider versions of these ULIRG templates in which the 11.3~$\mu$m PAH emission feature, which we do not see in our sources, is masked in the template; these are referred to as the ``trim [t]'' ULIRG templates. In total, there are 18 such astrophysical template versions considered in our fitting, with details about these templates provided in Table 8 of \citet{Aller2012}. We note that although matching such astrophysical templates to our data may not provide insight into the grain properties, it allows us to determine whether our observed profiles match those seen in other astronomical contexts, which may provide insight into the physical conditions in the absorber
sightline. 

The templates are presented in the format of the optical depth profile $\tau_{\rm norm}^{\rm template}(\lambda)$, normalized to have a maximum peak depth of 1.0 over the full spectral extent of the profile. Our absorption model assumes simple radiative transfer through the cloud, such that 
\begin{equation}
I(\lambda)/C(\lambda) = \exp\left[-\tau_{10}\cdot\tau_{\rm norm}^{\rm template}(\lambda)\right], 
\end{equation}
where $C(\lambda)$ is a quasar continuum. 

\subsection{Theoretical profiles}
We follow the approach of fitting the 10~$\mu$m silicate feature with the modified Drude profile and skewed Gaussian profiles used by \cite{Gordon2021} and \cite{Decleir2025}, respectively, in their analyses of silicate features in MW sightlines. As shown in Paper~I, the quasar MIR spectra show a significant variation depending on AGN morphology, redshift, and black hole mass. This complicates applying the method used  \citet{Gordon2021} for consistently fitting silicate features and continuum extinction. 
Therefore, we follow \cite{Decleir2025}, and fit the 10~$\mu$m silicate feature in the continuum-normalized quasar spectra using the theoretical profiles as follows:
\begin{equation}
I(\lambda)/C(\lambda) = \exp\left[-\tau_{10}\cdot \tau^{\rm theor}(\lambda)\right], 
\end{equation}
where $\tau^{\rm theor}(\lambda)$ is the modified Drude or the skewed Gaussian profile.
The modified Drude profile is parametrized as:
\begin{equation}
\tau^{\rm theor}(\lambda) = \frac{(\gamma/\lambda_0)^2}{\left(\lambda/\lambda_0 - \lambda_0/\lambda\right)^2+(\gamma/\lambda_0)^2},
\qquad
\gamma=\frac{2\gamma_0}{1+\exp[a\left(\lambda-\lambda_0\right)]}.,
\end{equation}
where $\lambda_0$, $\gamma_0$, and $a$ are fitting parameters representing the central wavelength, the unmodified width, and the asymmetry parameter, respectively. The skewed Gaussian profile is defined as:
\begin{equation}
\tau_{\rm theor}(\lambda) = N\frac{1}{\sqrt{2\pi}\omega}e^{-\frac{(\lambda-\xi)^2}{2\omega^2}}\left[1+{\rm erf}\left(\alpha\frac{\lambda-\xi}{\sqrt{2}\omega}\right)\right],
\end{equation}
where  ${\rm erf}$ is the error function,  $N$ is the normalization coefficient used to normalize $\tau_{\rm theor}(\lambda)$ to unity at the peak, and $\xi$, $\omega$, and $\alpha$ are additional fitting parameters representing the location and width of the peak, and asymmetry. As shown by \cite{Decleir2025}, the peak wavelength $\lambda_0$ and the profile width should be corrected for the profile's asymmetry, and can be calculated using their Eq.~6 and the FWHM.

\subsection{Correction for quasar spectral features}
The quasar continua were derived in Paper~I. For four quasars (J0900+0214, J0901+2044, J1007+2853, and J1017+4749), the 10~$\mu$m silicate absorption of the foreground galaxy overlaps with broad emission features from dust in the AGN (6.2~$\mu$m and 7.7~$\mu$m PAH lines or 10~$\mu$m silicate dust in emission). In these cases, a consistent fit to both the foreground silicate absorption and AGN emission features is essential for accurately determining the feature profiles.
Given that dust in the AGN is likely at a high temperature, we assume that it has a predominately amorphous dust grain structure, rather than a crystalline one. We, therefore, use an amorphous olivine template (from \citealt{Spoon2006}) to model the silicate dust emission line. The profile of the silicate emission line is calculated as $\exp(\tau^{\rm em}\cdot\tau_{\rm norm}^{\rm SpoonAmOliv}(\lambda)$, where $\tau^{\rm (em)}$  represents the strength of emission feature and is treated as a free parameter. In the case of the AGN PAH emission lines, we fit the lines with a Gaussian template with a dispersion of 0.2~$\mu$m centered at  the PAH wavelength (6.25~$\mu$m or 7.7~$\mu$m). 

We pay special attention to the potential for overfitting non-amorphous features in the silicate absorption profiles. For example, a real structure could be incorrectly modeled by introducing an emission feature stronger than what actually exists. We, therefore, consider two scenarios. 
In the first, the strengths of the AGN emission features are set the average values derived from the AGN composite spectrum, constructed for each source using SST/IRS spectra of AGNs with similar morphologies, redshifts, and black hole masses (see Paper~I). For J0900+0214, J0901+2044 and J1017+4749, the AGN composite shows pronounced PAH and silicate features,
and we just normalize the quasar spectra by the AGN composite spectrum. For J1007+2853, the AGN composite lacks any major spectral features, and we used a low-order polynomial fit with no silicate emission.

In the second scenario, we normalize the quasar spectra to the AGN continuum models without AGN emission features (see Paper~I), and then simultaneously fit the silicate absorption consistently with the AGN emission features.

Additionally, the quasar continuum around the 10~$\mu$m silicate feature may have small variations, for example, due to offsets in the continuum level. To account for such systematics, we
multiply the fitting model by a {\it scaling factor} of $(1+ \delta C)$, where  $\delta C$ is a wavelength-independent parameter
that can vary within one standard deviation of the normalized quasar flux at $\sim$7~$\mu$m in the absorption rest frame ($\delta C \leq3$\%, depending on the SNR of the data). 

To summarize, we fit the silicate feature in the MRS continuum normalized quasar spectra using a model:

\begin{equation}
\label{eq:sil_model}
M(\lambda) = A(\lambda, \tau_{10}^{\rm abs})\left(1+\delta C\right)\left(1+Em[\lambda, \tau^{\rm em}]\right),
\end{equation}
wherein  $A(\lambda, \tau_{10}^{\rm abs})$ is the silicate absorption profile shifted to $z_{\rm abs}$ and calculated using an astrophysical/laboratory template or a theoretical model, $(1+\delta C)$ is the continuum scaling factor, $Em[\lambda, \tau^{\rm em}]=\exp[\tau^{\rm em}\cdot\tau_{\rm norm}^{\rm em.template}(\lambda)]$ is the silicate or PAH emission profile shifted to $z_{\rm qso}$. 

\subsection{Fitting Results}
We use a Bayesian framework to estimate the posterior distribution functions for the fitting parameters. To sample posteriors, we use a Monte Carlo Markov Chain method (MCMC), as implemented in the python package {\sc emcee} \citep{Foreman2013}. Such a technique ensures that we confidently find the global maximum in the multi-variable parametric space, and provides reliable estimates of the statistical errors on the parameters. 
To report point and interval estimates on the parameters, we use maximum aposteriori probability and highest-posterior-density 68.3\% credible intervals, respectively. We note that reported uncertainties are statistical ones, derived for particular model assumptions.

\begin{figure*}
\begin{center}
    \includegraphics[width=1\textwidth]{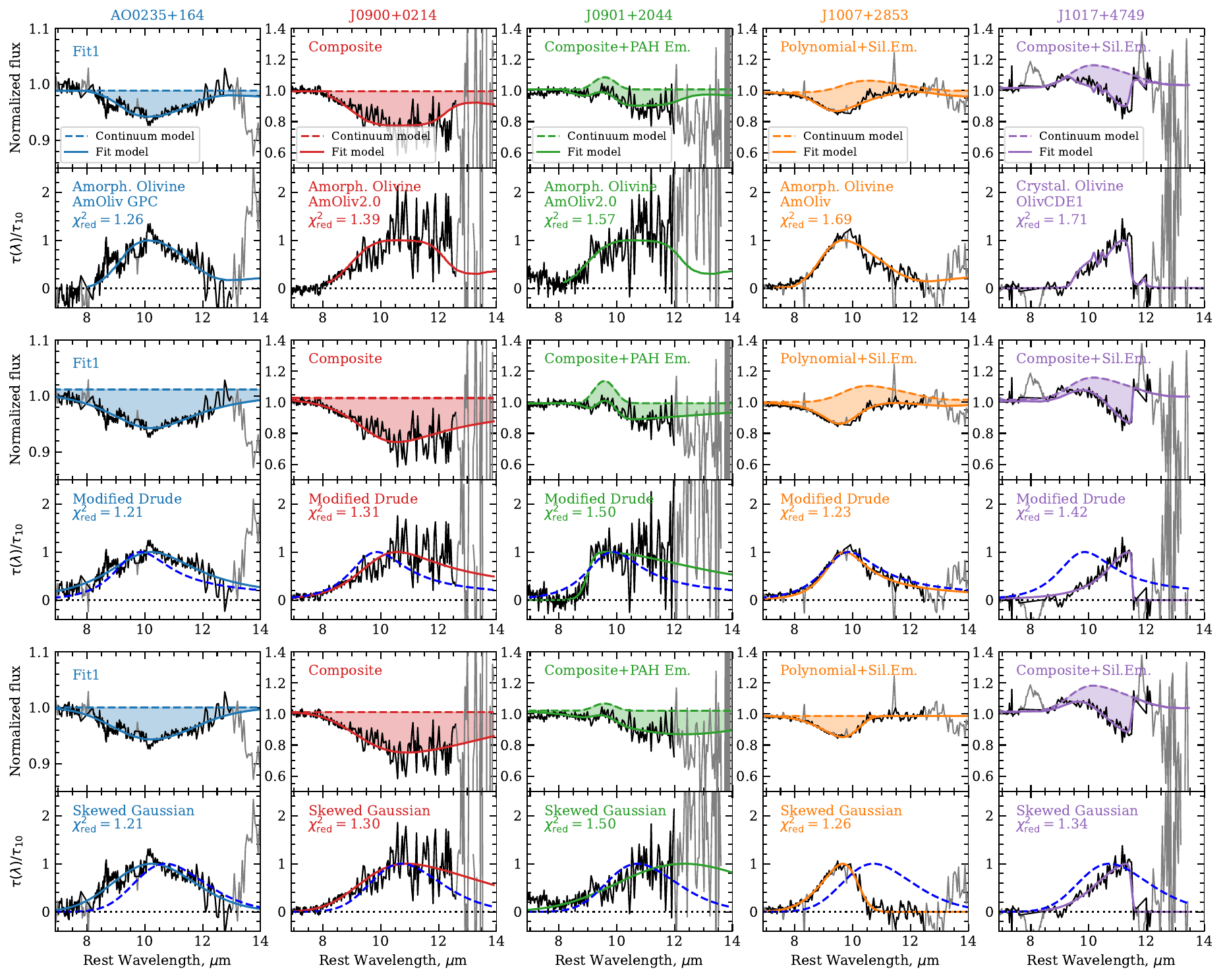}
    \caption{ Best fits to the 10~$\mu$m silicate features in quasars from our sample, derived using laboratory-based templates (first set of two rows), a modified Drude profile (second set of two rows), and a skewed Gaussian function (third set of two rows). For each set, we show fits to the normalized quasar spectra (top panel) and fits to the normalized optical depths (bottom panel). The MIRI MRS spectra are normalized to the same continuum template model (marked in the top left corner, see definitions in Paper~I). The dashed lines show the corrections for a local continuum by allowing different scaling parameters $(1+\delta C)$ and strengths of quasar emission features $(1+Em[\lambda, \tau^{\rm em}])$ - the PAH~6.2~$\mu$m and 10~$\mu$m silicate lines. Absorption profiles are shown by the solid lines and shaded areas. The wavelength is presented in the absorber rest frame, and the data are  binned by a factor of 7 for visual clarity. The spectral regions used for the fit are highlighted in black, while the remaining regions are shown in gray.
    The optical depth profiles are calculated using the corrected local continua and normalized to the peak optical depth $\tau_0$ (see Tables~\ref{tab:silictae-fit-results} and \ref{tab:extinction-results-2}). The template name and the corresponding values of the reduced $\chi^2_{\rm red}$  are shown in the top left corners of the bottom panels for each set.
    The blue dashed curves in the fourth and sixth rows show the model for the optical depth of a silicate feature in the MW sightlines, calculated using the median values of the parameters from \cite{Gordon2021} and \cite{Decleir2025}, respectively.}
    \label{fig:tau-profiles-best-fit}
\end{center}
\end{figure*}

We summarize the key results from our 10~$\mu$m profile fits for the astrophysical/laboratory templates and theoretical models in Table~\ref{tab:silictae-fit-results} and Table~\ref{tab:extinction-results-2}, respectively. 
Figure~\ref{fig:tau-profiles-best-fit} shows the fits to the silicate absorption and silicate optical depth profiles using the best-fit laboratory templates, as selected using the $\chi^2_{\rm red}$ criterion, and theoretical models. Other fits using the laboratory and astrophysical templates, which yield higher $\chi^2$ values, are shown in Appendix~\ref{app:b:fit-profiiles}.

Table~\ref{tab:silictae-fit-results} provides 
values for a reduced $\chi^2_{\rm red}$ criterion, $\tau^{\rm abs}_{\rm 10}$, and $\tau^{\rm em}$ (where the AGN emission feature was simultaneously fitted). For clarity, we only present results for the best-fitting template from each family of silicate templates in our library.\footnote{We show results for more reliable astrophysical templates (diffuse ISM, a Galactic center source, molecular clouds, embedded circumstellar material, comets and ULIRGs) and laboratory templates (Amorphous Olivine, Amorphous Pyroxene, Phyllosilicates, Crystalline Olivine, Crystalline Pyroxene and SiC). 
The results for the stellar material templates and silica are not shown. The stellar material is unlikely to be applicable to the dust probed along quasar sightlines. The fits using the amorphous silica template yield high $\chi^2$ values for all of our systems.}
The overall best-fitting mineralogies are identified in bold-face.
The last column in Table~\ref{tab:silictae-fit-results} shows the adopted value for $\tau^{\rm abs}_{\rm 10}$, determined by averaging the optical depths of the laboratory models with values of $\chi^2_{\rm red}\le\min(\chi^2_{\rm red})+0.2$. The uncertainty is calculated as the dispersion in the $\tau^{\rm abs}_{\rm 10}$ results, for those fits meeting our $\chi^2$ threshold, to estimate the systematic uncertainty associated with the template choice. 

We note that the adopted values of $\tau^{\rm abs}_{\rm 10}$ are derived from the fits using the laboratory templates, motivated by our goal to study silicate mineralogy, which remains poorly understood for some astrophysical templates. However, the comparisons with the astrophysical templates may potentially provide insight into the similarity of dust in distant galaxies to that in various astrophysical environments. We also note that for several of our sightlines, the $\chi^2_{\rm red}$ values for the best-fitting astrophysical templates are comparable to those for the best-fitting laboratory templates. 
In several of our sightlines (AO 0235+164, J0900+0214, J0901+2044), the best-fitting astrophysical templates are those for ULIRGs. However, the fits for comet templates are comparable in these sightlines. For J1007+2853, the Galactic ISM templates fit the best (with the diffuse ISM template fitting the best when the quasar silicate emission is considered). For J1017+4749, the 10 $\mu$m silicate profile has a very asymmetric shape that drops steeply on the red side, and is best-fitted by a comet template. The predominance of 
ULIRG templates in our sample is interesting, considering the indications of warmer dust and of resemblance to dust toward AGNs that we discuss in Section~\ref{sec:dicussion}. Similarity to comet dust is also interesting, but more difficult to understand given the vastly different physical conditions. However, we caution against overinterpreting the comparisons with astrophysical templates, especially given the small size of our sample and the limited SNR of our spectra. Higher SNR and higher spectral resolution observations for a larger number of quasar absorber sightlines would help to differentiate more reliably between the different astrophysical templates and to increase the statistical robustness of our findings.

Table~\ref{tab:extinction-results-2} shows values of the $\chi^2_{\rm red}$ and model fitting parameters for our theoretical fits. 
For most cases, we found $\tau_{10}^{\rm abs}$ to be consistent between the fits using the modified Drude and skewed Gaussian models, with differences for only 2 systems (J0901+2044 and J1007+2853), wherein the strength of AGN emission lines was varied. 
We also found the values of $\tau_{10}^{\rm abs}$ to be in good agreement with the adopted values of $\tau_{10}^{\rm abs}$ for fits using laboratory templates.
Although the peak optical depth values are similar for the fits using laboratory-based and theoretical models, the optical depth profiles show small variations in structure - mainly for J0900+0214 and J0901+2044. This may be explained by a degeneracy\footnote{The parameter likelihood distribution shows multiple peaks with similar probabilities.} in the parameter space due to low SNR near $\sim13~\mu$m in the absorber rest frame for these quasars ($\sim$4 in the MIRI MRS channel 4C) and the lack of data above $13~\mu$m.  

We also highlight the difference between our fits using theoretical models and the median of the fitted profiles for the silicate absorption observed in MW diffuse ISM sightlines, since \cite{Gordon2021} and \cite{Decleir2025} used the same fitting functions (the modified Drude and skewed Gaussian, respectively). The median models for the MW samples are shown by the blue dashed curve in the two bottom rows in Figure~\ref{fig:tau-profiles-best-fit}. 
We note differences between the extragalactic silicate profiles and the median models for each quasar absorber.
The difference is most evident in two systems, J1007+2853 and J1017+4749, where asymmetry of the silicate  profiles is opposite to the asymmetry observed in the MW diffuse ISM sightlines. This is also seen in the values of the asymmetry parameter in the theoretical models (see Table~\ref{tab:extinction-results-2}). The profiles show a steep decline on the long-wavelength side, potentially suggesting a significant fraction of crystallinity of silicate grains (see comments on these systems below), which is only $\leq1-5$\% in MW diffuse ISM sightlines \citep{Kemper2004, Li2007}.  
The median models for the MW samples also show variations, but on a significantly smaller scale (see the discussion in \citealt{Decleir2025}). For AO0235+164, J0900+0214, and J0901+2044, the median Drude model describes certain profiles better than the median Gaussian model, and vice versa.

For our sample, we do not observe a correlation between the peak wavelength and the width of the 10~$\mu$m feature, as detected for silicate features in MW sightlines \citep{Gordon2021, Shao2024}. This is likely due to small statistics and low SNR on the long-wavelength side, which may bias our results. We also note that for two objects in our sample (J0900+0214 and J0901+2044), we could not fit the width of the 10~$\mu$m feature reliably. 

\subsection{Comments for Individual Sources}
In the sections below, we provide comments on individual systems.

\subsubsection{AO~0235$+$164} 
In Paper~I, we showed that the spectrum of AO~0235$+$164 allows for two fits of the quasar continuum using a log-parabolic function, based on different fitting regions. Thus, we fit the silicate absorption using both versions of the quasar continuum. The best fit was obtained using the amorphous olivine template in each case, yielding 10~$\mu$m peak optical depths ($\tau_{10}$) of 0.05 and 0.06, respectively. Our adopted value is $0.06\pm0.01$, which is slightly lower than the $0.08 \pm0.02$ previously reported by \citet{Kulkarni2011}. However, the estimates are consistent within their uncertainties. Compared to the median models for the MW ISM, the silicate profile is significantly broader, and may exhibit sharp features that are not well-detected at the current SNR.

\subsubsection{J0900$+$0214}
The silicate absorption is the strongest in our sample, with a rest-frame equivalent width of 0.6~\AA. Unfortunately, this quasar is also the faintest in the sample, with a low SNR in MIRI MRS channel 4, decreasing from 13 to 3 at longer wavelengths, wherein the silicate absorption feature is detected. The best fit to the absorption is found for the amorphous olivine template with an optical depth of $\tau_{\rm 10}=0.25\pm0.01$. However, the crystalline olivine and crystalline pyroxene templates also provide a good fit with $\tau_{\rm 10}=0.51$ and 0.28. The absorption line may overlap with quasar PAH emission lines at 6.2~$\mu$m and 7.7~$\mu$m, although these lines are likely weak, as they are not prominent in the observed spectrum, Considering the low SNR, we limit our analysis to the simplest models with an absorption profile, and adopt a $\tau_{\rm 10}$ value of 0.30$\pm$0.10. Fits for the theoretical profiles are complicated by a lack of data at 13~$\mu$m. This affects the fitting results, because the width parameter ($\gamma_0$ or $\omega$) tends to have large values (often non-physical), which also biases the central wavelength. To correct for this, we constrain the width parameter to be 4, the maximum width found from laboratory templates. The peak optical depths for the fits using the modified Drude and skewed Gaussian profiles are $\tau_{\rm 10}=0.32\pm0.02$ and $0.30\pm0.02$, respectively, and are consistent with the adopted value.

\subsubsection{J0901$+$2044}
The foreground silicate absorption for this source overlaps with the AGN 6.2~$\mu$m PAH emission line. We consider two models for the quasar continuum: (1) the AGN composite continuum with a fixed PAH emission line strength (see details in Paper~I), and (2) a continuum component of the AGN composite spectrum, where we fit the PAH emission line strength together with the silicate absorption parameters.
In both cases, the silicate absorption can be well fitted using amorphous olivine and crystalline pyroxene/olivine (hypersthene/fayalite), with a slight preference for the amorphous model. This makes it difficult to reliably determine the silicate mineralogy in this quasar (mainly due to the low SNR in Channel 4, decreasing from 16 to 4).
The value of $\tau_{\rm 10}$ varies between 0.10 and 0.15. 
Thus, our adopted estimate is $0.13 \pm 0.03$. Fits with theoretical models give a value of $\tau_{10}=0.10\pm0.01$ for case~(1) and between 0.10 and 0.15 for case~(2). The fit in the case~(2) is shown in the third column in Figure~\ref{fig:tau-profiles-best-fit}. The best-fit strength of the quasar PAH emission varies slightly between fits with different absorption profiles $\tau^{\rm em}_{\rm PAH}\simeq0.07-0.15$, which introduces additional uncertainty in the derived silicate optical depth and profile shape. 

\subsubsection{J1007$+$2853}
J1007$+$2853 is the quasar with the most prominent foreground silicate absorption in our sample (see Paper~I). It is a bright quasar, and the silicate absorption falls within MIRI channel 3, where the SNR is $\sim60$, allowing for a detailed study of its mineralogy. However, because of a small difference in redshift between the quasar and the foreground absorber galaxy, the silicate 10~$\mu$m absorption overlays with the AGN 10~$\mu$m silicate emission line. Although the silicate emission line is not visible in the spectrum, it could be hidden by blending with the absorption line. Therefore, we consider two models: (1) without the AGN silicate emission line, and (2) including it, wherein the emission line is fitted consistently with the absorption line. 
In model (1), the phyllosilicate template fit is strongly preferred, resulting in $\tau_{10} = 0.16 \pm 0.01$. In model (2), a comparable fit is achieved with the amorphous olivine template, yielding a similar value of $\tau_{10} = 0.17 \pm 0.01$. 

The fit for model (1) is shown in Figure~\ref{fig:tau-profiles-best-fit}. It has a gentle decline on the short-wavelength side and a steep decline on the long-wavelength side. This contrasts sharply with a gentler decline on the long-wavelength side, which is typical for the amorphous silicates and the observed features in the MW ISM. 
The fit for model (2) is shown in Figure~\ref{fig:fit-silicate-j1007-em}. Including the emission feature in the fit allows us to correct for possible non-amorphous structure in the silicate absorption profile.

Fitting with theoretical profiles yields similar results. We find comparable values of $\tau_{10}$ and a change in the asymmetry of the silicate profile between models (1) and (2).

\subsubsection{J1017$+$4749}
J1017$+$4749 is the only quasar in our sample that exhibits strong dust emission in both the PAH and silicate lines. The silicate absorption feature produced by the foreground galaxy is primarily blended with the physically-unrelated, background-quasar, silicate emission line.  
Similar to J0901+2044, we fit the silicate absorption assuming two models for the quasar continuum: (1) - using the AGN composite (wherein the strength of AGN silicate emission is fixed), (2) - using a continuum component and an AGN silicate emission feature with variable strength.

The best fit using the laboratory-based templates is achieved for the crystalline olivine template,  yielding $\tau_{10}^{\rm abs} =0.29\pm0.01$ and $0.22\pm0.02$ for models (1) and (2), respectively.  Model (1) yields a higher value of $\tau_{10}^{\rm abs}$ and a higher $\chi^2$. This is likely due to an overestimated strength of the AGN silicate emission line in model (1). Therefore, we assume the best fit from model (2) is more appropriate. Among the astrophysical templates, a good fit is achieved only with the comet (Levy) template, which is known to have a crystalline silicate structure.

Fitting with the theoretical templates yields similar values of $\tau_{10}^{\rm abs}$ and $\tau^{\rm em}$ as fitting with the best-fitting, ``OlivCDE1'', template (model 2). Additionally, the silicate profile fits for this system show the highest degree of asymmetry among the observed profiles in our sample, perhaps consistent with crystallinity in the profile structure (see the rightmost panels in Figure~\ref{fig:tau-profiles-best-fit}). Unlike for the J1007+2853 absorber, the non-amorphous structure cannot be explained by a contribution from unresolved AGN emission lines or an instrumental artifact.

\setlength{\tabcolsep}{1pt}
\begin{table*}
\begin{center}
\caption{Fit results to the 10~$\mu$m absorption of foreground galaxies using astrophysical/laboratory templates.}
\begin{tabular}{|l|c|c|c|c|c||c|c|c|c|c|c|c|}
\hline
   & \multicolumn{11}{c|}{Silicate 10~$\mu$m Templates} &\\
   &  \multicolumn{5}{c}{Astrophysical}&  \multicolumn{6}{|c|}{Laboratory-based} &\\
   \hline
   Temp& Diffuse & Mol. & Embed & Extragal & Comet & Amorph.& Amorph. & Phyllo & Crystal. &Crystal. &Lab& Adopted\\

   late& ISM & Clouds & Circum & ULIRG &  & Olivine & Pyroxene & Silicate & Olivine&  Pyroxene &SiC& value\\
   \hline
\multicolumn{13}{|c|}{AO0235+164 (Silicate Absorption + Fit1)}\\
\hline
Name                 & WR118  & IC5146 & emb-$\tau$ & {\bf UL06301}& Levy        &  {\bf AmOliv}& AmP     & {Chamo} & Horton&  {Avg}     &B0-black&\\
                     &        &        &         & {\bf trim}   &             &  {\bf GPC   }& GPC     & {site}    & olite &  {Pyrox}   &aSiC&\\
$\chi^2_{\rm red}$    & 1.55  & 1.45 & 1.47      & {\bf 1.30}   & { 1.46}        &  {\bf 1.26} & 1.58    & {1.37}      & 2.13  &  { 1.46}  &3.77&\\
$\tau^{\rm abs}_{10}$ & 0.05  & 0.06 & 0.05      & {\bf 0.05}   & 0.05        &  {\bf 0.05} & 0.05    & {0.06}     & 0.06 &  {0.05}     & 0.05 & 0.06$\pm$0.01\\
\hline
\multicolumn{13}{|c|}{AO0235+164 (Silicate Absorption + Fit2)}\\
\hline
Name                  & WR118 & IC5146 & emb-$\tau$ & {\bf UL06301} &Brad  &  {\bf AmOliv}  & AmPyr & Chamo & Faya  &  Hyperst &B0-black&\\
                      &       &        &         & {\bf trim}    & field&  {\bf 2.0}     & GPC   & site    & lite    &  hene    &aSiC& \\
$\chi^2_{\rm red}$    & 2.84  & 1.95   & 2.12    & {\bf 1.47}    & 2.03 &  {\bf 1.25}    & 2.66  & 2.40   & 2.48   &  1.72    &5.78& \\
$\tau^{\rm abs}_{10}$ & 0.08  & 0.08   & 0.07    & {\bf 0.07}    & 0.08 &  {\bf 0.06}    & 0.07  & 0.09   & 0.09   &  0.08    &0.09 & 0.06$\pm$0.01 \\
\hline
\multicolumn{13}{|c|}{J0900$+${\bf 0214} (Silicate Absorption+Composite AGN Fit)}\\
\hline
Name                  & GCS3  & Tau16  & emb-$\tau$       & {\bf UL00397}  & Brad    &  {\bf AmOliv}       & AmPyr  & Chamo    & {Forst}   & {Hyperst}  & AJ& \\
                      &       &        &        &  {\bf trim}    & field   &  {\bf 2.0}         & GPC    & site   & {T300}    & {hene}     & aSiC& \\
$\chi^2_{\rm red}$    & 1.63  & 1.68   & 1.69      & {\bf 1.47}     & 1.56    & {\bf 1.39}       & 1.68   & 1.72    & {1.59}    & {1.55}     & 2.22& \\
$\tau^{\rm abs}_{10}$ & 0.26  & 0.16   & {\bf 0.16}     & {\bf 0.22}      & 0.33    &  {\bf 0.25}      & 0.18   & 0.31    & {0.51}    & {0.28}     & 0.68 & 0.30$\pm$0.10\\
\hline
\multicolumn{13}{|c|}{J0901$+$2044 (Silicate Absorption+Composite AGN Fit)}\\
\hline
Name                  & WR118 & IC5146       & emb-$\rho$  & {\bf UL18443}   &  Brad  & {\bf AmOliv}              & AmPyr       & {Chamo}    & {Forst}   &  {\bf Hyperst}  &{Lonza} &\\
                       &       &             & Oph         & {\bf trim}      &  field & {\bf 1.5}                 & GPC         & { site}     & {erite}    &  {\bf hene}   &{aSiC} &\\
$\chi^2_{\rm red}$    & 1.40   & 1.36        & 1.44        & {\bf 1.30} &  1.34  & {\bf 1.33}   & {1.59}      & {1.39}     & {1.42}    &  {\bf1.31}    & 1.85 & \\
$\tau^{\rm abs}_{10}$ & 0.11  & 0.12         & 0.10       & {\bf 0.11}  &  0.15  & {\bf 0.11}   & {0.09}      & {0.12}     & {0.20}   &  {\bf 0.13}   & 0.19 & 0.13$\pm$0.03\\
\hline
\multicolumn{13}{|c|}{J0901$+$2044 (Silicate Absorption+AGN PAH~6.2~$\mu$m Emission )}\\
\hline
Name                  & WR118 & IC5146       & emb-$\rho$ & {\bf UL06301} &  Brad    & {\bf AmOliv}  & AmPyr       & {Chamo}    & { \bf Faya} &  {\bf Hyperst}  &{G0-green}&\\
                      &       &             & Oph         & {\bf trim}    &  field    & {\bf 2.0}    & GPC         & {site}     & {\bf lite}    &  {\bf hene}   &{ aSiC}&\\
$\chi^2_{\rm red}$    & 1.73  & 1.68         & 1.78       & {\bf 1.59 }   &  1.61    & {\bf 1.57}   & {1.89}      & {1.70}     & {\bf 1.60}    &  {\bf 1.62}    & {1.66}&\\
$\tau^{\rm abs}_{10}$ & 0.12 & 0.12          & 0.08       & {\bf 0.11}    &  0.14       & {\bf 0.10}   & {0.07}     & {0.12}    & {\bf 0.15}   &  {\bf 0.13}   & { 0.18}& 0.13$\pm$0.03\\ 
$\tau^{\rm em}_{PAH}$  & 0.09  & 0.02          & 0.07       & {\bf 0.08}   &  0.08     &  {\bf 0.07}  & 0.03      & {0.09}    & {\bf 0.10}     &  {\bf 0.08}   &{ 0.00} &  \\
\hline
\multicolumn{13}{|c|}{J1007$+$2853 (Silicate Absorption+Low-Order Polynomial Fit)}\\
\hline
Name                  & GCS3  & $\rho$Oph       & emb-$\tau$     & UL00397 & Brad  & AmOliv & AmPyr  & {\bf Mont} & Faya   &  Avg    &G0-green &\\
                      &       &                 &               &          & field &        & GPC    & {\bf mor}   &  lite  &  Pyrox  &aSiC &\\
$\chi^2_{\rm red}$    & 2.74  & {\bf 2.29}      & 3.19           & 3.39    & 6.55  & 2.50   & 2.67   & {\bf 1.89}    & 10.3   &  2.67   &25.8&\\
$\tau^{\rm abs}_{10}$ & 0.15  & {\bf 0.14}      & 0.11           & 0.13    & 0.19  & 0.14   & 0.11   & {\bf 0.16}    & 0.20   &  0.12   &0.16& 0.16$\pm$0.01\\ 
\hline
\multicolumn{13}{|c|}{J1007$+$2853 (Silicate Absorption + AGN Silicate Emission)}\\
\hline
Name                  & {\bf WR118}    & {IC5146}      & emb-$\rho$    & UL00397 & Brad  & {\bf AmOliv}     & AmPyr & { Mont}      & Faya   &  Hyperst &G0-green&\\
                      &           &             &  Oph              &         & field &            & GPC   &    { mor}          &  lite    &  hene    &aSiC&\\
$\chi^2_{\rm red}$    & {\bf 1.58}  & {1.84}      & 2.30        & 2.40    & 3.41  & {\bf 1.69} & 2.56  & {1.87}    & 10.3   &  2.12    &27.5&\\
$\tau^{\rm abs}_{10}$ & {\bf 0.21} & {0.19}       & 0.18         & 0.16    & 0.35  & {\bf 0.17} & 0.12  & {0.17}   & 0.20    &  0.25    &0.05 &0.17$\pm$0.01 \\
$\tau^{\rm em}_{10}$ & {\bf 0.10}  & {0.10}       & 0.08         & 0.09    & 0.22  &  {\bf 0.07}& 0.00  & { 0.01}   & 0.00    &  0.12    &0.00& \\
\hline
\end{tabular}
\label{tab:silictae-fit-results}
\begin{tablenotes}
      \small
     \item{ {\bf Notes:} The best-fitting templates are shown in {\bf boldface}. In the last column, we present the average and dispersion of $\tau_{10}^{\rm abs}$ for the laboratory templates with values of $\chi^2_{\rm red}\le \min(\chi^2_{\rm red}) + 0.2$.
     }
\end{tablenotes}
\end{center}
\end{table*} 

\setlength{\tabcolsep}{1pt}
\begin{table*}
\begin{center}
\caption*{Table 1 (continued)}
\begin{tabular}{|l|c|c|c|c|c||c|c|c|c|c|c|c|}
\hline
   & \multicolumn{11}{c|}{Silicate 10~$\mu$m Templates} &\\
   &  \multicolumn{5}{c}{Astrophysical}&  \multicolumn{6}{|c|}{Laboratory-based} &\\
   \hline
   Temp& Diffuse & Mol. & Embed & Extragal & Comet & Amorph.& Amorph. &  Phyllo & Crystal. &Crystal. &Lab& Adopted\\

   late& ISM & Clouds & Circum & ULIRG &  & Olivine & Pyroxene & Silicate & Olivine&  Pyroxene &SiC& value\\
   \hline
\multicolumn{13}{|c|}{J1017$+$4749 (Silicate Absorption+Composite AGN Fit)}\\
\hline
Name                  & WR118  & IC5146         & emb         & {UL06301}   & {\bf Levy}  &  AmOliv & AmPyr & Chlor   & {\bf Oliv}    & Hyperst & B0-black & \\
                      &        &                & $\rho$Oph   & {trim}     &              &  2.0    & GPC   & ite     & {\bf CDE1}    & hene    & aSiC & \\
$\chi^2_{\rm red}$    & 7.09   & 6.90           & 7.27        & {5.32}    & {\bf4.28}     &  4.55   & 7.38  & 5.35    & {\bf 1.86}   & 5.50   & 4.50 & \\
$\tau^{\rm abs}_{10}$ & $0.11$   & $0.12$ & $0.10$         &  0.13     & {\bf0.18} &  0.13  & 0.09 & 0.23    & {\bf 0.29}   & $0.16$   & 0.33 & 0.29$\pm$0.01\\
\hline
\multicolumn{13}{|c|}{J1017$+$4749 (Silicate Absorption + AGN Silicate Emission)}\\
\hline
Name                  & WR118  & IC5146         & emb         & {UL06301}   & {\bf Levy}  &  AmOliv & AmPyr & Chlor   & {\bf Oliv}    & Bronz & Lonza & \\
                      &        &                & $\rho$Oph   & {trim}     &              &  2.0    & GPC   & ite     & {\bf CDE1}    & ite    & aSiC & \\
$\chi^2_{\rm red}$    & 2.67   & 2.67           & 2.67        & {2.60}    & {\bf1.99}     &  2.35   & 2.73  & 2.49    & {\bf 1.71}   & 2.63   & 2.24 & \\
$\tau^{\rm abs}_{10}$ & $<0.02$   & $<0.02$ & $<0.02$         & {$<0.02$}     & {\bf0.38} &  0.17  & $<0.02$& 0.05    & {\bf 0.22}   & $<0.02$   & 0.16 & 0.22$\pm$0.02\\
$\tau^{\rm em}_{10}$ & 0.06   & 0.07             & 0.030      & {0.07}    & {\bf0.34}  &  0.21  & 0.07    & 0.06   & {\bf 0.13}   & 0.07 & 0.05 & \\
\hline
\end{tabular}
\end{center}
\end{table*} 

\setlength{\tabcolsep}{1pt}
\begin{table*}
\small
\caption{Fit results to the 10~$\mu$m absorption of foreground galaxies using theoretical models.}
\label{tab:extinction-results-2}
\begin{center}
\begin{tabular}{|l|c|c|c|c|c|c|c|c|c|c|c|c|}
\hline
  Quasar & $\chi^2_{\rm red}$ & $\tau_{\rm 10}^{\rm abs}$& $\tau^{\rm em}$ & $\lambda_0$ & $\gamma_{0}$ & $a$ &$\chi^2_{\rm red}$ & $\tau_{\rm 10}^{\rm abs}$ & $\tau^{\rm em}$ &$\lambda_0$ & FWHM & $\alpha$\\
  &  & & ($\mu$m) & ($\mu$m)  & &  & &($\mu$m) & ($\mu$m) &  ($\mu$m) & &\\
\hline
& \multicolumn{5}{c|}{Modified Drude model} &  \multicolumn{7}{c|}{Modified Gaussian model}\\
\hline
AO0235+164 & $1.21$ & $0.07^{+0.01}_{-0.01}$ & - & $10.22^{+0.03}_{-0.04}$ & $3.82^{+0.15}_{-0.14}$ & $-0.02^{+0.12}_{-0.15}$ &  $1.21$ & $0.06^{+0.01}_{-0.01}$ & - & $10.26^{+0.03}_{-0.03}$ & $3.37^{+0.18}_{-0.15}$ & $1.15^{+0.54}_{-0.52}$ \\ 
J0900+0214 & $1.31$ & $0.32^{+0.02}_{-0.02}$ & - & $10.54^{+0.07}_{-0.08}$ & $4.0$ & $-0.30^{+0.11}_{-0.14}$ & 1.30 & $0.30^{+0.02}_{-0.02}$ & - & $10.90^{+0.11}_{-0.11}$ & $3.90^{+0.06}_{-0.06}$ & $3.68^{+0.50}_{-0.90}$ \\
J0901+2044 & $1.23$ & $0.10^{+0.01}_{-0.01}$ & - & $9.54^{+0.07}_{-0.07}$ & $4.0$ & $-2.84^{+0.46}_{-0.50}$ & 1.23 & $0.10^{+0.01}_{-0.01}$ & - & $9.52^{+0.09}_{-0.09}$ & $4.02^{+0.04}_{-0.04}$ & $18.9^{+3.90}_{-3.06}$ \\
.. & $1.50$ & $0.10^{+0.06}_{-0.06}$ & $0.12^{+0.03}_{-0.03}$  & $9.60^{+0.11}_{-0.10}$ & $4.0$  & $-2.54^{+0.56}_{-0.67}$ & 1.50 & $0.15^{+0.03}_{-0.01}$ & $0.04^{+0.01}_{-0.02}$ & $11.88^{+0.97}_{-0.68}$& $3.30^{+0.30}_{-0.35}$ & $1.80^{+0.80}_{-1.20}$ \\
J1007+2853 & $1.26$ & $0.16^{+0.01}_{-0.01}$ & - & $9.53^{+0.04}_{-0.04}$ & $1.53^{+0.03}_{-0.03}$ & $0.25^{+0.08}_{-0.08}$ & 1.26  & $0.15^{+0.01}_{-0.01}$ & - & $9.66^{+0.03}_{-0.03}$ & $1.45^{+0.03}_{-0.03}$ & $-2.79^{+0.35}_{-0.35}$ \\
  .. & $1.23$ & $0.21^{+0.03}_{-0.04}$ & $0.10^{+0.03}_{-0.03}$ & $9.73^{+0.07}_{-0.05}$ & $1.95^{+0.10}_{-0.11}$ & $-0.36^{+0.09}_{-0.12}$ & 1.26 & $0.15^{+0.01}_{-0.01}$ & $<0.01$ & $9.66^{+0.02}_{-0.03}$ & $1.45^{+0.07}_{-0.07}$ & $-2.72^{+0.60}_{-0.80}$ \\ 
J1017+4749 & $1.92$ & $0.34^{+0.05}_{-0.06}$ & - & $11.28^{+0.05}_{-0.05}$ & $1.04^{+0.15}_{-0.15}$ & $5^{+3}_{-2}$ & 1.63 & $0.32^{+0.03}_{-0.03}$ & - & $11.33^{+0.18}_{-0.17}$ & $1.43^{+0.09}_{-0.10}$& $-12^{+3}_{-7}$\\  
.. & $1.42$ & $0.23^{+0.05}_{-0.06}$ & $0.13^{+0.02}_{-0.02}$ & $11.41^{+0.12}_{-0.14}$ & $1.01^{+0.13}_{-0.13}$ & $22^{+20}_{-5}$ & 1.34 & $0.22^{+0.05}_{-0.02}$ & $0.13^{+0.02}_{-0.02}$ & $11.44^{+0.05}_{-0.05}$ & $1.31^{+0.15}_{-0.15}$& $-60^{+20}_{-30}$\\  
\hline
Diffuse MW$^a$ & .. & .. & .. & $9.86\pm0.04$ & $2.51^{+0.12}_{-0.11}$ & $-0.23^{+0.06}_{-0.06}$ &  & & & 9.74 & 1.85 & 2.1 \\
\hline
\end{tabular}
\begin{tablenotes}
      \small
     \item{ {\bf Notes:} $^a$ Average values for silicate profiles in the diffuse ISM of the MW are taken from \cite{Gordon2021} and \cite{Decleir2025} for \textbf{the} modified Drude and skewed Gaussian models, respectively.}
\end{tablenotes}
\end{center}
\end{table*}

\section{Fitting the extinction curve}
\label{sec:extinction}
In extragalactic studies, extinction curves are typically derived in a different way than for Local Group galaxies, where the pair method \citep{Massa1983} is commonly used. 
For distant galaxies, extinction curves have been measured by analyzing the reddening of background quasars caused by dust in foreground galaxies \citep[e.g.,][]{York2006} and by studying reddening in the afterglow spectra of Gamma-Ray Bursts (GRBs) to probe dust in their host galaxies \citep[e.g.,][]{Zafar2012}.
These methods rely on certain assumptions about the intrinsic, unreddened source spectrum. Another important difference between the Galactic and extragalactic approaches is the effect of cosmological redshift. This results in different galaxy rest-frame wavelength coverage for the extinction curves of local and distant objects, for a given observed wavelength range. For example, the commonly used amounts of extinction in the B and V filters are often not available for extragalactic objects. Although extragalactic studies have demonstrated notable variations in extinction/attenuation curve properties relative to those observed in Local Group galaxies \citep[e.g.,][]{Zafar2012, Shivaei2025}, the overall picture remains poorly understood.

Previous studies of dust extinction in our targets focused primarily on the observed-frame UV/optical range, using quasar spectra obtained with SDSS (primarily) or HST (when available): see \cite{Junkkarinen2004, Kulkarni2011} for AO~0235+164, \cite{Jiang2011} for J0900+0214 and J1017+4749, \cite{Ma2017} for J0901+2044, and \cite{Zhou2010} for J1007+2853. For J0900+0214 and J1017+4749, the extinction curve parameters were not fully determined, as their studies especially focused on the 2175~\AA\ bump feature, and did not derive the absolute extinction curve. For J0901+2044, the constraints on $A_V$ and $A_{\rm 2175}$ from \cite{Ma2017} are quite uncertain due to the degeneracy between these parameters and variations of the quasar composite model (see discussion in \citealt{Ma2017}).

We aim to improve the extinction curve fits by analyzing the quasar spectra over a broader wavelength range, including IR data (near-IR [NIR] and MIR photometry and MIR spectral data), which were missing in previous works. IR data are essential for reconstructing the rest-frame optical/IR portions of the extinction curve and for   deriving the intrinsic quasar brightness. 
For example, \cite{Zafar2012} found that including the IR data in the fitting of the extinction curve for GRB absorbers has significant impacts on the derived shape, and can change the estimate of $A_V$ by a factor of 2-3.

Additionally, we expand our sample by including four quasar sight lines from our earlier studies of the 10 and 18~$\mu$m silicate features in distant galaxies observed with the Spitzer IRS \citep{Kulkarni2011}:  J0852+3435, J0937+5628, J1203+0634, and 3C196 (hereafter the ``archival sample''). This ensures that extinction curves are derived in a uniform and consistent manner across the sample.

\subsection{Data set}
We analyze our MIRI MRS data and the archival data using  spectra from the HST/STIS (\cite{Junkkarinen2004} for AO0235+164) and SDSS \citep{Abdurro2022}, along with photometric data from GALEX \citep{Bianchi2017}, SDSS, the Two Micron All-Sky Survey (2MASS, \citealt{2MASS}), the UKIRT Infrared Deep Sky Survey (UKIDSS, \citealt{UKIRT}), and the Wide-Field Infrared Survey Explorer (WISE, \citealt{WISE}). The data were first corrected for Galactic extinction using  an average extinction curve for the MW diffuse ISM (\citealt{Gordon2009, Fitzpatrick2019, Decleir2022, Gordon2021, Gordon2023};  together hereafter G23) and $A_V({\rm MW})$ values from \cite{Schlafly2011}. The spectral regions and photometric bands significantly affected by absorption features or broad emission lines were excluded from the analysis. The GALEX NUV fluxes are considered as lower limits on quasar flux, since the NUV band probes the spectral region below the quasar Ly$\alpha$ emission, i.e., a region affected by intergalactic absorption that effectively reduces the quasar flux within the band.

Comparing spectroscopic and photometric data  requires scaling the spectroscopic data - measured within small apertures (e.g., $\sim$3" for SDSS and $\sim$1.5" for MIRI MRS) - to match the corresponding quasar photometric fluxes. For SDSS, we use the photometric flux in the z-band, while for MIRI MRS we measure the average flux in the 1A channel of the MRS cube using a 4"-diameter aperture, corresponding to a radius of $\sim5\sigma$ of the MRS point-spread function (PSF).

{However,  we still observe offsets between photometric and spectroscopic data. This offset may be caused by (i) the difference in the PSF FWHM of the instruments (4.9" for GALEX (NUV), 1.4" for SDSS, $\sim$0.8" for UKIRT, $\sim$2.5–3" for 2MASS, 6.1–6.5" for the WISE W1–W3 bands (and 12" for W4), and $\sim$0.4" for MIRI MRS 1A) and extraction methods used in different surveys (e.g., the elliptical "Kron" aperture for GALEX, and PSF-fitting, Petrosian aperture, or fixed aperture photometry for SDSS, 2MASS, UKIRT, and WISE), and (ii) a temporal variability of the quasar brightness. 

Optical quasar variability was studied by \citet{Guo2016}, who compared quasar spectra from SDSS Data Releases 7 (DR7) and 9 (DR9). They found that the integrated quasar flux between faint and bright epochs can vary by up to a factor of 2 ($\sim$0.7~mag) (for $\lambda_{\rm rest}=1300-6000$~\AA), and that the variability can be described by a power-law component, with less variability at longer wavelengths. We have inspected the variability for our two targets that have 2 datasets in SDSS (J0901+2044 and J1007+2853) and found variations of the brightness by about 30\% and similar power-law behavior for the variable component.   

In order to correct for these offsets, we first exclude from the analysis WISE 3 and 4 bands, which have FWHMs and diffraction limits\footnote{The FWHM of W3 and W4 bands is slightly above their diffraction limits, 7.5\arcsec and 13.8\arcsec, respectively, while W1 and W2 are not diffraction-limited (2.1\arcsec and 2.9\arcsec, respectively).} much larger than the WISE 1 and 2 bands. Additionally, the quasar template at the redshift of the quasars only covers bands 1 and 2 (see Section~\ref{app:c:composite}). Secondly, we introduce scaling parameters for our datasets (GALEX, SDSS, and 2MASS/UKIRT, and MIRI MRS, if available), using WISE as the reference (since it covers the longest wavelengths\footnote{MIRI MRS data are not available for quasars in the archival sample, while the available Spitzer IRS data $\lambda_{\rm obs}>7.4~\mu$m (SL1) are not covered by the quasar template $\lambda_{\rm rest}<3.5~\mu$m shifted to the quasar redshift $z\sim 1$.}), to simultaneously scale their fluxes while varying the fitting parameters.
We note that the ratios between bands within each dataset are kept fixed.

It is worth noting that the UKIDSS photometric measurements in the Y, J, H, and K bands were not obtained on the same dates as the WISE and 2MASS observations. For two of our targets, J0900+0214 and J1203+0634, the Y and J bands, and the H and K bands, were observed in pairs with a time gap of about one year. Therefore, we scale the fluxes in the UKIRT Y+J and H+K bands separately, taking into account possible offsets due to quasar variability.

\subsection{Quasar intrinsic model (composite) from UV to mid-IR}
\label{app:c:composite}

Deriving the extinction of a quasar by dust in a foreground galaxy requires an accurate estimate of the quasar intrinsic spectrum. This is typically obtained from composite spectra constructed by averaging large samples of quasars that are considered to be free of host galaxy contamination, strong associated absorption systems, and reddening in the MW.
Quasar composite spectra have been determined in the literature for various wavelength ranges and samples, differing in selection criteria, quasar luminosities, and redshifts, as well as in the corrections applied for intergalactic medium absorption and host galaxy contamination. \citet{VandenBerk2001}, \citet{Glikman2006}, \citet{Jiang2011}, \citet{Lusso2015} and \citet{Selsing2016} focus on the UV-optical range, while \citet{Hatziminaoglou2005}, \citet{Glikman2006}, \citet{Hill2014}, and \citet{Hernan2016} study the near-IR  to mid-IR range. For a comparison of the composites, and a discussion of systematic effects, see \citet{Selsing2016} and \citet{Hernan2016}.

We consider a quasar composite that spans the full range from the UV to the NIR ($0.08~\mu$m$\le\lambda\le3.5~\mu$m in the AGN rest frame) using a combination of models from  \cite{VandenBerk2001} ($\lambda\le0.11~\mu$m), \cite{Selsing2016} ($0.11~\mu{\rm m}\le\lambda\le1.03~\mu$m), and \cite{Glikman2006} ($1.03~\mu{\rm m}\le\lambda\le3.5~\mu$m), hereafter the VSG template. Above 3.5~$\mu$m, quasar composite templates are less consistent across different samples \cite[see, e.g.,][]{Hatziminaoglou2005, Hernan2016}. These differences are  likely caused by non-zero contribution from the quasar  host galaxies' stellar emission \citep{Xu2015} and by distinct AGN properties (morphology, redshift, and black hole mass, see  Paper~I). 

For one quasar in our sample, AO0235+164 (a blazar), the situation is different, because its BL Lac spectrum is dominated by synchrotron emission and lacks emission features. In this case, we assume that the composite is described by the synchrotron emission model, and determine the parameters of the synchrotron emission model consistently by fitting an extinction curve to the data.

\subsection{Correction for quasar spectral variability}
\label{app:c:variability}

\cite{Guo2016} analyzed the variability of 1876 quasar spectra at UV/optical wavelengths observed in two SDSS data releases and found that about 94\% of quasars exhibit a trend whereby they become bluer when brighter, whereas the other 6\% become redder when brighter. By comparing composite spectra from bright and faint epochs, they found that the spectral difference is well described by a power-law relation with an index of approximately~$\simeq-2$. 

To account for the effects of spectral  variability for the quasar composite, we allow the quasar composite spectrum to become bluer or redder by adding a variable continuum component, and define the corrected quasar composite  as follows:
\begin{equation}
C_{\rm new}(\lambda, f_V) = C_{\rm VSG}(\lambda) + f_V \times \lambda^{-2}\times\frac{\int_{\lambda_{\rm min}}^{\lambda_{\rm max}} C_{\rm VSG}(\lambda)  d\lambda}{\int_{\lambda_{\rm min}}^{\lambda_{\rm max}} \lambda^{-2} d\lambda}, \end{equation}
where $C_{\rm VSG}$ is the original VSG template,  
and $f_V$ is the relative difference\footnote{The parameter $f_V$ is calculated as follows: 
$f_V = \left({\int_{\lambda_{\rm min}}^{\lambda_{\rm max}} C_{\rm new}(\lambda)  d\lambda - \int_{\lambda_{\rm min}}^{\lambda_{\rm max}} C_{\rm VSG}(\lambda)  d\lambda}\right)/{\int_{\lambda_{\rm min}}^{\lambda_{\rm max}} C_{\rm VSG}(\lambda)  d\lambda}$} between the integrated fluxes of the corrected and original  composite spectra in the range of $1300-6000$~\AA.
We vary $f_V$ within the range of $-0.3$ to $0.3$, which corresponds to the amplitude of optical variability observed in two quasars from our sample. The parameter $f_V$ is varied consistently with other fitting parameters.

\subsection{Dust extinction model from UV to mid-IR}
\label{app:c:extinction}
We parameterize the extinction curve in distant galaxies using the same approach as the dust extinction model for the MW diffuse ISM (see G23). We analyze the extinction in absolute units of magnitude, $A(\lambda)$, and divide the UV to near-IR wavelength range into 3 parts: UV ($\lambda<0.3~\mu$m), optical ($\lambda=0.3-1~\mu$m), and Near-IR ($\lambda=1-5~\mu$m).
The UV part is described using the formalism from \cite{FM1990, FM2007}, $A_{\rm uv}(\lambda)={\rm FM(\lambda)}$,  with 6 fitting parameters ($C_1$, $C_2$, $C_3$, $C_4$, $x_0$ and $\gamma$).\footnote{We vary $C_4$ for the quasars, where GALEX NUV data is available, while assuming it to be 0 in other cases.} 
The optical part is described by a fourth-order polynomial in the inverse wavelength: $A_{\rm opt}(\lambda)= \sum_{j=0}^{4} E_j x^j$, where $x=1/\lambda[{\rm\mu m}^{-1}]$. We do not include in the fit the Intermediate Scale Structure (ISS) features \citep{Massa2020}, due to the limited spectral coverage of the 2MASS/UKIDSS photometry. The IR part is fitted using  a power-law relation, $A_{\rm ir}(\lambda) = B\lambda^{-\alpha}$. The VSG quasar template at the quasar redshift only covers $\lambda<5~\mu$m in the galaxy rest frame. Additionally, two continuity constraints between the fitting functions for the UV, optical, and NIR parts allow us to reduce the number of free parameters by two. Thus, we determine $E_1$ and $E_2$ from these boundary conditions:
\begin{equation}
\label{eq:lam1}
\sum_{j=0}^{4} E_j \lambda_1^{-j} = {\rm FM}(\lambda_1; C_1, C_2, C_3, C_4, \gamma, x_0)
\qquad  
\sum_{j=0}^{4} E_j \lambda_2^{-j} = B \lambda_2^{-\alpha}
\end{equation}
\color{black}
where $\lambda_1 = 0.3~\mu$m and $\lambda_2 = 1~\mu$m. Unfortunately, for quasars that lack 2MASS/UKIDSS data, we are not able to constrain the optical parameters well, and therefore approximate the optical part with a first-order polynomial, where $E_0$ and $E_1$ are determined from the boundary conditions.

We also applied prior distributions to the fitting parameters to ensure a physically realistic extinction curve. Specifically, we set $C_i \ge 0$ for $i=2$–$4$, $0.6 < \gamma < 1.5$, and $4.4 < x_0 < 4.8$, covering the range observed in the MW \citep{Gordon2009}. We also required that $A(\lambda_1) \ge A(\lambda_2) > 0$ and $A(5~\mu{\rm m})/A_V < 0.1$, ensuring that the infrared extinction is not too high, consistent with the predictions for large grains in the MW ISM model by \cite{Hensley2023}.

\subsection{The fitting method}

Taken together, our model for the reddened quasar composite is given by 
\begin{equation}
    F(\lambda)=f_0 \times C_{\rm new}(\lambda, f_V)\times\exp\left[-0.92 A(\lambda, C_{1-4}, x_0, \gamma, E_{0-4}, B, \alpha)\right],
    \label{eq:reddened_composite}
\end{equation}
where $f_0$ and $f_V$ are the constant scaling factors for the intrinsic quasar brightness and the quasar spectral variability parameter, and $C_{1-4}$, $x_0$, $\gamma$, $E_{0-4}$, $B$, $\alpha$ are the fitting parameters for the extinction in the foreground galaxy. Additionally, there are four scaling coefficients  ($f_{\rm GALEX}$, $f_{\rm SDSS}$, $f_{2MASS}$, $f_{MRS}$)\footnote{For quasars that lack 2MASS data and have UKIDSS data, instead of $f_{2MASS}$ there are two scaling parameters, as described in Section~\ref{app:c:variability}. For AO0235+164, which lacks SDSS data, the $f_{SDSS}$ parameter is instead used to scale the HST STIS spectrum.} accounting for temporal variability between the observations. We fit the underlying intrinsic continuum flux, the extinction parameters,  and the data scaling parameters simultaneously, using the MCMC method.

To ensure similar weighting for all spectral data, the spectra from the HST/SDSS/MRS were rebinned to a common resolution of $\lambda/\delta\lambda=1000$. In order to give the photometry a more appropriate weight compared to the densely sampled spectral data, we
calculate an effective number of pixels per filter. We calculate this quantity by integrating the filter transmission curves scaled to a maximum of 1 and interpolated onto a grid with the same sampling as the spectral data. The uncertainty for each filter is then divided by the square root of this number.

Because of the lack of near-UV spectral data (since GALEX NUV photometry provides only a lower limit on the quasar flux and an upper limit on the extinction curve), we found a degeneracy between the UV slope and the strength of the UV bump (mainly $C_2$, $C_3$, and $\gamma$), which affects the estimates of the total $V$-band extinction $A_V$ and the bump strength $A_{2175} = C_3 / \gamma^2$. This problem was discussed earlier by e.g., \cite{Ma2017}. To obtain a conservative constraint on the UV bump strength (and to exclude fits with extremely broad or strong bumps, or unrealistically narrow ones), we follow a similar two-step fitting approach to that of \cite{Ma2017}.

In the first step, we varied all parameters simultaneously, except for $x_0$ and $\gamma$, which were fixed to the average MW ISM values ($x_0 = 4.6$ and $\gamma = 1.0$, \citealt{Gordon2009}). In the second step, we fixed the other parameters to the values determined in the first step and derived the best-fit values for $C_3$, $x_0$ and $\gamma$.


\subsection{Results}

Our determinations of the extinction fit parameters are listed in Table~\ref{tab:extinction-parameters-uv} and Table~\ref{tab:extinction-parameters-opt-ir}. The fits to the observed quasar spectra and the corresponding extinction curves for the foreground absorbers are shown in Figure~\ref{fig:fit-dust-ext-oursample} for both our quasar sample, and the archival sample. The best fit using the reddened quasar template is shown together with the unreddened quasar spectrum. In most cases, we were able to consistently fit the data across the UV to IR range. 

We note that the observed extinction curves shown in Figure~\ref{fig:fit-dust-ext-oursample} in gray and black are calculated   using the following equation:
\begin{equation}
A(\lambda) = -2.5 \log\left(\frac{F_{\rm obs}(\lambda)}{F_{\rm intrinsic}(\lambda)}\right),
\label{eq:obs_extinction}
\end{equation}
where $F_{\rm intrinsic} = f_0 \times C_{\rm new}(\lambda, f_V)$ is the intrinsic quasar flux, derived for the best fitting parameters $f_0$ and $f_V$. Our method does not fit the observed extinction curve using the UV–optical–IR functional form (like G23). Instead, all fitting parameters are derived self-consistently and simultaneously by fitting the observed quasar spectra with a reddened composite model using Eq.~\eqref{eq:reddened_composite}. Red lines in Figure~\ref{fig:fit-dust-ext-oursample} show the modeled extinction curves, constructed using the best-fitting parameters, that well match the observed extinction curves derived using Eq.~\eqref{eq:obs_extinction}.

We found strong differences between the derived extinction curves and the average extinction curve for the MW diffuse ISM from G23 for $R_V = 3.1$ in six of the nine sources. For better illustration of these differences, we show the fit to the observed quasar spectrum using the MW extinction template with the same $A_V$ value as derived from our fit (by the green dashed line in Figure~\ref{fig:fit-dust-ext-oursample}). 
In four cases (J0900+0214, J0901+2044, J1007+2853, and J1017+4749), the extinction curves show an excess in the rest-frame optical/IR part (probed by 2MASS and WISE photometric points and MIRI MRS spectra) and a flattening in the UV part compared to the MW extinction curve. In two other cases (J1203+0634 and 3C196), the extinction curve has little, if any, optical/IR extinction (i.e., the observed spectrum is close to the composite template in this part), while it is extremely steep in the UV part. Similar peculiar extinction curves, showing little or no reddening in the optical/IR and significant reddening in the UV, were previously found in a few quasars \citep[][]{Krogager2016}. 
It is worth noting that several systems (J1017+4749, J0937+5628) lack 2MASS/UKIDSS data, and therefore lack constraints on the extinction in the optical range\footnote{In such cases, the shape of the extinction curve in the optical range cannot be determined, and the fit shown in Figure~\ref{fig:fit-dust-ext-oursample} is derived from the
continuity conditions at  $0.3~\mu$m and $1~\mu$m. i.e., between the UV/optical and optical/NIR parts. These constraints may change if additional data become available.}, while constraints are still available in the UV and NIR.

Compared to previously published fits for the UV portion of the extinction curves using SDSS quasar spectra and assuming MW/SMC/LMC extinction curves (e.g., see Table~1 in Paper~I, and \citealt{Kulkarni2011} for the archival sample) or the Fitzpatrick \& Massa parametrization \citep[e.g.,][]{Zhou2010, Ma2017}, the new values of $A_V$ differ depending on the shape of extinction curve in the optical/IR part. For J0852+3425 and J0937+5628, our values are similar to those derived using a fit with the LMC Supershell template \citep{Kulkarni2011} - this is because the derived extinction curves are similar to extinction in local galaxies. However, in the case of excess  optical/IR extinction (as for J0901+2044) or small optical/IR extinction (as for J1203+0634) the differences in the $A_V$ values are significant ($\sim$4 times higher and lower, respectively). 
This is explained by the large uncertainty in the intrinsic quasar brightness, which is difficult to constrain from fits to the UV portion alone. 
For example, fits to the SDSS spectra of  J1203+0634 using the LMC Supershell template and the VSG quasar composite can well describe the SDSS portion of the quasar spectrum \citep[see, e.g.,][]{Kulkarni2011}. However, such a fit significantly overestimates the observed quasar flux in the NIR range compared to the photometric measurements from 2MASS and WISE.\footnote{Fitting the SDSS data alone, without IR constraints where extinction is low, leaves the intrinsic quasar brightness uncertain. While fit with the LMC Supershell extinction template can reproduce the SDSS data, it requires a higher intrinsic brightness and larger extinction $A_V$ (as in the case of J1203+0634). As a result, such fits overpredict the quasar flux in the IR bands and the amount of extinction.}
Robust estimates of the quasar’s intrinsic brightness and the total amount of extinction, $A_V$, can be obtained only when IR observations are included in the analysis. 

The general properties of the extinction curves are presented in Table~\ref{tab:extinction-results}. The values of the total extinction in the $V$ band ($A_V$), the strength of the UV bump ($A_{2175} = C_3 / \gamma^2$), the ``area under" the UV bump ($S_{2175} = \pi C_3 / 2\gamma$), and the ratio of total-to-selective extinction ($R_V = A_V / E(B-V)$) are calculated from the extinction curve based on the derived distribution of extinction parameters. The parameters $A_V$ and $R_V$ were calculated by integrating the obtained best fit model for $A(\lambda)$ over the B and V bands. Panels (a), (b), and (c) in Figure~\ref{fig:extinction-comparison} shows the comparison of these parameters in our sample with those measured in the MW ISM \citep{Rieke1985, Gordon2009, Gordon2021}, the LMC \citep{Gordon2003}, and the SMC \citep{Gordon2003, Gordon2024}. 

The extinction values $A_V$ in the absorption systems along the quasar sightlines in our sample and the archival sample range from 0.1 to 1.9~mag, indicating that the sightlines probe the diffuse gas. Similar values of $A_V$ are observed in the diffuse ISM of the LMC and SMC, while $A_V$ values in the MW diffuse ISM are slightly higher. 
On the other hand, the relative strengths of the UV bump ($A_{2175}/A_V$) in the quasar sightlines also resemble those seen in the LMC sightlines, and lie in between those seen in the SMC sightlines and the MW sightlines. 

These findings suggest that the quasar sightlines may probe the diffuse ISM of foreground galaxies similar to the LMC, whose metallicity ($Z({\rm LMC})\sim 0.5 Z_{\odot}$ on average) lies between the  metallicity of the MW ($\sim Z_{\odot}$), and the SMC ($Z({\rm SMC}) \sim 0.2 Z_{\odot}$, \citealt{Russel1992}).  
This is generally consistent with measurements of the dust-corrected metallicity for strong quasar absorbers \citep{DeCia2018}, which show $Z \sim 0.5_{-0.2}^{+0.4} Z_{\odot}$ at $z_{\rm abs} \sim 0.8$. 
Accurate metallicity measurements from future UV/optical follow-up observations for our sightlines are essential to verify this conclusion more robustly.

The distributions of $R_V$ and $A_V$ are also more consistent with those in the LMC and SMC, although several quasar sightlines show $R_V \sim 6$–$13$, much higher than the values observed in Local Group galaxies. Such sightlines may lack small grains, which can be explained if they probe dust in the circumgalactic medium of foreground galaxies, where small grains may be destroyed \citep[e.g.,][]{Hirashita2024, Richie2025}.  The comparison of $R_V$ and $A_{2175}/A_V$ shows that larger $R_V$ values tend to correspond to lower $A_{2175}/A_V$, indicating an anticorrelation between the grain size and the fraction of carbonaceous PAH grains responsible for the 2175~\AA\ feature. This is consistent with the PAH grains being smaller in size compared to the silicate grains. Note that such an anticorrelation is also seen in the MW sample, although on a lower $R_V$ scale.

The contribution of large grains to the extinction can be probed by the ratio of extinction in the K and V bands, $A_K/A_V$. Quasar sightlines span a wider range of $A_K/A_V$ values than those observed in local galaxies and roughly follow the trend reported by G23 for the MW diffuse ISM, showing an increase of $A_K/A_V$ with $R_V$, as illustrated by the gray line in panel (d) of Figure~\ref{fig:extinction-comparison}. For the LMC and SMC sightlines, we found the average values of $A_K/A_V$ from \cite{Gordon2003, Gordon2024} are in good agreement with our low-$R_V$ sightlines. For the MW, IR extinctions have been measured by \cite{Decleir2022}.  We used their IR extinctions curves to calculate $A_K/A_V$ which closely matches the relation reported by G23.
The quasar sightlines with high $R_V\sim6-13$ values show high $K$-band extinction ($A_K/A_V\sim0.25$), consistent with these sightlines probing absorbing clouds rich in large dust grains.

Panels (e) and (f) in Figure~\ref{fig:extinction-comparison} illustrate the range of extinction curve shapes found in our quasar sightlines, in comparison with the average extinction curves\footnote{ See \url{https://dust-extinction.readthedocs.io/en/latest/dust_extinction/model_flavors.html}.} of the MW (G23), the LMC \citep{Gordon2003}, and the SMC \citep{Gordon2024}. These panels highlight the differences in the UV bump strength and the relationships among $R_V$, $A_{2175}/A_V$ and $A_K/A_V$ discussed earlier.
}

\begin{figure}[h]
    \centering
    \includegraphics[width=0.32\textwidth]{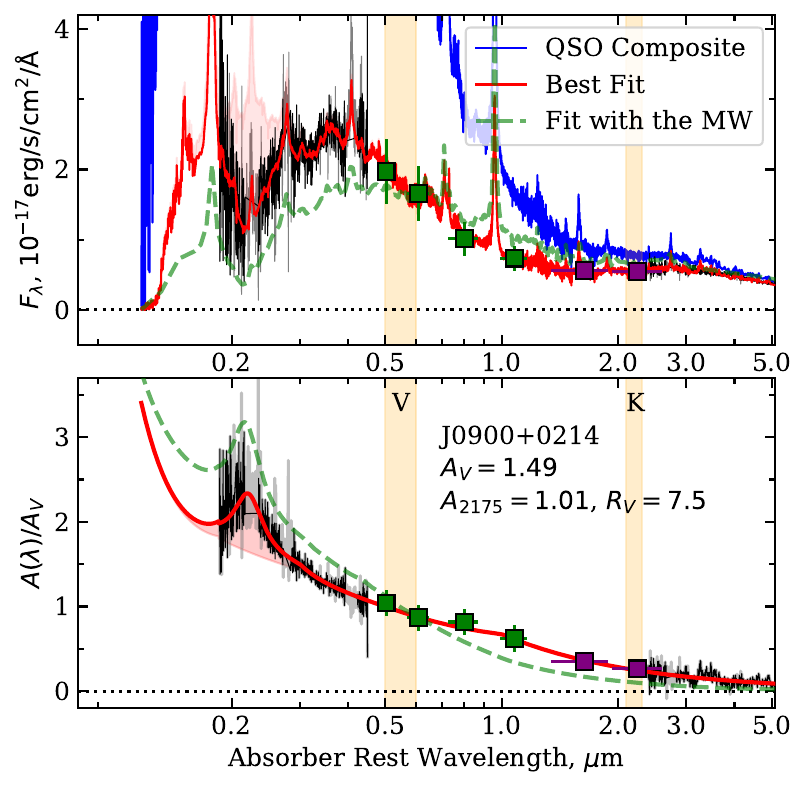}
    \includegraphics[width=0.32\textwidth]{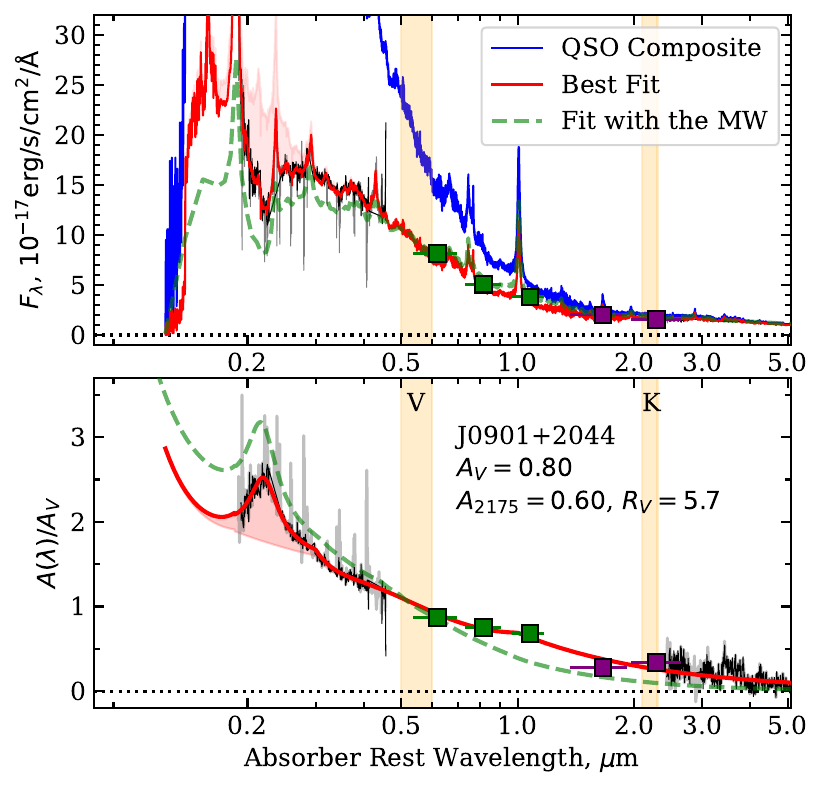}
    \includegraphics[width=0.32\textwidth]{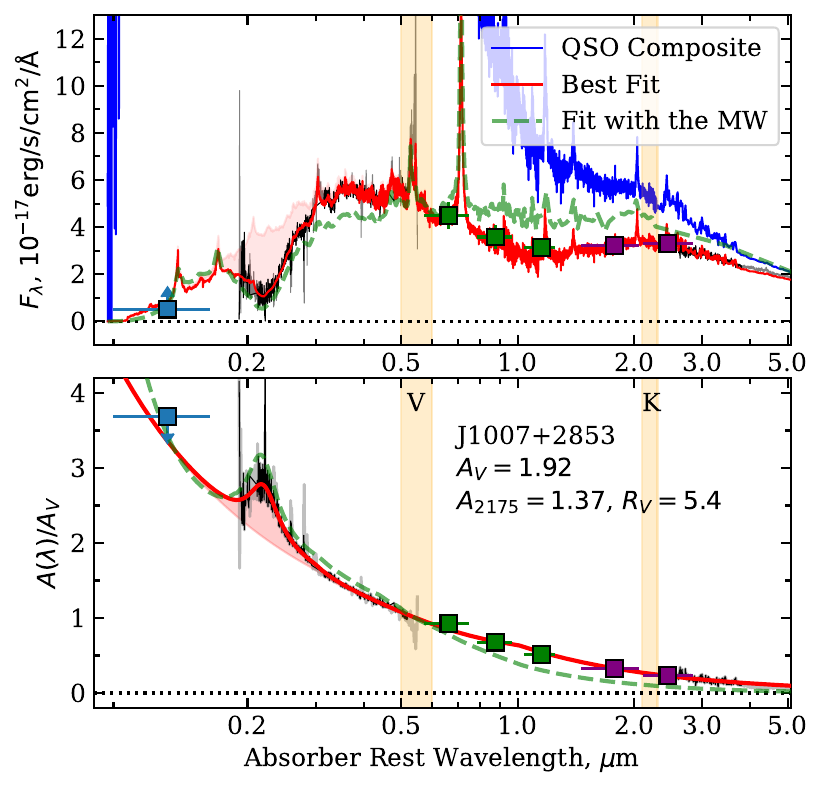}
     \includegraphics[width=0.32\textwidth]{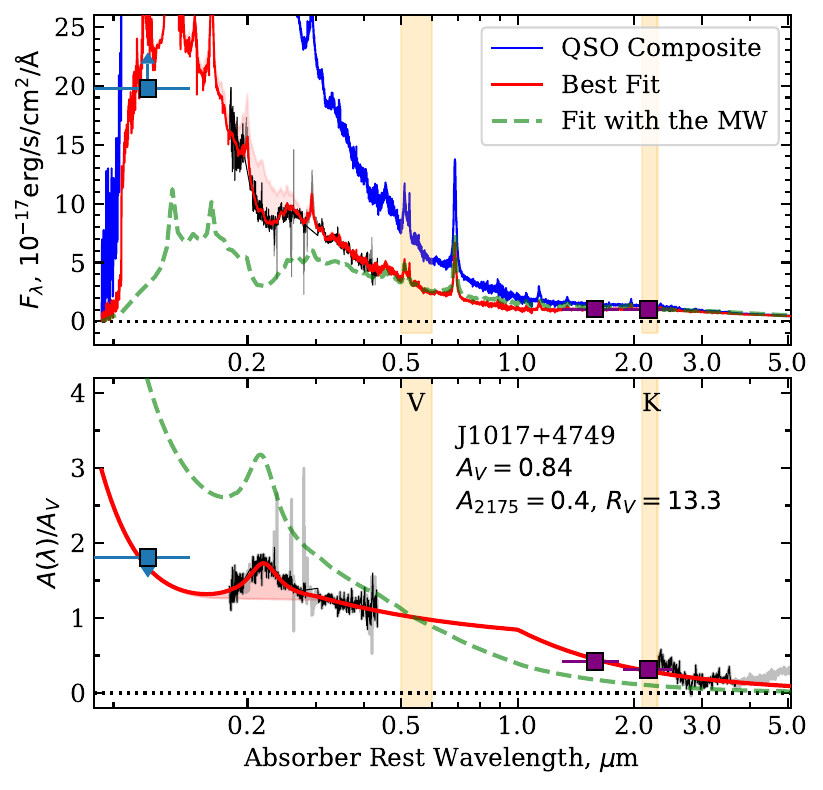}
     \includegraphics[width=0.32\textwidth]{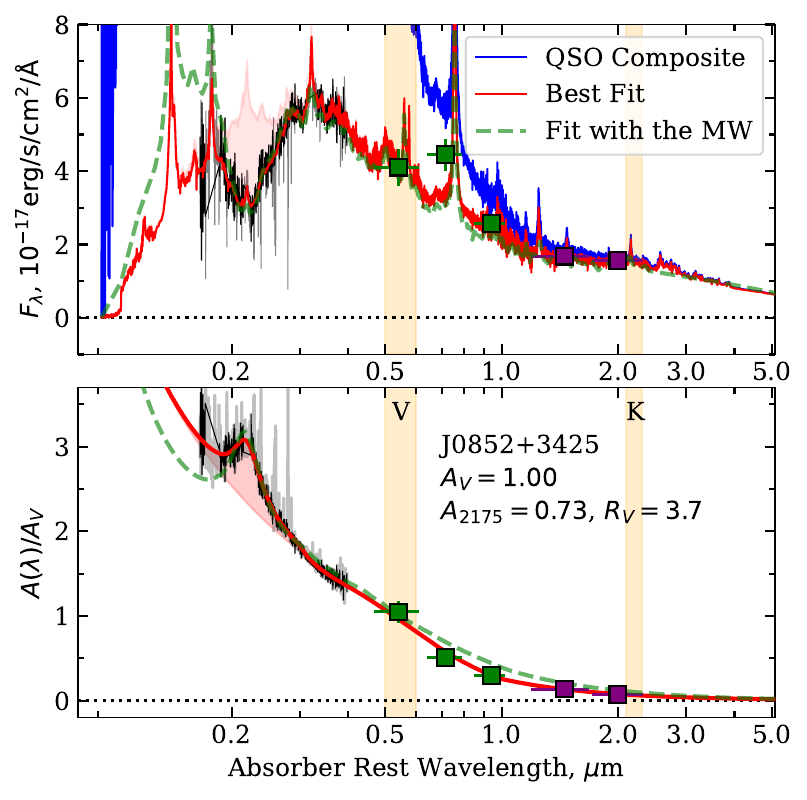}
     \includegraphics[width=0.32\textwidth]{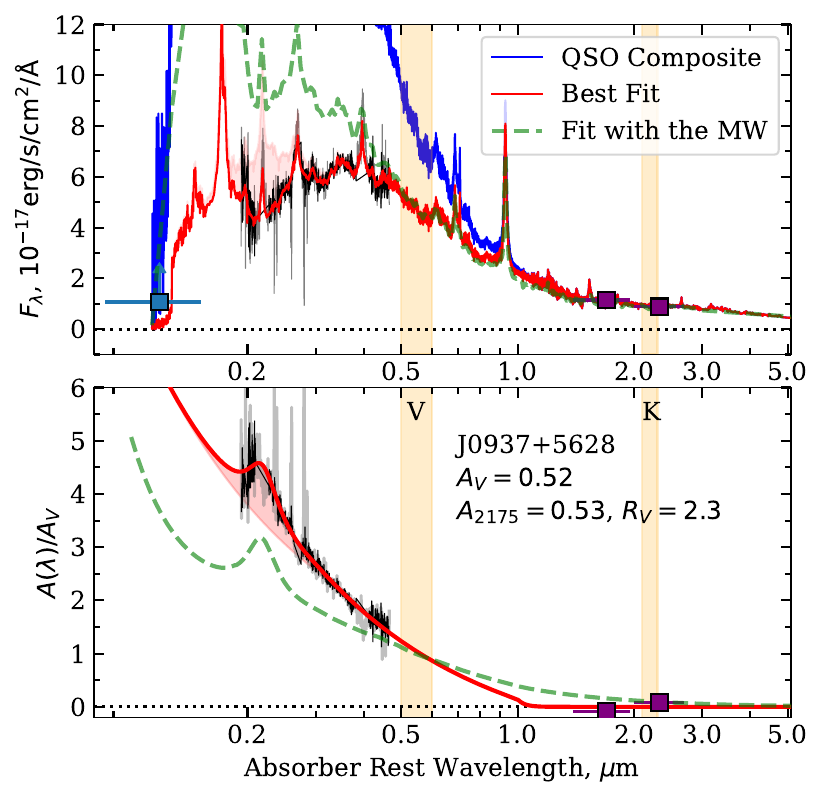}
    \includegraphics[width=0.32\textwidth]{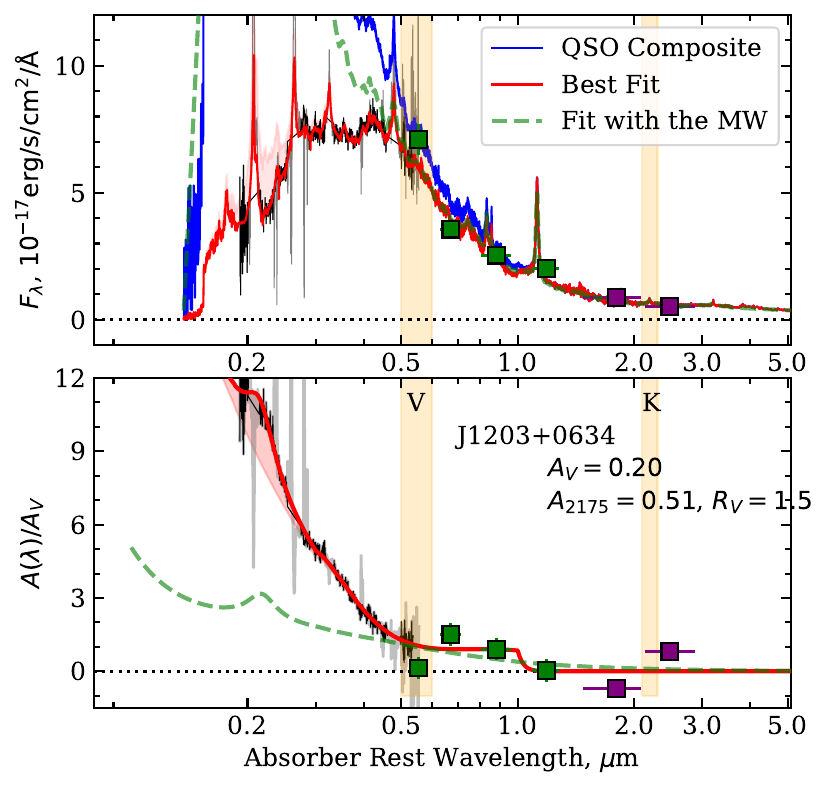}
    \includegraphics[width=0.32\textwidth]{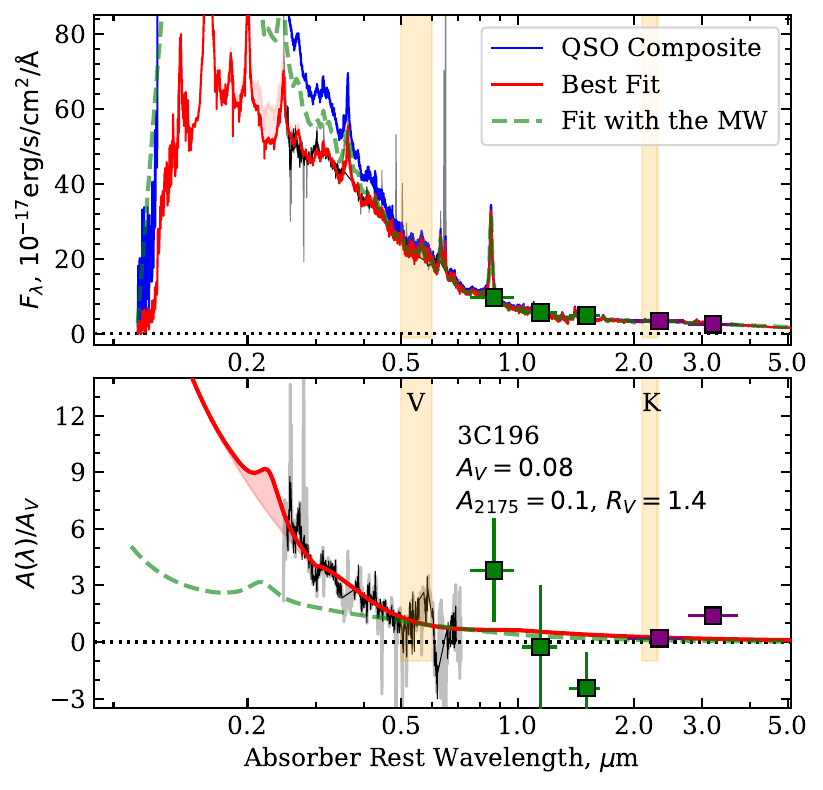}
    \includegraphics[width=0.32\textwidth]{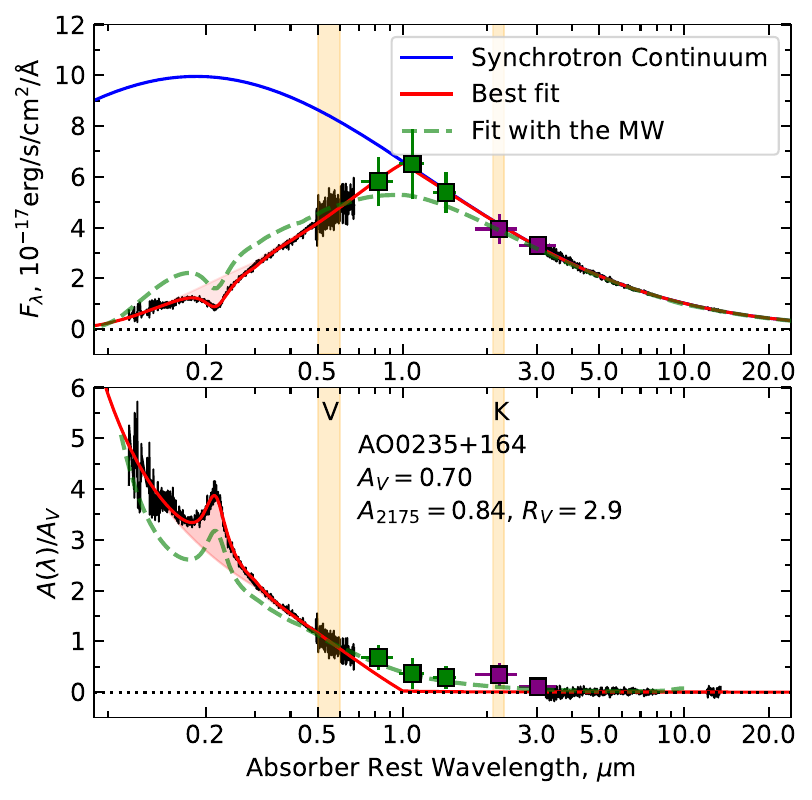}
    \caption{Extinction curve fits for quasar spectra from our sample and the archival sample. For each target, the top and bottom panels show the fit to the observed quasar spectrum using the reddened quasar composite model and the derived best-fit extinction extinction curve, respectively. The observed spectrum is shown in black (highlighting the fitting regions) and gray and presents a combination of SDSS (or HST/STIS) and MIRI MRS (available only for our sample) data. The blue, green and purple squares indicate the GALEX (NUV), 2MASS/or UKIDSS, and WISE (W1, W2) photometric data points, respectively. The blue line in the top panels represents the unreddened quasar composite template corrected for quasar variability (see text for details). The red line shows the composite template reddened by the best-fit extinction curve at the absorber's redshift in the upper panels, and the best-fit extinction curve in the lower panels. The red shaded area marks the contribution of the 2175~\AA\ bump. To highlight the differences between the derived extinction curves and the MW extinction, the green dashed curve shows the quasar template reddened by the MW ISM extinction curve from G23 (for $R_V=3.1$), assuming the same $A_V$ value as in the best-fit model. The vertical yellow bands indicate the positions of the photometric V and K bands.  }
     \label{fig:fit-dust-ext-oursample}
\end{figure}

\begin{figure*}
\begin{center}
        \includegraphics[width=1\textwidth]{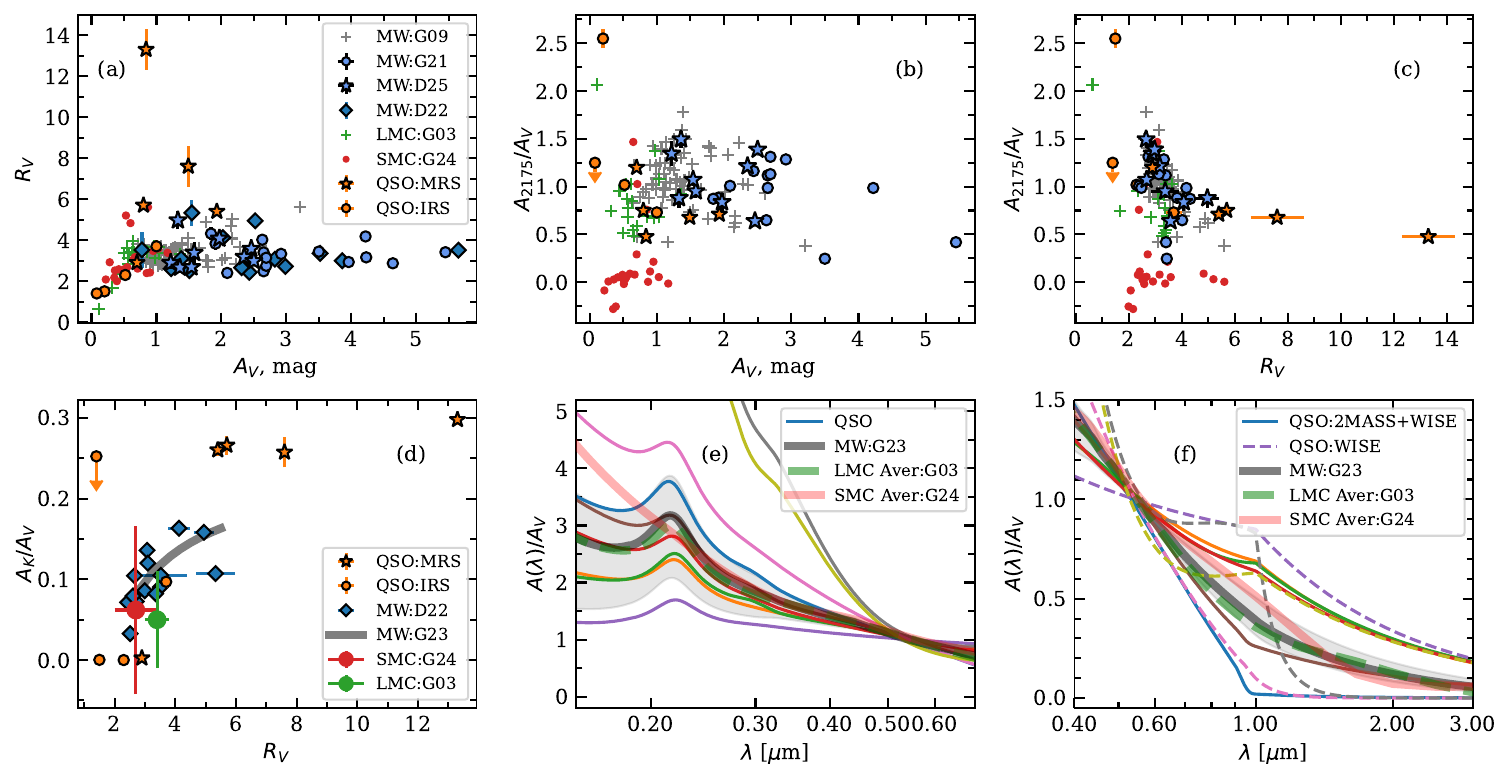}
        \caption{
        The general extinction parameters for quasar sightlines in our sample and the archival sample are compared with those measured across the Local Group galaxies: the MW, LMC, and SMC. Panels (a)-(c) show  the relations between $R_V$, $A_V$, and $A_{2175}/A_V$ using data from our sample and the archival sample (orange stars and orange circles, respectively), the samples of MW sightlines from \cite{Gordon2009}, \cite{Gordon2021}, \cite{Decleir2022} and \cite{Decleir2025} (gray crosses, blue circles, blue diamonds, and blue stars, respectively), and the sightlines in the LMC (green crosses, \citealt{Gordon2003}), and the SMC  (red circles, \citealt{Gordon2024}). Panel (d) presents the relation between $A_K/A_V$ and $R_V$. Our data are shown by orange stars and circles, the MW sample from \cite{Decleir2022} by blue diamonds (see text for details), and the universal relation of $A(\lambda)-R_V$ from G23 by the gray line. For the LMC and the SMC, the average values from \cite{Gordon2003, Gordon2024} are shown by green and red circles. Panels (e) and (f) illustrate the diversity of extinction curves in quasar sightlines (our sample and the archival sample) in comparison with that in the MW ISM (G23, gray line for $R_V=3.1$, and gray shaded area indicates the range of $R_V=2.5-5.5$), the LMC (green dashed line, \citealt{Gordon2003}), and the SMC (red line,  \citealt{Gordon2024}). The solid and dashed curves in panel (f) represent quasar sightlines with and without 2MASS/UKIRT photometric observations, which are used to constrain the extinction around $\sim1~\mu$m.
        }
        \label{fig:extinction-comparison}
\end{center}
\end{figure*}

\setlength{\tabcolsep}{2pt}
\begin{table*}
\caption{UV Extinction parameters}
\label{tab:extinction-parameters-uv}
\begin{center}
\begin{tabular}{|l|c|c|c|c|c|c|}
\hline
  Quasar & C$_1$ & C$_2$ & C$_3$ & C$_4$ &  $x_{0}$ & $\gamma$  \\ 
  & [$A(\lambda)/A_V$] & [$A(\lambda)/A_V$] & [$A(\lambda)/A_V$] & [$A(\lambda)/A_V$]  & ($\mu$m$^{-1}$)  &  ($\mu$m$^{-1}$) \\ 
\hline
AO0235+164  & $0.68^{+0.02}_{-0.03}$ & $0.41^{+0.02}_{-0.03}$ & $1.19^{+0.04}_{-0.07}$ & $0.02^{+0.02}_{-0.02}$ & $4.62^{+0.03}_{-0.03}$ & $1.00^{+0.02}_{-0.02}$  \\ 
J0900+0214 & $0.86^{+0.09}_{-0.10}$ & $0.19^{+0.03}_{-0.03}$  & $0.61^{+0.06}_{-0.07}$& None & $4.54^{+0.02}_{-0.02}$ & $0.95^{+0.05}_{-0.03}$ \\
J0901+2044 & $1.12^{+0.03}_{-0.03}$ & $0.14^{+0.01}_{-0.01}$  & $0.65^{+0.04}_{-0.04}$& None & $4.54^{+0.01}_{-0.01}$ & $0.93^{+0.01}_{-0.01}$ \\ 
J1007+2853 & $0.47^{+0.05}_{-0.05}$ & $0.35^{+0.02}_{-0.02}$  & $0.68^{+0.02}_{-0.02}$& $0.00^{+0.02}_{-0.00}$ & $4.55^{+0.01}_{-0.01}$ & $0.96^{+0.02}_{-0.02}$ \\ 
J1017+4749 & $1.18^{+0.03}_{-0.02}$ & $0.01^{+0.02}_{-0.01}$ & $0.47^{+0.03}_{-0.03}$ & $0.06^{+0.03}_{-0.02}$ & $4.54^{+0.01}_{-0.01}$ & $1.00^{+0.02}_{-0.02}$ \\ 
\hline
J0852+3425 & $0.27^{+0.05}_{-0.05}$ & $0.47^{+0.02}_{-0.03}$ & $0.66^{+0.03}_{-0.02}$ & None & $4.57^{+0.01}_{-0.01}$ & $0.94^{+0.02}_{-0.02}$  \\ 
J0937+5628  & $0.09^{+0.11}_{-0.10}$ & $0.72^{+0.05}_{-0.08}$ & $1.12^{+0.08}_{-0.10}$ & $0.00^{+0.02}_{-0.00}$ & $4.58^{+0.01}_{-0.01}$ & $1.04^{+0.02}_{-0.02}$ \\ 
J1203+0634 & $-5.04^{+0.90}_{-0.80}$ & $2.85^{+0.35}_{-0.42}$ & $2.55^{+0.40}_{-0.30}$ & None & $4.65^{+0.02}_{-0.02}$ & $1.02^{+0.02}_{-0.02}$ \\ 
3C196  & $-5.42^{+0.12}_{-0.21}$ & $2.80^{+0.08}_{-0.06}$ & $<1.25$ & None & $4.60$ & $1.0$  \\ 
\hline
Diffuse MW$^a$ &$0.81[{0.61},{1.23}]$ & $0.27[0.01,0.40]$ & $1.06[0.63,1.27]$ & $0.11[0.02,0.16]$ & $4.60[4.55,4.68]$ & $1.0[0.8,1.3]$  \\
\hline
\end{tabular}
\begin{tablenotes}
      \small
     \item{ {\bf Notes:} $^a$ The typical values of UV extinction parameters for the MW diffuse ISM  from G23. The central values and ranges (shown in brackets) correspond to the extinction-curve parametrization, $a_{\mathrm{uv}}+b_{\mathrm{uv}} \times (1/R_V - 1/3.1)$, for a mean corresponding to $R_V$ of 3.1 and an observed range corresponding to $R_V=2.5–5.5$, respectively.}
\end{tablenotes}
\end{center}
\end{table*}

\setlength{\tabcolsep}{1pt}
\begin{table*}
\caption{Optical and IR extinction parameters}
\label{tab:extinction-parameters-opt-ir}
\begin{center}
\begin{tabular}{|l|c|c|c|c|c|c|c|}
\hline
  Quasar & $E_0$ & $E_1$ & $E_2$ & $E_3$& $E_4$  & $B$ & $\alpha$\\
  & [$A(\lambda)/A_V$] & [$A(\lambda)/A_V$] & [$A(\lambda)/A_V$] & [$A(\lambda)/A_V$]  &  [$A(\lambda)/A_V$] & [$A(\lambda)/A_V$] & \\ 
\hline
AO0235+164 &  $-1.37^{+0.37}_{-0.22}$ & $1.21^{+0.25}_{-0.60}$ & $0.42^{+0.38}_{-0.22}$&  $-0.31^{+0.10}_{-0.09}$ & $0.05^{+0.01}_{-0.02}$ & $0.02^{+0.04}_{-0.02}$ & $2.65^{+0.54}_{-0.18}$\\ 
 J0900+0214 &  $1.02^{+0.30}_{-0.60}$ & $-1.04^{+1.00}_{-0.80}$ & $1.21^{+0.55}_{-0.90}$ & $-0.37^{+0.20}_{-0.20}$ &  $0.04^{+0.02}_{-0.02}$ & $0.69^{+0.05}_{-0.03}$ & $1.24^{+0.05}_{-0.05}$\\ 
J0901+2044 &  $2.35^{+0.12}_{-0.12}$ & $-4.22^{+0.22}_{-0.25}$ & $3.69^{+0.12}_{-0.15}$ & $-1.20^{+0.04}_{-0.04}$ &  $0.14^{+0.01}_{-0.01}$ & $0.68^{+0.02}_{-0.03}$ & $1.19^{+0.02}_{-0.02}$\\ 
 J1007+2853  & $1.01^{+0.04}_{-0.07}$ & $-1.24^{+0.15}_{-0.10}$ & $1.25^{+0.08}_{-0.12}$ & $-0.39^{+0.05}_{-0.05}$ &  $0.05^{+0.01}_{-0.01}$ & $0.66^{+0.02}_{-0.02}$ & $1.18^{+0.02}_{-0.02}$\\ 
 J1017+4749 & $0.66^{+0.02}_{-0.02}$ & $0.18^{+0.02}_{-0.02}$& None & None & None & $0.85^{+0.02}_{-0.02}$ & $1.33^{+0.04}_{-0.04}$\\ 
\hline
 J0852+3425   & $0.43^{+0.30}_{-0.12}$ & $-1.97^{+0.50}_{-0.40}$& $2.58^{+0.25}_{-0.39}$& $-0.95^{+0.07}_{-0.07}$ &  $0.12^{+0.01}_{-0.01}$ & $0.26^{+0.02}_{-0.02}$ & $1.25^{+0.20}_{-0.20}$\\ 
 J0937+5628  & $-0.97^{+0.12}_{-0.05}$ & $1.08^{+0.02}_{-0.03}$ & None & None & None & $0.10^{+0.07}_{-0.02}$ & $>10$\\ 
 J1203+0634 &  $-4.60^{+0.80}_{-0.80}$ & $13.8^{+2.0}_{-1.9}$ & $-12.5^{+1.7}_{-1.7}$  & $4.70^{+0.66}_{-0.66}$ & $-0.58^{+0.10}_{-0.10}$& $0.83^{+0.03}_{-0.04}$ & $>10$\\
3C196  & $-1.30^{+0.18}_{-0.41}$ & $6.32^{+0.64}_{-0.95}$ & $-6.72^{+0.90}_{-0.96}$ & $2.91^{+0.15}_{-0.31}$ & $-0.39^{+0.04}_{-0.02}$ & $0.64^{+0.02}_{-0.03}$ & $1.18^{+0.03}_{-0.03}$\\ 
\hline
Diffuse MW$^a$ & -0.36[-0.37,-0.35] & 0.71[0.53,1.08] & 0.09[-0.19,0.23] & -0.05[-0.08,0.00] & 0.01[0.00,0.02] & 0.38[0.31,0.52] & 1.68[1.60,1.83]\\ 
\hline
\end{tabular}
\begin{tablenotes}
      \small
     \item{ {\bf Notes:} $^a$ The typical values of UV extinction parameters for the MW diffuse ISM from G23. The central values and ranges (shown in brackets) correspond to extinction-curve parametrization, $a_{\mathrm{uv}}+b_{\mathrm{uv}} \times (1/R_V - 1/3.1)$, for a mean at $R_V$ of 3.1 and an observed range at $R_V=2.5–5.5$, respectively.}
\end{tablenotes}
\end{center}
\end{table*}

\setlength{\tabcolsep}{2pt}
\begin{table*}
\caption{Best-fitting parameters of the extinction curves for the sample of quasar absorption systems}
\label{tab:extinction-results}
\begin{center}
\begin{tabular}{|l|c|c|c|c|c|c|c|c|c|c|}
\hline
  Quasar & $A_V$ & $A_{2175}$ & $S_{2175}$ & $R_V$  &  \multicolumn{6}{c|}{$A_{\lambda}/A_V$}\\
  & mag & mag &  &  &   0.45~$\mu$m & 0.80~$\mu$m & 1.24~$\mu$m & 1.65~$\mu$m & 2.15~$\mu$m & 4.45~$\mu$m \\ 
\hline
 AO0235+164& $0.70^{+0.05}_{-0.05}$ & $0.84^{+0.02}_{-0.03}$ & $1.32^{+0.02}_{-0.03}$ & $2.9^{+0.2}_{-0.1}$ & $1.33^{+0.04}_{-0.03}$ & $0.36^{+0.04}_{-0.03}$ &  $<0.03$ & $<0.01$ & $<0.01$ & $<0.01$  \\
 J0900+0214 & $1.49^{+0.08}_{-0.07}$ & $1.01^{+0.07}_{-0.07}$ & $1.51^{+0.11}_{-0.11}$ & $7.5^{+1.0}_{-1.0}$ & $1.13^{+0.02}_{-0.02}$ & $0.78^{+0.04}_{-0.03}$ &  $0.53^{+0.03}_{-0.03}$ & $0.37^{+0.02}_{-0.02}$ & $0.26^{+0.02}_{-0.02}$ & $0.09^{+0.01}_{-0.01}$  \\
 J0901+2044 & $0.80^{+0.04}_{-0.04}$ & $0.60^{+0.01}_{-0.01}$ & $0.88^{+0.03}_{-0.03}$ & $5.7^{+0.1}_{-0.2}$ & $1.17^{+0.01}_{-0.02}$ & $0.72^{+0.02}_{-0.02}$ & $0.51^{+0.02}_{-0.02}$ &  $0.37^{+0.02}_{-0.02}$ & $0.26^{+0.02}_{-0.02}$ & $0.09^{+0.01}_{-0.02}$  \\ 
 J1007+2853  & $1.92^{+0.02}_{-0.02}$ & $1.37^{+0.03}_{-0.03}$ & $2.11^{+0.05}_{-0.05}$ & $5.4^{+0.1}_{-0.2}$ & $1.18^{+0.01}_{-0.02}$ & $0.75^{+0.02}_{-0.02}$ & $0.50^{+0.02}_{-0.02}$ &  $0.36^{+0.02}_{-0.02}$ & $0.26^{+0.02}_{-0.02}$ & $0.09^{+0.01}_{-0.01}$  \\ 
 J1017+4749 & $0.84^{+0.03}_{-0.03}$ & $0.40^{+0.02}_{-0.02}$ & $0.63^{+0.02}_{-0.02}$ & $13.3^{+0.6}_{-1.0}$ & $1.07^{+0.02}_{-0.02}$ & $0.89^{+0.02}_{-0.02}$ & $0.63^{+0.02}_{-0.02}$ &  $0.43^{+0.02}_{-0.02}$ & $0.29^{+0.02}_{-0.02}$ & $0.09^{+0.01}_{-0.01}$  \\ 
\hline
 J0852+3425 & $1.00^{+0.03}_{-0.05}$ & $0.73^{+0.02}_{-0.02}$ & $1.09^{+0.02}_{-0.02}$ & $3.7^{+0.3}_{-0.3}$ & $1.26^{+0.03}_{-0.02}$ & $0.50^{+0.02}_{-0.02}$ & $0.20^{+0.04}_{-0.04}$ &  $0.14^{+0.05}_{-0.02}$ & $0.09^{+0.04}_{-0.02}$ & $0.02^{+0.02}_{-0.01}$  \\ 
 J0937+5628 & $0.52^{+0.05}_{-0.02}$ & $0.53^{+0.02}_{-0.02}$ & $0.90^{+0.04}_{-0.05}$ & $2.3^{+0.1}_{-0.1}$ & $1.44^{+0.02}_{-0.02}$ & $0.37^{+0.04}_{-0.04}$ & $<0.01$ &  $<0.01$ & $<0.01$ & $<0.01$\\ 
 J1203+0634  & $0.20^{+0.03}_{-0.03}$ & $0.51^{+0.02}_{-0.02}$ & $0.82^{+0.03}_{-0.03}$ & $1.5^{+0.2}_{-0.2}$ & $1.66^{+0.10}_{-0.10}$ & $0.86^{+0.04}_{-0.02}$ & $<0.01$ &  $<0.01$ & $<0.01$ & $<0.01$\\ 
3C196 & $0.08^{+0.01}_{-0.01}$ & $<0.10$ & $<0.31$ & $1.4^{+0.1}_{-0.1}$ & $1.72^{+0.03}_{-0.03}$ & $0.65^{+0.03}_{-0.03}$ & $<0.50$ &  $<0.35$ & $<0.25$ & $<0.1$\\ 
\hline
Diffuse MW$^a$ & ..& ..& .. & $3.1^{+2.4}_{-0.6}$ &  $1.32^{+0.08}_{-0.08}$ & $0.55^{+0.03}_{-0.03}$ & $0.27^{+0.02}_{-0.02}$ & $0.17^{+0.02}_{-0.02}$ & $0.11^{+0.02}_{-0.02}$ & $0.04^{+0.01}_{-0.01}$\\ 
\hline
\end{tabular}
\begin{tablenotes}
      \small
     \item{ {\bf Notes:} The typical values of $R_V$ for the MW diffuse ISM from \cite{Gordon2009}, and $A_{\lambda}/A_V$ for the average MW diffuse ISM extinction from \cite{Fitzpatrick2019} at 0.45~$\mu$m, \cite{Decleir2022} at 0.80$~\mu$m and \cite{Gordon2021} at 1.24-4.45$~\mu$m.}
\end{tablenotes}
\end{center}
\end{table*}

\section{Discussion}
\label{sec:dicussion}
We now discuss our results and compare the strength of the 10~$\mu$m feature, silicate mineralogy, and dust extinction in our sample of quasar absorption systems with studies of silicate dust grains in the MW ISM (\citealt{Gordon2021, Decleir2025}; representing cold dust) and with dust in nearby AGN tori and ULIRGs (warm dust) from the literature. Particularly, we examine (i) the implications for silicate mineralogy implied by the results of our profile fits; (ii) trends in silicate strength versus  other dust and gas properties, such as $A_{V}$ (assuming that silicate grains are responsible for the UV/visible extinction at least in part), the strength of the 2175 {\AA} bump feature (in order to investigate whether  silicate and carbonaceous grains correlate with each other in any way), and 
the gas-phase metal content (especially Mg and Fe which are important constituents of the silicate grains in the MW ISM).

\setlength{\tabcolsep}{2pt}
\begin{table*}
\begin{center}
\caption{Dust grain properties and gas-phase metal content for the sample of quasar absorption systems }
\label{tab:sightlines-results}
\begin{tabular}{|l|c|c|c|c|c|c|c|c|c|}
\hline
  Quasar & $z_{\rm abs}$ & $\tau_{10}^{\rm abs}$ & $\tau_{10}^{\rm abs}/A_V$ & $\tau_{10}^{\rm abs}/A_{2175}$        & $W^{\rm rest}_{\rm MgII2796}$   & $W^{\rm rest}_{\rm FeII2382}$     & $\lambda_c$ & \multicolumn{2}{c|}{Best-Fit Silicate Template$^a$}\\
   & &  &   & &  $\AA$  & $\AA$  & $\mu$m  & Fit1 & Fit2  \\
\hline
\multicolumn{10}{|c|}{Our sample (JWST MIRI MRS)}\\
\hline
AO0235+164 & 0.52 & 0.06$\pm$0.01 & $0.08^{+0.02}_{-0.02}$  & $0.07^{+0.01}_{-0.01}$  & 2.42$\pm$0.20 & 1.46$\pm$0.18 &  10.17  & Am.Oliv.GPC & Am.Oliv.2.0 \\ 
 J0900+0214& 1.05 & 0.30$\pm$0.10  & $0.20^{+0.08}_{-0.08}$ & $0.29^{+0.11}_{-0.11}$   & 3.95$\pm$0.24 & 1.92$\pm$0.53   & 10.17  &   Am.Oliv.2.0  & -  \\ 
 J0901+2044 & 1.02 &0.13$\pm$0.03 &$0.16^{+0.05}_{-0.05}$ & $0.22^{+0.06}_{-0.06}$  & 2.04$\pm$0.05 & 1.48$\pm$0.03 & 10.55  & Hypersthene & Am.Oliv.2.0 \\ 
 J1007+2853 & 0.88 & 0.17$\pm$0.03&  $0.09^{+0.02}_{-0.02}$ & $0.12^{+0.03}_{-0.03}$  & 3.34$\pm$0.17 &2.61$\pm$0.35  & 9.70  & Montmorillonite & Am.Oliv.  \\ 
 J1017+4749 & 1.12 & 0.22$\pm$0.02 & $0.26^{+0.03}_{-0.03}$ & $0.54^{+0.08}_{-0.08}$  & 2.78$\pm$0.13 & 1.69$\pm$0.11 & 11.20  & \multicolumn{2}{c|}{Crystal.Oliv.CDE1}\\ 
\hline
\multicolumn{10}{|c|}{Archival sample (SST IRS), \cite{Kulkarni2011,Aller2014}}\\
\hline
 J0852$+$3435 & 1.31  & 0.17$\pm$0.01 &  $0.17^{+0.02}_{-0.02}$ & $0.22^{+0.02}_{-0.02}$  & 2.86$\pm$0.12 & 1.66$\pm$0.20  &  10.17  & Am.Oliv.GPC & - \\ 
 J0937$+$5628 & 0.98 & 0.19$\pm$0.02  & $0.37^{+0.07}_{-0.07}$ & $0.36^{+0.05}_{-0.05}$ & 4.94$\pm$0.12 & 2.50$\pm$0.12  & 10.17  & Am.Oliv.GPC& -\\ 
 J1203$+$0634 & 0.86 & 0.28$\pm$0.25  &$1.35^{+0.40}_{-0.40}$ & $0.52^{+0.50}_{-0.50}$  & 5.30$\pm$0.15& 3.85$\pm$0.11   &  9.17  & Am.Pyrox& -\\ 
  3C~196       & 0.44 & 0.10$\pm$0.01  & $1.28^{+0.24}_{-0.24}$ & $>1.0$ & 1.88$\pm$0.07 & 1.57$\pm$0.30  & 10.55  & Am.Oliv.2.0& -\\ 
  \hline
\end{tabular}
\begin{tablenotes}
      \small
     \item{ {\bf Notes:} $^a$ The last two columns present the names of the best-fit laboratory-based templates for the 10~$\mu$m silicate feature from Table~\ref{tab:silictae-fit-results} using different continuum models (see text for details).} 
\end{tablenotes}
\end{center}
\end{table*}

To quantify the gas-phase metal content, we measured the rest-frame equivalent widths ($W_r$) of the Mg~{\sc ii}~2796~\AA\ and Fe~{\sc ii}~2382~\AA\ lines in SDSS quasar spectra at the redshift of quasar absorbers (see also Paper I and \citealt{Kulkarni2011}). For a sample of MW sightlines from \cite{Gordon2021}, we measured $W_r$ for these features using HST/STIS spectra obtained with the E230M grating from program PID~11516 (PI J.~Green) and derived it for cases where both the Mg~{\sc ii}~2796~\AA\ and Fe~{\sc ii}~2382~\AA\ lines are covered. For the sample from \cite{Decleir2025}, only the Fe~{\sc ii}~2382~\AA\ line is covered by existing STIS data: PID 8462 (PI M. Pena), PID 9465 (PI U. Sofia), PID 15335 (PI A.~De~Cia), PID 16285 (PI M. Decleir), PID 17703 (PI A.~De~Cia). In order to estimate $W_r$ for Mg~{\sc ii}~2796~\AA,  we  simulated Voigt profiles using the oscillator strength for Mg~{\sc ii}~2796~\AA\ and column densities and Doppler parameters of Mg~{\sc ii}  derived from fitting to Mg~{\sc ii}~$\lambda\lambda$1239,1240 absorption lines by \cite{JS2007}.

\begin{figure*}
\begin{center}
        \includegraphics[width=1\textwidth]{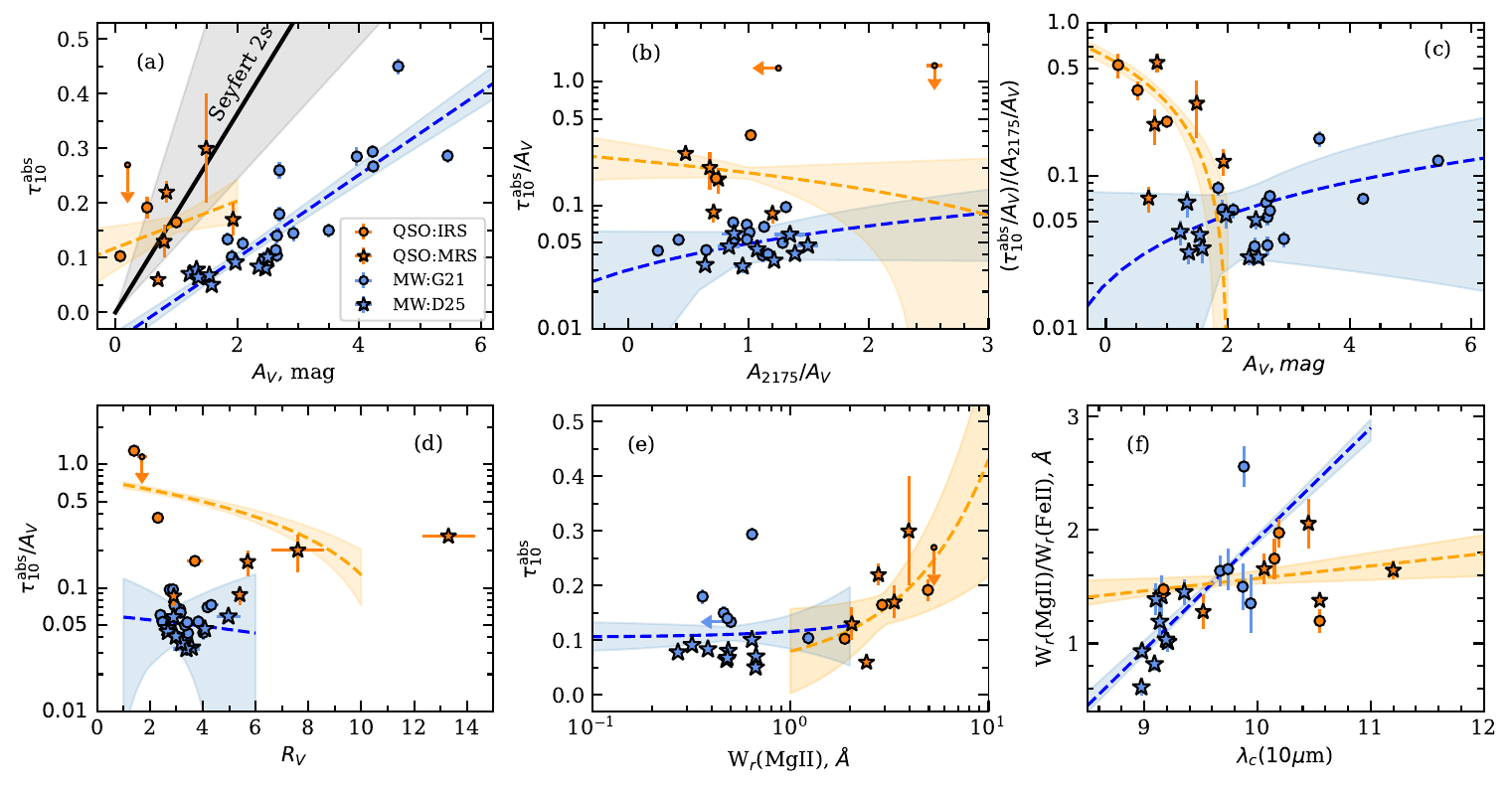}
        \caption{\textit{Top row}: (a) the peak optical depth of 10~$\mu$m silicate absorption ($\tau_{\rm 10}^{\rm abs}$) versus  V-band extinction $A_{V}$; (b) the ratio of peak optical depth to $A_V$ ($\tau_{\rm 10}^{\rm abs}/A_V$) versus the ratio  of the 2175 {\AA} bump strength to $A_V$ ($A_{2175}/A_V$); and (c) the ratio of the strengths of the 10~$\mu$m silicate and 2175~\AA\ features      
        $\tau_{\rm 10}^{\rm abs}/A_{2175}$ versus $A_V$. Orange stars and circles represent data from quasar sightlines from our sample (using observations with MIRI MRS) and the archival sample (using SST IRS;  \citealt{Kulkarni2011, Aller2014}). The small blue circles and stars represent data for the local diffuse ISM in studies using SST IRS  \citep{Gordon2021}, and MIRI MRS (\citealt{Decleir2025}). The black line and gray shaded area in panel (a) represent the average ratio of $A_V/\tau_{10}=5.5\pm2.7$ in for dust in tori of Seyfert~II galaxies from \cite{Lyu2014}. 
        The blue and orange dashed lines represent the relations for the MW ISM and our sample, respectively, fitted with a linear function that has a variable intercept.
        The shaded regions around the fits show the $\pm1\sigma$ uncertainty.        \textit{Bottom row}:  (d) $\tau_{\rm 10}^{\rm abs}/A_V$ versus  the ratio of total-to-selective extinction $R_V$,  (e) $\tau_{\rm 10}^{\rm abs}$ versus Mg II $\lambda$2796 rest-frame equivalent width,  (f) the ratio of Mg II $\lambda$2796 and Fe II $\lambda$2382 rest-frame equivalent widths versus the peak wavelength of the 10 $\mu$m silicate feature. The symbols are the same as in the top panels.
        }
        \label{fig:correlations-vs-MW}
\end{center}
\end{figure*}

Table~\ref{tab:sightlines-results} summarizes the dust properties 
(the peak optical depth of the 10~$\mu$m silicate absorption, the ratios of the silicate peak optical depth to $A_V$ and $A_{2175}$, 
the rest equivalent widths of Mg~{\sc ii}~2796\AA\ and Fe~{\sc ii}~2382\AA\ absorption lines, the peak wavelength ($\lambda_c$) of the optical depth profile, and the best-fit laboratory-based silicate templates) for our sample and the archival sample. The comparison with dust properties in the MW samples and nearby AGN tori is shown in Figure~\ref{fig:correlations-vs-MW} and discussed below.

\subsection{Silicate Grain Structure and Mineralogy}
\label{sec:minearology}
Differences in silicate dust grain mineralogy and structure, such as composition, stoichiometry, geometry, size, temperature, and crystallinity, impact the shape and substructure within the $\sim 10~\mu$m absorption feature. This is illustrated in the right panels of Figure~\ref{fig:tau-profiles-comparison}, which depicts amorphous olivine and pyroxene, a crystalline olivine (Mg$_{1.9}$Fe$_{0.1}$SiO$_4$) with a continuous distribution of ellipsoidal (CDE) particles \citep{Fabian2001}, the crystalline pyroxene natural-orthohypersthene \citep[Mg$_{0.65}$Fe$_{0.35}$SiO$_3$][]{Jager1998}, and the phyllosilicate montmorillonite \citep{Dorschner1978}. Amorphous grains produce broad features which lack substructure, but differ in their peak absorption wavelength with composition and stoichiometry (e.g., 9.0-9.5~$\mu$m for pyroxenes vs. 9.7-10.5~$\mu$m for olivines, see Table~10 in \citealt{Aller2012}). By contrast, crystalline olivine and pyroxene profiles typically peak at longer wavelengths ($>10~\mu$m), and include narrower substructures. Past SST IRS studies of QASs found spectral structures suggestive of variations in silicate grain properties, such as crystallinity \citep[e.g.,][]{Aller2012, Kulkarni2016}, and noted that the peak wavelength and breadth of some 10~$\mu$m absorption features are shifted to longer wavelengths and are too broad to be consistent with the Mg-rich, amorphous olivine templates that typically best-fit the MW ISM \citep{Kemper2004}. Our JWST MRS spectra provide the first opportunity to examine the 10 $\mu$m feature from interstellar silicate dust in distant, normal galaxies at moderately-high resolution.  

We find that the silicate absorption feature shapes and peak wavelengths vary from system-to-system (Figure~\ref{fig:tau-profiles-comparison}, left) 
resulting in differences in the best-fitting mineralogies, as illustrated in Figure~\ref{fig:tau-profiles-best-fit}. Varying species of amorphous olivine, crystalline olivines and pyroxenes, and phyllosilicates yield acceptable fits, but neither pure amorphous pyroxene nor silicon carbide (SiC) produce the best fits in any systems (see Table~\ref{tab:silictae-fit-results}). Based on the reduced chi-squared ($\chi_{red}^2$) values, amorphous olivine fits are better than amorphous pyroxene fits for all sources.\footnote{For J1007+2853, when the quasar silicate emission contribution is not included in the fitting, the amorphous olivine and pyroxene fits are comparable but both poor ($\chi_{red}^2\geq2.5$).} For comparison, in the MW diffuse ISM, \citet{Chiar2006} found that when fitting the observed 10 and 18~$\mu$m silicate features simultaneously, they were well-reproduced by olivine grains mixed with pyroxenes, but when fitting over just the 10~$\mu$m feature,  it was best-matched by a pure olivine species. Similarly, \citet{Min2007} found that MW interstellar silicates have a stoichiometry in-between pyroxenes and olivines (O/Si $\approx 3.5$), with a small percentage of the material ($\sim$3\%) present in the form of silicon carbide. In our analysis, we lack coverage of the 18~$\mu$m feature to investigate whether a combination of olivines and pyroxenes are required to simultaneously fit the 10 and 18~$\mu$m features for our absorbers.  We summarize the key results from our 10~$\mu$m profile fits for the various templates in Table~\ref{tab:silictae-fit-results} and Figs.~\ref{fig:fit-silicate-ao0235-fit1}-\ref{fig:fit-silicate-j1017}, and discuss below the implications and key results of these fits. For each mineralogical category in Table~\ref{tab:silictae-fit-results}, we present the best-fitting template, as selected using the $\chi_{red}^2$ measured over the fitting region. The overall best-fitting mineralogies are identified in bold-face in Table~\ref{tab:silictae-fit-results}. 

The 10~$\mu$m feature in the absorbers toward AO~0235+164 and J0900+0214 is best-fit by amorphous olivine templates, as in the MW diffuse ISM, albeit with possibly different grain properties. For J0900+0214 and AO~0235+164 (Fit 2), a larger grain size ($\sim$2~$\mu$m) is implied by the fits.
The spectral characteristics of large grain size are a broadened 10~$\mu$m band with the peak shifted to a slightly longer wavelength \citep{Dorschner1995}. The alternate best-fit for AO~0235+164 (Fit~1) is an amorphous olivine (MgFeSiO$_4$) template based on porous (50\% porosity) grains with a continuous distribution of ellipsoids (CDE); such profiles are also broader than for solid CDEs \citep{Chiar2006}. Thus, we find that even with the higher spectral resolution of MIRI MRS, the 10~$\mu$m features are broader than for typical MW diffuse ISM sightlines, and are shifted to longer peak absorption wavelengths. Slightly poorer fits ($\Delta \chi_{red}^2\leq0.2$) are provided by crystalline olivines, crystalline pyroxenes, and/or phyllosilicates. For AO~0235+164 (Fit~1) both chamosite (a phyllosilicate with Fe$\cdot\cdot_4$Al$[$Si$_3$AlO$_{10}][$OH$]_6\cdot$ nH$_2$O; \citealt{Dorschner1978}) and an average crystalline pyroxene (blending enstatite, bronzite, hypersthene, anthophyllite, dopside, pyroxene-pigeonite, wollastonite, and augite; \citealt{Koike2006}) provide viable fits. \citet{Dorschner1978} also found consistency between chamosite and the 10~$\mu$m absorption feature in MW sightlines. For AO~0235+164 (Fit~2), amorphous olivine provides the best fit, while the next best-fit ($\Delta\chi_{red}^2=0.47$) is produced by natural orthohypersthene (hypersthene; Mg$_{0.65}$Fe$_{0.35}$SiO$_3$, \citealt{Jager1998}), a crystalline pyroxene species with a longer central peak wavelength ($\sim$10.6~$\mu$m) and relatively-broad feature, and features just blueward ($\sim$10.4~$\mu$m) and redward ($\sim$11.4~$\mu$m) of the central peak. Hypersthene also produces the second-best fit in J0900+0214, with a warm version of the Jena crystalline forsterite, a Mg-rich crystalline olivine, (Mg$_2$SiO$_4$), at T=300 K, \citep{Koike2006} producing a comparable match. 

For the absorber toward J0901+2044, a range of plausible mineralogies are consistent with the observed 10~$\mu$m absorption profile. When including 6.2~$\mu$m PAH emission from the physically-unrelated background AGN in the fitting, the best fit is produced by amorphous olivine with a large (2~$\mu$m) grain size, but statistically comparable fits result from a Mg-poor, crystalline olivine (synthetic fayalite, Fe$_2$SiO$_4$, \citealt{Jager1998}) with a relatively long ($\sim11.4$~$\mu$m) peak wavelength and a nearby feature at 12.1~$\mu$m, and by hypersthene, with SiC and chamosite producing slightly poorer fits. We note that fayalite and hypersthene have the longest peak wavelengths of the considered crystalline olivine and pyroxene species.  When the PAH emission is ignored during the fitting,  hypersthene provides the best-fit, with amorphous olivine, chamosite, and Mg-rich synthetic forsterite (Mg$_2$SiO$_4$, \citealt{Jager1998}) producing viable fits ($\Delta\chi_{red}^2\leq0.2$). Depending on whether the AGN 6.2~$\mu$m PAH feature is accounted for in the fitting the best-fitting mineralogy can shift from Mg-poor fayalite (when accounted for) to Mg-rich forsterite (when omitted). 

In the absorber toward J1007+2853, the identified grain properties likewise depend on whether or not quasar emission features are accounted for when fitting the shape of the feature, although the derived peak optical depths ($\tau_{10}$) are similar. When silicate emission attributed to the dusty torus in the physically-unrelated background AGN is accounted for in the fit, amorphous olivine similar to the MW diffuse ISM and montmorillonite provide good fits. When this feature is ignored, the best-fit is produced by only the phyllosilicate montmorillonite (($m$\{Mg$_3[$Si$_4$O$_{10}][$OH$]_2$\}$\cdot\cdot$ $p$\}(Al, Fe $\cdot\cdot\cdot$)$_2[$Si$_4$O$_{10}][$OH$]_2$\}$\cdot$ nH$_2$O with $m:p\sim0.8-0.9$; \cite{Dorschner1978}). 

For the absorber toward J1017+4749, regardless of whether we include the contribution from the AGN silicate dust emission, the best-fitting template is Mg-rich crystalline olivine \citep[Mg$_{1.9}$Fe$_{0.1}$SiO$_4$ with CDE particles of sizes in the Rayleigh limit;][]{Fabian2001}). Amorphous olivine does not produce a comparable fit to crystalline olivine, even when allowing for larger grain sizes. This is largely because the peak absorption wavelength occurs redward of 11~$\mu$m, in contrast to the peak at $\sim$9.7-10.5~$\mu$m expected from amorphous olivine (see Figures~\ref{fig:tau-profiles-comparison}-\ref{fig:tau-profiles-best-fit}).

For the majority (80\%) of absorbers, we identify a species of amorphous olivine as producing the best-fit, similar to the MW diffuse ISM. However, for all but J1007+2853, the implied best-fitting grain silicate size would be larger ($\sim1.5-2~\mu$m). Large-grain fits better match due to the breadth of the 10~$\mu$m absorption feature and its slightly longer peak absorption wavelengths. Based on the MW diffuse ISM, dust grain sizes are typically characterized by a power-law distribution with $dn/da=a^{-3.5}$ (\citealt{Mathis1977}, MRN distribution) where $a$ is the dust grain radius, reaching a maximum size of 0.25~$\mu$m for silicate grains. In the diffuse Galactic ISM, grains are typically sub-micron sized, Mg-rich, amorphous olivinic silicates \citep[e.g.,][and references therein]{Bromley2024}, but large grained silicates (up to $\sim$1-2~$\mu$m) have been associated with stellar environments, including some planetary nebulae and AGB outflows \cite[e.g.,][]{Hofner2008,Bladh2017}.  Comparing the observed surface brightness profiles in the near-IR with radiative transfer models of scattering by dust grains for a low-mass molecular core, \citet{Andersen2013} found that interstellar silicate grains larger than 0.25~$\mu$m are needed, but the largest grains are limited to 1-1.5~$\mu$m sizes. Thus, the implied grain sizes in our systems are at or slightly-above this upper limit. Furthermore since a large contribution of such grains would be required, the size distribution appears to be skewed toward larger grain sizes relative to the MW. For comparison, \citet{Andersen2013} found that only $4.6\times10^{-6}$ of the silicate dust grains by number (albeit 25\% of the grains by mass) would be larger than $0.5~\mu$m. Alternatively, rather than 
larger grains, the broader features could be the result of a mixture of mineralogical species combining to produce a broader feature and shift the peak to longer wavelengths. For instance, alumina dust (Al$_2$O$_3$), is commonly present in AGB stars and other stellar objects such as dusty red novae \citep{Steinmetz2025,Speck2000,Takigawa2017,Sargent2018}, where it can broaden the 10~$\mu$m feature due to a peak near 11-12~$\mu$m. In some such sources, the alumina to silicate ratio can be as high as 3:2 \citep{Steinmetz2025}. We also note that for several of our sources the 10~$\mu$m feature on the long-wavelength side has a low SNR value due to the high thermal noise in the Channel 4 of the MRS detector, which could be artificially broadening the silicate feature. 

A Mg-rich crystalline olivine species yields the best fit for the absorber toward J1017+4749, while the best-fitting species of the other sources range from  Mg-rich (e.g., forsterite) to Mg-poor (e.g., fayalite) species. In the MW diffuse ISM, amorphous silicates are Mg-rich ($>90$\% Mg, $<$10\% Fe), comprised of a blend of olivines and pyroxenes \citep{Min2007}, unlike some of our fits. However, as discussed above, the identified mineralogy can strongly depend on the adopted shape of the continuum and inclusion of background AGN emission features. 

Crystalline silicate species are able to account for the longer peak absorption wavelengths and breadths of the features in many of our sources, as well as some of the possible substructure. For J1017+4749, crystalline silicates provide the best fit, while for the absorbers toward AO~O235+164, J0900+0214, and J0901+2044, some substructure is not accounted for with the amorphous olivine fits (see Figure~\ref{fig:tau-profiles-best-fit}). Silicate crystallinity has been identified in comets, stellar environments, AGN tori, and merging, star-forming galaxies such as ULIRGs (e.g., \citealt{Spoon2006}), but in the MW diffuse ISM, grains are predominately amorphous ($<$5\% crystallinity; \citealt{Kemper2004,Li2007}). This has been attributed to rapid amorphization by mechanisms such as cosmic-ray bombardment in the ISM \citep{Kemper2004,Bringa2007}. Higher crystallinity (e.g. up to 15\% in ULIRGs, \citealt{Spoon2006}) has been attributed to recent star-formation with a lag in the amorphization proccess relative to the formation and injection of new dust into the ISM. However, \citet{Kemper2011} suggest that even with the high star formation rates (SFRs) found in ULIRGs and efficient dust production, it is challenging to explain crystallinities up to 15\%. In diffuse ISM environments such as probed in this study, we would expect little silicate crystallinity, but amorphous olivines alone cannot account for the residual substructure, nor do amorphous silicates adequately fit the absorber toward J1017+4749. This suggests that there may be some degree of silicate crystallinity in these sources, which could be indicative of variations in interstellar processing with environment or redshift, such as lower cosmic ray spallation rates or weaker radiation fields. Alternatively, the substructure could be due to due to noise features in the longer-wavelength MIRI MRS data or nanosilicates which can produce such small sub-structures \citep{Bromley2024}. To definitively identify whether the substructure features are associated with crystalline silicates, we would need to investigate whether longer wavelength resonance features are present in the spectra; for instance, \citet{Spoon2006,Spoon2022} identified crystalline forsterite silicate features at 16-33 $\mu$m in the SST IRS spectra of $\sim$24\%  of the galaxies in the Infrared Database of Extragalactic Observables from Spitzer (IDEOS). Those features are beyond our MIRI MRS rest-frame wavelength coverage, but the large fraction of galaxies with some (albeit low, $\sim2.7$\% on average) crystallinity \citep{Spoon2022}, means that crystalline structure may be present in the dust grains for our sources. \citet{Tsuchikawa2021} found that the crystallinity and olivine-to-pyroxene ratio are higher for silicates in heavily obscured AGNs compared to those in the local diffuse ISM, and \citet{Spoon2022} found that the crystalline band strengths increase with the amorphous silicate strength. Considering the stronger silicate features in our non-local sightlines compared to the MW ISM (see Sec. 5.2), it would not be surprising to find crystallinity in some of our sightlines. Since in this analysis we do not consider fits blending amorphous and crystalline silicates, as in \citet{Aller2012,Aller2014}, we are unable to address whether the reason for the substructure is a blend of crystalline and amorphous grains, or quantify the percent crystallinity of our sources. 

For several of the sources (AO~0235+164 Fit1, J0901+2044, J1007+2853), species of sub-micron-sized phyllosilicates (chamosite and monotmorillonite) produce viable fits; in the case of J1007+2853 when not including AGN emission in the fits this species produces the best fit. These more complex minerals include a combination of Fe, Al, Si, O, hydroxide, and H$_2$O, as well as Mg for montmorillonite. Phyllosilicates were originally considered as possible sources for the 10~$\mu$m feature in the MW diffuse ISM, since they are constituents of Type I and Type II carbonaceous chondrites, and may thus be similar to the primitive material found in the solar nebula \citep{Zaikowski1975,Dorschner1978}. These materials are also found in environments such as ancient terrains on Mars \citep{Mustard2008}. Furthermore, olivines and pyroxenes can be transformed into phyllosilicates at low temperatures in environments with H$_2$O, such as may exist in some stellar environments and dense molecular clouds \citep{Dorschner1978}. When compared with interstellar dust, such as in the Orion Trapezium region, \citet{Dorschner1978} found the phyllosilicates produced narrower bands, which was attributed to an interstellar broadening mechanism, with chamosite providing the best-fit to the MW ISM. In subsequent studies, however, features originally attributed to phyllosilicates have since been attributed to water ice and silicates such as olivines (e.g., \citealt{Gibb2004,Potapov2021}). Thus, should phyllosilicates be producing the absorption features detected in our absorbers, it would be quite unusual.

\subsection{Silicate Absorption vs. Extinction}

It is of much interest to understand how the strength of the silicate absorption correlates with the extinction in the sightline. While one expects stronger silicate absorption in dustier sightlines, it is not essential that such sightlines also have higher extinction and reddening (since silicate features can arise in small as well as large grains, and the large grains do not contribute much to reddening).

\setlength{\tabcolsep}{2pt}
\begin{table}
\begin{center}
\caption{Results of the correlation tests between the strength and peak wavelength of the silicate absorption, the dust extinction (A$_V$), the strength of the 2175~\AA\ bump, and the rest-frame equivalent widths of the Mg~{\sc ii}~$\lambda$2796 and Fe~{\sc ii}~$\lambda$2382 absorption lines in samples of the MW sightlines and the absorbers in quasar spectra}
\label{tab:correlation-results}
\begin{tabular}{|l|c|c|c|c|c|c|}
\hline
   &   & \multicolumn{2}{c|}{Correlation }& \multicolumn{3}{c|}{Linear fit: $f(x)=k(x-x_{ref}) + b$ }\\
  Parameters & Sample & $r_S$ & $P$   & $k$ & $b$ & $x_{ref}$\\
   \hline
   \multirow{2}{*}{$\tau_{10}^{\rm abs}$ vs. $A_V$} & MW & 0.89 &$9.2\times10^{-10}$ & $0.076\pm0.006$ & $0.100\pm0.008$ & 2 \\
   & QSO & 0.35 & 0.43 & $0.025\pm0.041$ & $0.161\pm0.016$  & 1\\
    \hline
   \multirow{2}{*}{$\tau_{10}^{\rm abs}/A_V$ vs. $A_V$} & {MW} & {$0.29$} & {$0.14$} & $0.007\pm0.006$ & $0.050\pm0.010$ & 2 \\
   & {QSO} & {$-0.21$} & {$0.64$} & $-0.122\pm0.040$ & $0.185\pm0.016$ & 1 \\
    \hline
   \multirow{2}{*}{$\tau_{10}^{\rm abs}/A_V$ vs. $A_{2175}/A_V$} & {MW} & {$0.07$} & {$0.73$} & $0.006\pm0.050$ & $0.052\pm0.018$ & 1 \\
   & {QSO} & {$0.35$} & {$0.43$} & $-0.040\pm0.120$ & $0.182\pm0.033$ & 1  \\
    \hline
   \multirow{2}{*}{$\tau_{10}^{\rm abs}$ vs. $A_{2175}$} & {MW} & {$0.52$} & {$0.01$} & $0.043\pm0.020$ & $0.123\pm0.018$ & 2 \\
   & {QSO} & {$-0.23$} & {$0.57$} & $-0.048\pm0.090$ & $0.175\pm0.040$ & 1  \\
    \hline
   \multirow{2}{*}{$\tau_{10}^{\rm abs}/A_{2175}$ vs. $A_V$} & {MW} & {$0.36$} & {$0.09$} & $0.020\pm0.018$ & $0.050\pm0.020$ & 2 \\
   & {QSO} & {$-0.51$} & {$0.15$} & $-0.300\pm0.042$ & $0.300\pm0.034$ & 1  \\
    \hline
      \multirow{2}{*}{$\tau_{10}^{\rm abs}/A_V$ vs. $R_V$} & {MW} & {$-0.13$} & {$0.50$} & $-0.003\pm0.023$ & $0.052\pm0.020$ & 3  \\
   & {QSO} & {$-0.45$} & {$0.22$} & $-0.062\pm0.008$ & $0.561\pm0.030$ & 3  \\
    \hline
   \multirow{2}{*}{$\tau_{10}^{\rm abs}$ vs. $W_r({\rm MgII})$} & {MW} & {$-0.10$} & {$0.71$} & $0.011\pm0.042$ & $0.111\pm0.014$  & 0.5\\
    & {QSO} & {$0.78$} & {$0.01$} & $0.040\pm0.028$ & $0.158\pm0.031$  & 3\\
    \hline
      \multirow{2}{*}{$\tau_{10}^{\rm abs}/A_V$ vs. $W_r({\rm MgII})$} & {MW} & {$-0.28$} & {$0.31$} & $-0.005\pm0.043$ & $0.050\pm0.010$ & 0.5  \\
   & {QSO} & {$-0.11$} & {$0.77$} & $-0.012\pm0.025$ & $0.158\pm0.030$ & 3  \\
    \hline
      \multirow{2}{*}{$W_r({\rm MgII})$ vs. $A_V$} & {MW} & {$0.06$} & {$0.82$} & $0.050\pm0.015$ & $0.530\pm0.020$ & 2  \\
   & {QSO} & {$0.73$} & {$0.03$} & $2.168\pm0.100$ & $3.150\pm0.040$ & 1  \\
    \hline
   \multirow{2}{*}{$W_r({\rm MgII})/W_r({\rm FeII})$ vs. $\lambda_c$} & {MW} & {$0.76$} & {$1.5\times10^{-3}$} & $0.967\pm0.027$ & $1.680\pm0.020$ & 9.7  \\
    & {QSO} & {$-0.03$} & {$0.93$} & $0.095\pm0.051$ & $ 1.554\pm0.035$  & 9.7\\
  \hline
\end{tabular}
\begin{tablenotes}
      \small 
     \item{{\bf Notes:}} Columns description: $r_S$ represents the Spearman rank-order correlation coefficient; $P$ denotes the $p$-value indicating the probability that the observed $r_S$ value arises purely by chance; $k$ and $b$ are the slope and intercept parameters of the linear fit, respectively.  $x_{ref}$ represents a reference value for the linear fit function and was chosen to be close to the average value of the $x$ parameter for the studied sample.
    \end{tablenotes}
\end{center}
\end{table}

Figure~\ref{fig:correlations-vs-MW}a shows the comparison of the peak optical depth of the 10 $\mu$m silicate absorption feature versus the extinction $A_{V}$ in our new and archival quasar absorbers, and in the diffuse sightlines in the MW from observations with SST IRS \cite{Gordon2021} and MIRI MRS \cite{Decleir2025}.
We also compare these results with the $\tau_{10}/A_V$ ratio found for dust in the tori of Seyfert~2 galaxies \citep{Lyu2014}, although in that study $A_{V}$ values were inferred from the H-$\alpha$/H-$\beta$ emission line ratio.
While systems in  quasar sightlines do not show a significant correlation between $\tau_{10}$ and $A_V$, the values of $\tau_{10}$ are significantly larger than the values for comparable $A_V$ in  the MW diffuse ISM. 
This confirms our previous findings from  SST IRS-based studies \citep[e.g.,][]{Kulkarni2011, Aller2014}.
We fit this relation using a linear function, allowing the intercept to vary (i.e., considering the strength of the silicate feature to be independent of the V-band extinction, since $\tau_{10}$ and $A_V$ can be sensitive to different type of grains in terms of compositions and sizes). The fitting results and Spearman rank-order correlation coefficients are presented in Table~\ref{tab:correlation-results}.

The median value of $\tau_{10}/A_{V}$ for our sightlines\footnote{We do not include J1203+0634 in the calculation because of the poor signal-to-noise ratio in the SST/IRS spectrum and the only 1$\sigma$ significance of the detection of the 10~$\mu$m feature.} ($\tau_{10}/A_{V}=0.17\pm0.09$) is $\sim$3 times  higher than that for the MW ISM sightlines ($\tau_{10}/A_{V}=0.059\pm0.006$).\footnote{We have calculated an average value of $\tau_{10}/A_V$ for the combined sample of MW sight lines from \cite{Gordon2021} and \cite{Decleir2025}. We note that the values of $\tau_{10}/A_V$ for the two samples differ slightly: \cite{Decleir2025} found an average of $\tau_{10}/A_V = 0.043$, while \cite{Gordon2021} reported $\tau_{10}/A_V = 0.064$. This difference likely arises from  the different fitting method used, see \cite{Decleir2025}.}
A similar behavior has been reported for Seyfert~2 galaxies, which show $\tau_{10}/A_V = 0.18^{+0.16}_{-0.06}$.

It thus appears that the dust grains in normal galaxies at the intermediate (z$\sim$1) redshifts considered may have higher quantities of silicate dust than expected for the amount of extinction in the sightline (more silicate-rich grains). This may be consistent with the expectation that the dust produced in stellar outflows and core-collapse SNe in younger galaxies may be more dominated by {\it silicates} \citep[e.g.,][]{Dwek1998, Dwek2011, Zhukovska2014, Asano2014, Dubois2024}.
Indeed such a possibility has been invoked to explain the $\tau_{10}/A_{V}\sim 0.11$ (or $A_{V}/\tau_{10} \sim 9$) observed toward the center of the MW 
by \citet{Roche1985}, who suggested that the presence of more O-rich stars near the center of the MW would increase the abundance of silicate grains. Alternatively, the higher $\tau_{10}/A_V$ values may suggest that the silicate dust grains in distant galaxies are associated with smaller $A_{V}$ than expected for their optical depths, possibly suggesting {\it a lower fraction of  small grains}.

The fact that the silicate optical depth to $A_V$ ratios for our sightlines  are in good agreement with Seyfert 2 AGN dust ratios suggests that the dust grains in our sightlines may be warmer than dust grains in the MW ISM (although not as warm as the AGN dust grains). Given that the spatial extent of the diffuse ISM is much smaller than that of the circumgalactic medium (CGM), random quasar sightlines are more likely to probe dust in the CGM rather than the ISM of foreground galaxies. This favors a scenario in which grains have higher temperatures and lower $A_V$ values, since small grains can be destroyed by sputtering in shocks driven by galactic winds or supernova ejecta, or by exposure to a stronger radiation field in the CGM. Indeed, we note that one of the lowest $R_{V}$ values in our sample is found in the sightline to 3C196, which passes only $\sim$8.5 kpc from the center of a star-forming spiral galaxy \citep{Gharanfoli2007}, consistent with small grains surviving in the ISM. For the other sightline (J1203+0634) with a low $R_V$ value, a galaxy 3$\arcsec$ from the quasar is seen in SDSS data, but does not have a spectroscopic confirmation \citep{Kulkarni2011}. We also note that the sightline to AO0235+164 also traces the ISM, passing  $\sim7$ kpc from a spiral galaxy \citep[e.g.,][]{Chun2006} and shows an $R_{V}$ value comparable to that in the MW. It would be interesting to determine whether the  sightlines with the large $R_{V}$ values pass at larger impact parameters from the centers of the absorbing galaxies.

Dust grains are expected to absorb and scatter light most efficiently at wavelengths comparable to their sizes ($2 \pi a/\lambda \sim$1). 
Assuming compact spherical grains with dielectric functions of astronomical silicates from \cite{Draine1984}, \cite{Lyu2014} calculated the distribution of $A_{V}$/$\tau_{10}$ values expected for grains of different sizes from Mie theory. For dust in the AGN circumnuclear environment, they derived $a < 0.09~\mu$m or $a \geq 0.44~\mu$m, giving priority to large grains ($a>0.44~\mu$m), since small grains are unlikely to survive in the hostile AGN environment.
In contrast, the median $A_{V}$/$\tau_{10}$ value of 16.9 in the MW diffuse ISM suggests a typical grain size of about $0.2~\mu$m. By comparing our median $A_{V}$/$\tau_{10} = 5.9$ to this distribution, we infer that the typical grain size in our extragalactic sightlines is either small ($a \sim 0.08~\mu$m) or large ($a \sim 0.4~\mu$m), with the latter scenario being more likely. 
This is consistent with the indications of larger grain sizes deduced from the shapes of the silicate absorption profiles (Sec.~\ref{sec:minearology}). 

The slope of the extinction curve also provides information about the dust-grain size distribution. Figure~\ref{fig:correlations-vs-MW}d shows the distribution of $\tau_{10}/A_V$ vs. $R_V$, an indicator of the relative abundance of large versus small grains. Although our sample spans a wide range of $R_V$ values ($1.4$–$13.3$), we do not find a significant correlation between $R_V$ and $\tau_{10}/A_V$. This may indicate that our sightlines probe dust in a wide variety of environments - where only small or only large grains survive - or that some other factors significantly affect the $\tau_{10}/A_V$ ratio.
However, we note that for the two sightlines (AO0235+164 and J1007+2853) where $\tau_{10}/A_V\simeq0.09$ is closest to the MW ISM value, the corresponding $R_V$ values are also consistent with those observed in MW sightlines. In contrast, the remaining sightlines, which have much larger $\tau_{10}/A_V$, show $R_V$ values that differ significantly, being either lower or higher than the typical $R_V$ in the MW diffuse ISM. More robust conclusions will require modeling the extinction curves using various dust compositions and grain size distributions (Klimenko et al., in prep.).

\subsection{ Silicate Absorption vs. Carbonaceous Absorption}

It is also of great interest to examine whether the strength of the silicate 10 $\mu$m absorption feature correlates with the strength of the 2175 \AA\ bump believed to arise in carbonaceous dust. At the most basic level, comparing the strengths of these features can help assess the independence of the grains producing these features. Moreover, the ratio of carbonaceous and silicate dust abundances can discriminate between dust formed in C-rich and O-rich environments, and therefore, between the relative contribution of O- and C-rich AGB stars and core-collapse SNe in producing materials for dust formation.

The absolute strengths of the 10~$\mu$m silicate and the 2175 \AA\ features (see Table~\ref{tab:correlation-results}) show a correlation in the diffuse ISM sample with an average $\tau_{10}/A_{2175}\sim0.05$.  In contrast,  lack of correlation is seen for distant galaxies, consistent with earlier results from SST IRS quasar spectra \citep{Kulkarni2011}. 
This suggests that the increase in both $\tau_{10}$ and $A_{2175}$ in the local ISM is driven by the corresponding rise in $A_V$ and that the formation processes of the two types of grains may be independent. Such a conclusion is consistent with the results of \citet{Gordon2021, Decleir2025}. For dust in distant galaxies probed by quasar sightlines, 
the abundances of silicate and carbonaceous grains may be unrelated to their formation processes.  
For example, if quasar sightlines cross the CGM of distant galaxies, the relative abundances of silicate and carbonaceous grains may reflect survival effects rather than formation processes.

Figure~\ref{fig:correlations-vs-MW}b shows the relation between relative strengths of 10~$\mu$m silicate and the 2175 \AA\ features (normalized to $A_V$).  Both samples do not show significant correlations between $\tau_{10}/A_V$ and $A_{2175}/A_V$, even though the median $\tau_{10}/A_V$ in quasar sightlines is about three times larger than in local diffuse ISM sightlines, while $A_{2175}/A_V$ is nearly the same - indicating that the relative strengths of the silicate feature are stronger than the carbonaceous features.

However, it is important to also consider the $A_V$ parameter, since the quasar sightlines probe gas and dust with smaller $A_V$ values than sightlines in the local ISM. Figure~\ref{fig:correlations-vs-MW}c shows the ratio of the strengths of the 10~$\mu$m silicate and the 2175 \AA\ features versus $A_V$. The $\tau_{10}/A_{2175}$ ratio in quasar sightlines appears to decrease with increasing $A_V$, approaching the value observed in the local diffuse ISM ($\tau_{10}/A_{2175} \simeq 0.05$) at $A_V\simeq1.5-2$. The local ISM sightlines probe the range of $A_V=1.2-5.4$ and show a tentative positive correlation of $\tau_{10}/A_{2175}$, but with a much shallower slope. We note however that these trends should be treated with caution, since the p values are large for both the MW sample and the QSO absorber sample. This can have several interpretations, which can be tested with future observations of the 10~$\mu$m and 2175~\AA\ features in quasar sightlines with larger $A_V$.  While QSOs with large $A_V$ would be harder to observe in the UV and measure the 2175~\AA\ bump for, IR observations would enable measurements of their silicate absorption features and potentially PAH emission features in the associated galaxies. If $\tau_{10}$, $A_{2175}$, and $A_V$ in such sightlines are found to be similar to those observed in the local ISM, then the observed difference in our sample of quasar sightlines and local ISM sightlines may primarily result from probing different physical environments. A significant fraction of our current sample of quasar sightlines with relatively small $A_V$ likely probes gas/dust in the outer regions of the absorber galaxy or in the CGM in such distant galaxies, where the dust content is low and silicates and larger grains can survive, while small carbonaceous  grains are largely destroyed. Quasar sightlines with larger $A_V$ may instead probe dust in the ISM of distant galaxies, where the relations could be comparable to those of the MW ISM.
Alternatively, if $\tau_{10}/A_{2175}$ continues to decrease at higher $A_V$, this would suggest that the dust chemistry in the ISM of distant galaxies differs fundamentally from that in the local ISM.
Of course, the current sample size is small, and larger samples are needed to more definitively establish the presence or absence of such correlations.

\subsection{ Trends with Gas-phase Metals}

Figure~\ref{fig:correlations-vs-MW}e shows the optical depth of the 10~$\mu$m silicate feature vs. the rest equivalent width of the Mg II~$\lambda 2796$ absorption line, $W_{r}({\rm Mg II})$. A positive correlation between $\tau_{10}$ and $W_{r}({\rm Mg II})$ is clearly visible for extragalactic sightlines ($r_S=0.78$ and $p=0.01$).  
However, we note  that higher  $W_{r}({\rm Mg II})$ values do not necessarily indicate higher Mg~{\sc ii} column densities in the gas, since at the high equivalent widths, the Mg~{\sc ii} lines are often saturated, and their strength is an indication of the velocity spread of the absorbing gas (with higher velocity spreads denoting either higher galaxy masses, lower impact parameters, or stronger outflows). Thus, the positive correlation in Fig.~\ref{fig:correlations-vs-MW}e may suggest that stronger silicate features arise in more massive galaxies or in galaxies driving stronger outflows (potentially due to higher star formation rates or more powerful central supermassive black holes). 
We plan to investigate these issues in a future paper searching for the galaxies associated with the absorption systems (Klimenko et al., in prep). 

However, no correlation between $\tau_{10}$ and $W_{r}({\rm Mg II})$ is observed in the sample of MW sightlines ($r_S=-0.10$ and $p=0.71$).
Furthermore, the MW sightlines clearly have much lower $W_{r}({\rm Mg II})$ than the extragalactic sightlines (our sample). This suggests that the extragalactic sightlines trace gas through not just the disk of the galaxy but also its outer disk/circumgalactic medium \citep[e.g.,][]{Bowen1996}, supporting our earlier discussion of observational bias  in the sense of probing different environments between the samples.

\subsection{Mg:Fe ratio vs. silicate feature location}

The peak wavelength of the 10~$\mu$m silicate feature is known to correlate with the Mg:Fe ratio \citep[e.g.,][and references therein]{Draine2003}. Laboratory measurements \citep[e.g.,][]{DeMyk2017a} show that amorphous  pyroxenes with more Mg (compared to Fe) have silicate features peaking at longer wavelengths. 
On the other hand, for olivines, \cite{Pitman2010} reported that as the percentage of Mg increases, the peak wavelength decreases. Likewise \cite{Jager1998} show that the peak is at 10.2 $\mu$m for Mg-rich olivines, and increases to 10.6~$\mu$m for Mg-poor olivines.

While we do not know the Mg:Fe column density ratio for our extragalactic sightlines, we can examine trends with equivalent width ratios.  The rest equivalent widths of Fe~{\sc ii}~$\lambda$2382 were measured in SDSS quasar spectra and archival HST STIS spectra for the MW sightlines. 
Figure~\ref{fig:correlations-vs-MW}f shows the ratio of the Mg~{\sc ii}~$\lambda$2796 and  Fe~{\sc ii}~$\lambda$2382 rest equivalent widths vs. the central wavelength of the 10 $\mu$m silicate feature. For the MW sight lines values of the central wavelength are taken from \cite{Gordon2021, Decleir2025}.
A positive trend is seen for the diffuse ISM sightlines ($r_S=0.76$ and $p=1.5\times10^{-3}$), whereas our extragalactic sightlines show no strong correlation ($r_S=-0.03$ and $p=0.93$). This difference would indicate different chemical compositions (consistent with the differences in silicate mineralogy noted in Section~\ref{sec:minearology}), if the $W({\rm MgII})/W({\rm FeII})$ ratio were indicative of the gas-phase Mg~{\sc ii}/Fe~{\sc ii} column density ratio. However, this is not necessarily the case, since Mg~{\sc ii} and /or Fe~{\sc ii} lines with high equivalent widths are saturated (in many of the quasar absorber sightlines, and even in some of the MW ISM sightlines). 

High-resolution UV/optical spectroscopy is essential to more accurately determine the gas-phase Mg~{\sc ii} and Fe~{\sc ii} column densities in each sightline. High-resolution  UV spectroscopy of lines from less depleted elements (elements that condense less readily from the gas phase to the solid phase, e.g., O, S, Zn) would also help to determine the line-of-sight depletion factors and the depletion-corrected metallicities (following the prescription of \citealt{Jenkins2009}), and thereby infer the dust-phase Mg~{\sc ii}/Fe~{\sc ii} column density ratios to investigate whether that ratio is indeed anti-correlated with the peak wavelength of the 10~$\mu$m silicate absorption feature. 

\section{Conclusions}
\label{sec:conclusions}
We have performed the first high spectral resolution observations of silicate features in extragalactic sightlines to background quasars arising in intervening galaxies along the sightlines, using the JWST MIRI MRS.  
We have detected and measured the silicate 10 $\mu$m absorption features in five quasar absorbers at $0.5 < z < 1.2$ that are known to show the 2175 \AA\ bump. The high resolution of the MRS allows us to accurately measure the strength and profile shape of the 10 $\mu$m silicate absorption. 
Combining results of our analysis with past studies based on Spitzer IRS spectroscopy, we examine the trends between various dust and metal absorption properties for quasar absorbers and compare them with trends for local ISM and AGN sightlines from the literature. Our main results are as follows:

(1) We find that the peak optical depth of the silicate absorption in distant galaxies is about three times larger than expected for diffuse ISM sightlines in the MW over the range $A_V=0.1-2.0$. Within this range of $A_V$, the $\tau_{10}$ shows no strong correlation with the extinction in the sample of quasar sightlines, in contrast to the positive correlation observed in the MW diffuse ISM at higher $A_V=1.2-5.4$.

(2) The extinction-curve fits for our galaxies indicate substantial diversity compared to the extinction curves in the MW ISM. About half of the systems show extinction curves that are steeper in the UV/optical, similar to those observed in the LMC in terms of $R_V$, $A_V$, $A_{2175}$, and $A_K/A_V$. The remaining systems show much flatter extinction, with $R_V$ values of $5–13$ and a significant excess of IR extinction, indicating a prevalence of large grains. If the optical/IR extinction is caused mainly by silicate grains, then the larger $R_V$ would suggest larger silicate grains, and the larger $\tau_{10}/A_V$ values could be explained in part because of the larger silicate grains, although the grains may also be more silicate-rich.  We plan to compare the inferred extinction curves in the distant galaxies in our sample with predictions from theoretical models for various grain compositions and size distributions in a future study.

(3) The relative strength of the 2175~\AA\ bump ($A_{2175}/A_V$) is similar to the average value observed in  MW diffuse sightlines. A comparison of the relative strengths of the 10~$\mu$m silicate feature and the 2175~\AA\ bump shows no strong correlation.
However, the relation of $\tau_{10}/A_{2175}$ and $A_V$ is different for dust probed by quasar sightlines and in the local ISM. For distant galaxies, we find that $\tau_{10}/A_{2175}$ gradually decreases over the range $A_V = 0.1$–$2.0$, approaching the average value measured in the local ISM. In contrast, the sightlines in the local ISM shows a slight increase in $\tau_{10}/A_{2175}$ with $A_V$  over the range $A_V=1.2$–$5.4$. Future observations of the 10 $\mu$m and 2175 \AA bump in larger samples are needed to establish these trends more robustly.

(4)A possible interpretation of results (1)–(3) likely involves an interplay of several factors. First, some of our quasar sightlines may probe dust in the CGM, where the relative fractions of silicate and carbonaceous grains are determined to a greater degree by survival processes of grains. Second, the dust grains in our galaxies may have undergone significant reprocessing in the ISM and CGM under conditions that may differ from those in the local ISM, which may give rise to different dominant dust-processing mechanisms.

(5) 
The 10 $\mu$m profile fits for three of our sightlines show detection of possible non-amorphous (crystalline) silicate grains. The 10 $\mu$m profile fits in the remaining  cases suggest the presence of amorphous silicates. Compared with the predominance of amorphous silicates in MW ISM sightlines \citep[$> 95\%$, e.g., ][]{Draine2003}, this suggests that the distant galaxies traced by our sightlines may have  conditions (e.g., lower cosmic ray fluxes), making it easier for crystalline silicates to survive.

(6) We find the silicate absorption strength to be strongly correlated with the gas-phase Mg~{\sc ii} absorption strength for the extragalactic sightlines. This suggests that stronger silicate absorption is found in our sightlines (which have strong Mg~{\sc ii} absorbers) arising in more metal-rich or more massive galaxies, or in galaxies with stronger outflows. The outflow scenario may be more likely, given also that the $\tau_{10}/A_V$ ratio is consistent with AGN dust, potentially indicating warmer dust grains. The MW sightlines show significantly lower Mg~{\sc ii} equivalent widths  (by a factor of 3-5), suggesting that MW sightlines and our extragalactic sightlines probe dust in different environments, e.g., ISM versus outer disk/CGM.

(7) 
We find a positive correlation between the Mg~{\sc ii}/Fe~{\sc ii} equivalent width ratio and the central wavelength of the 10~$\mu$m silicate feature in the local diffuse ISM sightlines, and no significant correlation for the sample of extragalactic sightlines.  We caution, however, that this difference cannot be interpreted definitively with the existing data. The Mg~{\sc ii} and Fe~{\sc ii} lines in extragalactic sightlines are often saturated; so the $W_r({\rm MgII})/W_r({\rm FeII})$ ratio does not directly reveal the gas-phase Mg/Fe abundance ratio, but is also affected by the kinematics of the absorbing gas. 

\section{Data availability}

Data directly related to this publication and its figures can be requested from the authors. The {\it JWST} data used in this paper are publicly available and can be found in the MAST (Mikulski Archive for Space Telescopes) portal under program ID 2155,  DOI: 10.17909/br31-0120. 

\section*{Acknowledgements}
We thank an anonymous referee for constructive comments that have helped to improve this paper. We would like to thank Marjorie Decleir for providing the data ahead of publication. This work is supported by a 
grant from the Space Telescope Science Institute for JWST GO program 2155 and by NASA grant 80NSSC24K1162 (PI V. Kulkarni). Support from NASA grant 80NSSC20K0887 (PI V. Kulkarni) is also gratefully acknowledged. This work is based in part on observations made with the NASA/ESA/CSA James Webb Space Telescope. The data were downloaded from the Mikulski Archive for Space Telescopes at the Space Telescope Science Institute which is operated by the Association of Universities for Research in Astronomy, Inc., under NASA contract NAS 5-03127 for JWST. Our 
observations are associated with program 2155. Support for program 2155 was provided by NASA through a grant from the Space Telescope Science Institute, which is operated by the Association of Universities for Research in Astronomy, Inc., under NASA contract NAS 5-03127.

\bibliographystyle{mnras}
\bibliography{Library} 

\newpage
\appendix

\section{fit to silicate absorptions}
\label{app:b:fit-profiiles}

\begin{figure*}
\begin{center}       
\includegraphics[width=0.9\textwidth]{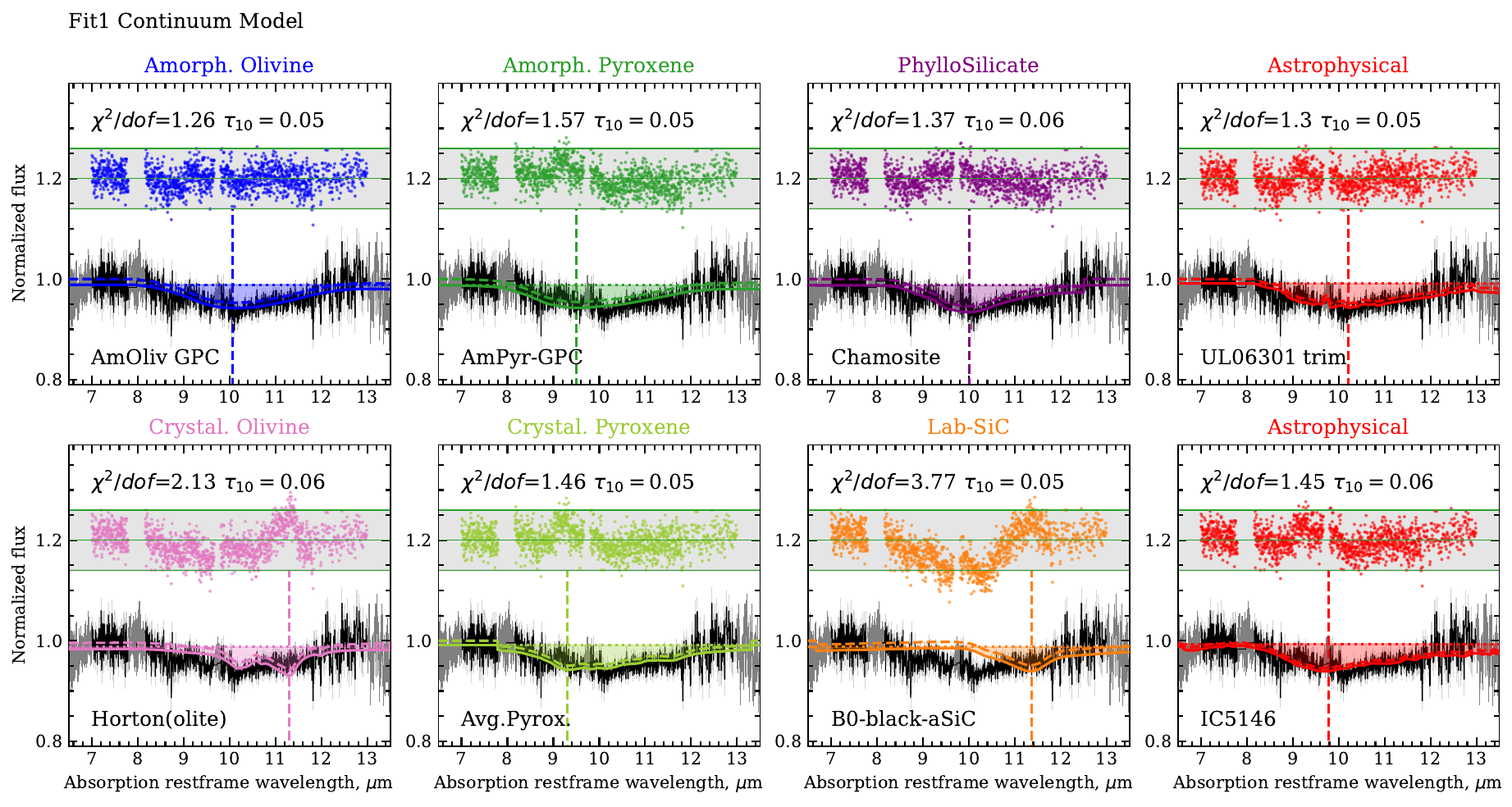}
 \includegraphics[width=0.9\textwidth]{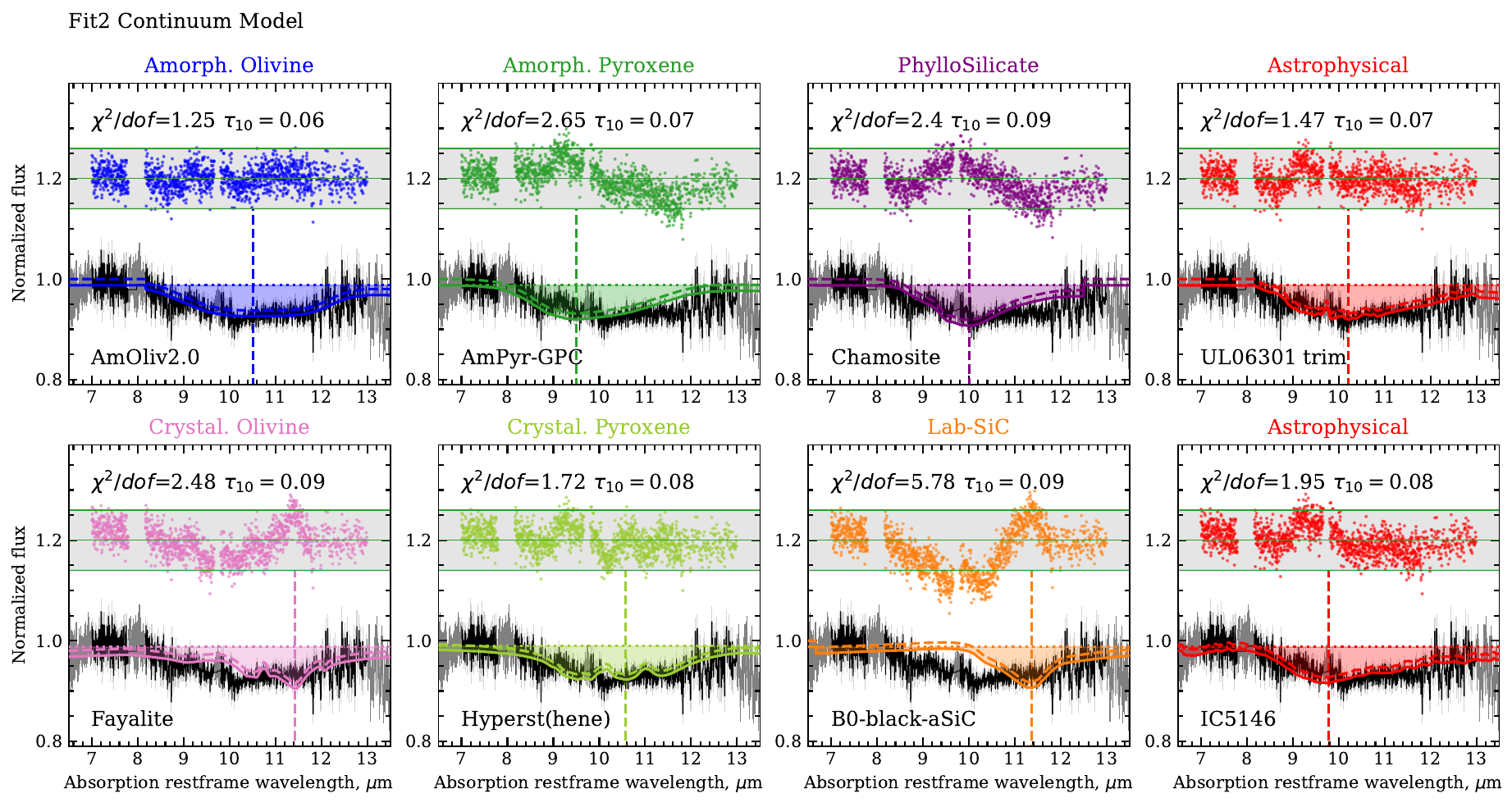}
        \caption{\rm Fit to the silicate absorption in the spectrum of AO0235$+$164, normalized with the `Fit1' model (top panels) and with the `Fit2' model (bottom panels); (see text for details).  The  quasar normalized spectrum and its uncertainty (flux error bars) are shown in dark and light gray, respectively. The black line highlights the fitting regions. Dashed, dotted and solid colored curves represent the normalized absorption line profile ($A(\lambda,\tau_{\rm 10}^{\rm abs})$,  see Equation~\ref{eq:sil_model}), the continuum model ($(1+\delta C)\times(1+Em(\lambda,\tau^{\rm em})$), and the combined fit ($M(\lambda)$), respectively. The panels in the three left columns show the best fits with the laboratory templates, while the rightmost panels present the best fits using the astrophysical templates: ULIRGs (top) and a molecular cloud (bottom) for AO0235+164.   The small filled circles at the top of each panel show the residuals, and the $\pm3\sigma$ errors levels are shown by green lines and the gray shaded area. The dashed vertical line indicates the peak wavelength of the template.
        }
        \label{fig:fit-silicate-ao0235-fit1}
\end{center}
\end{figure*}

\begin{figure*}
\begin{center}
\includegraphics[width=0.95\textwidth]{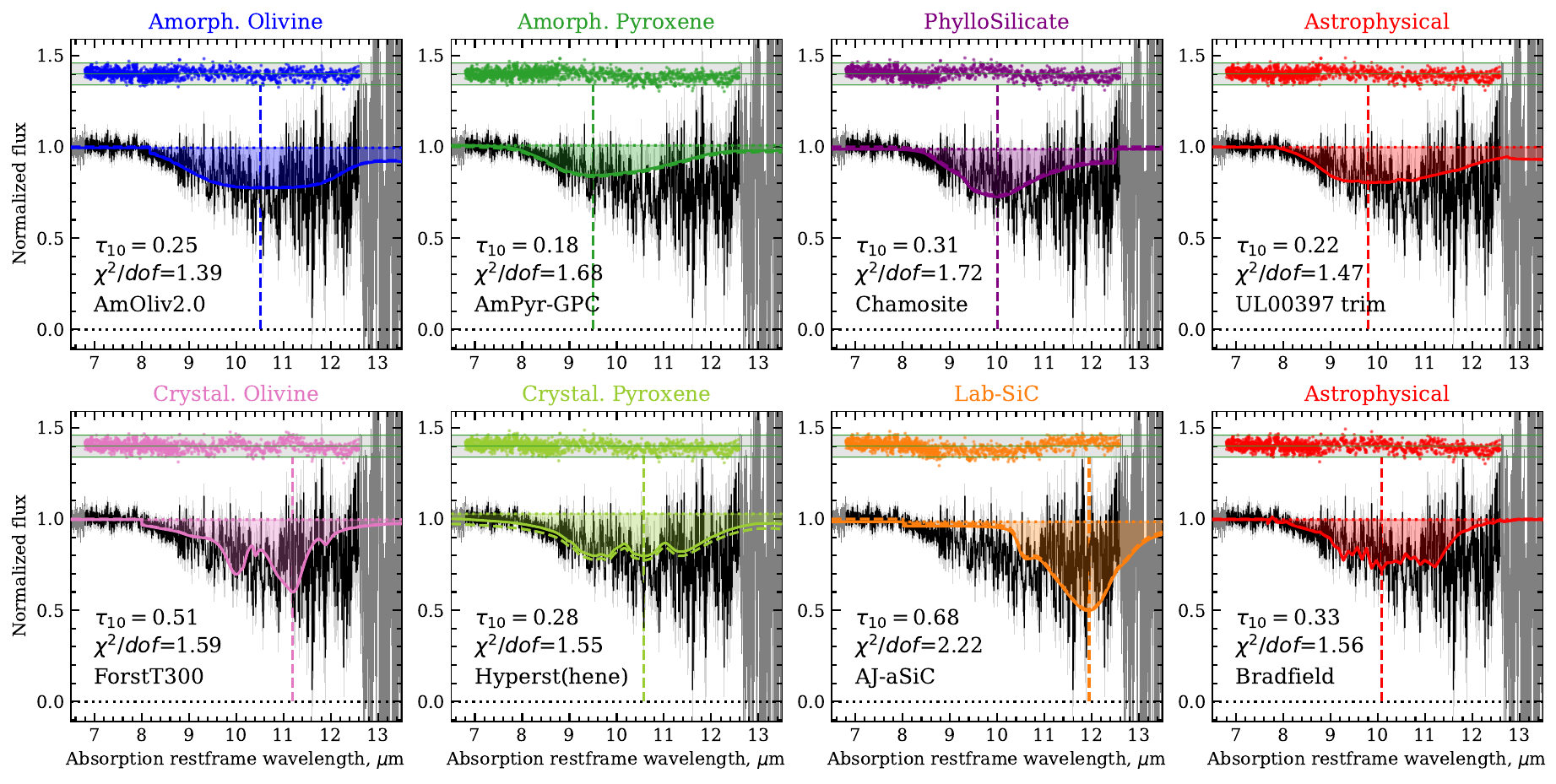}
        \caption{\rm Fit to the silicate absorption in the spectrum of J0900$+$0214. The description of the panels are the same as in Fig.~\ref{fig:fit-silicate-ao0235-fit1}.
        }
        \label{fig:fit-silicate-j0900}
\end{center}
\end{figure*}

\begin{figure*}
\begin{center}
\includegraphics[width=0.95\textwidth]{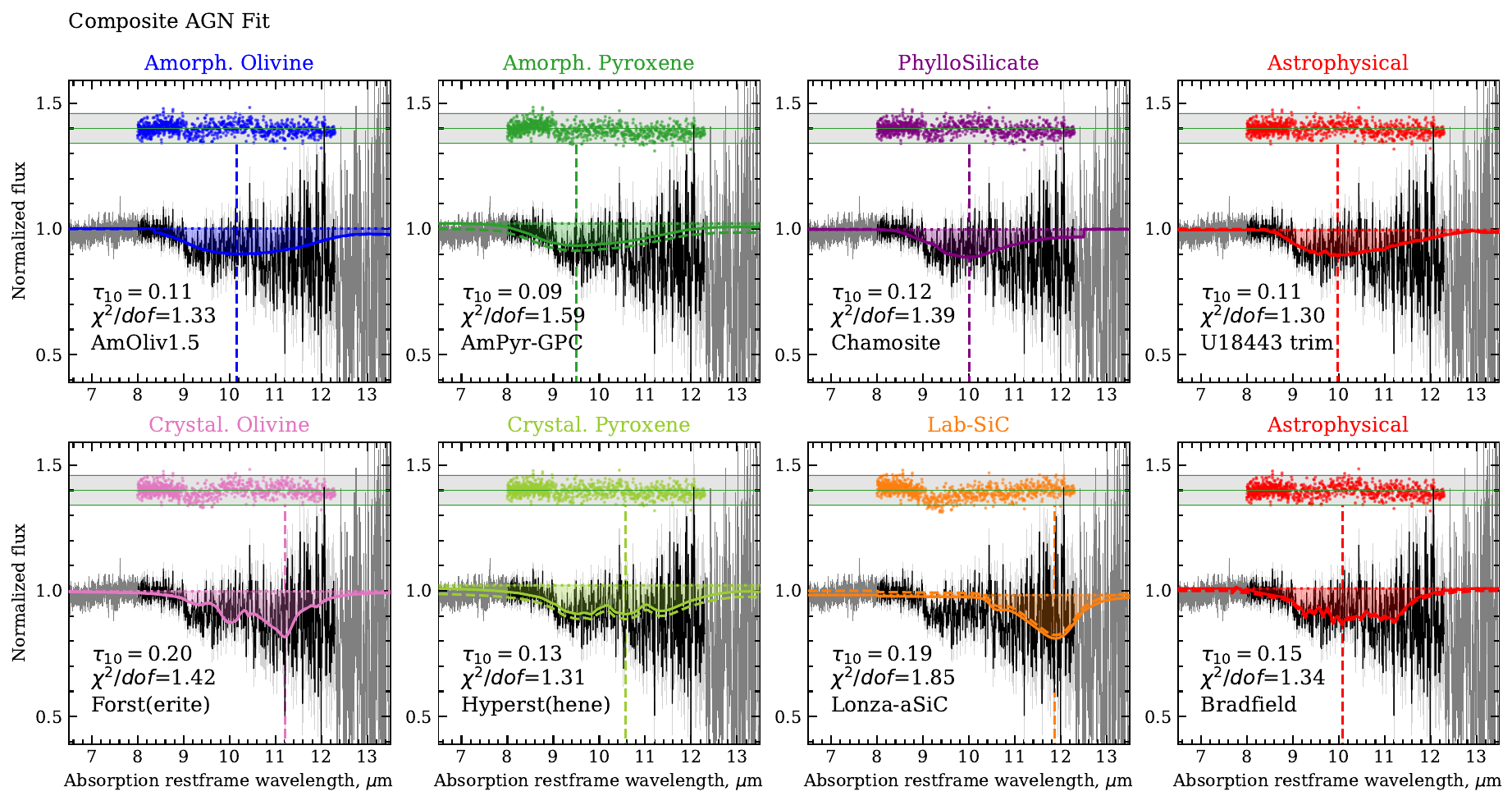}
\includegraphics[width=0.95\textwidth]{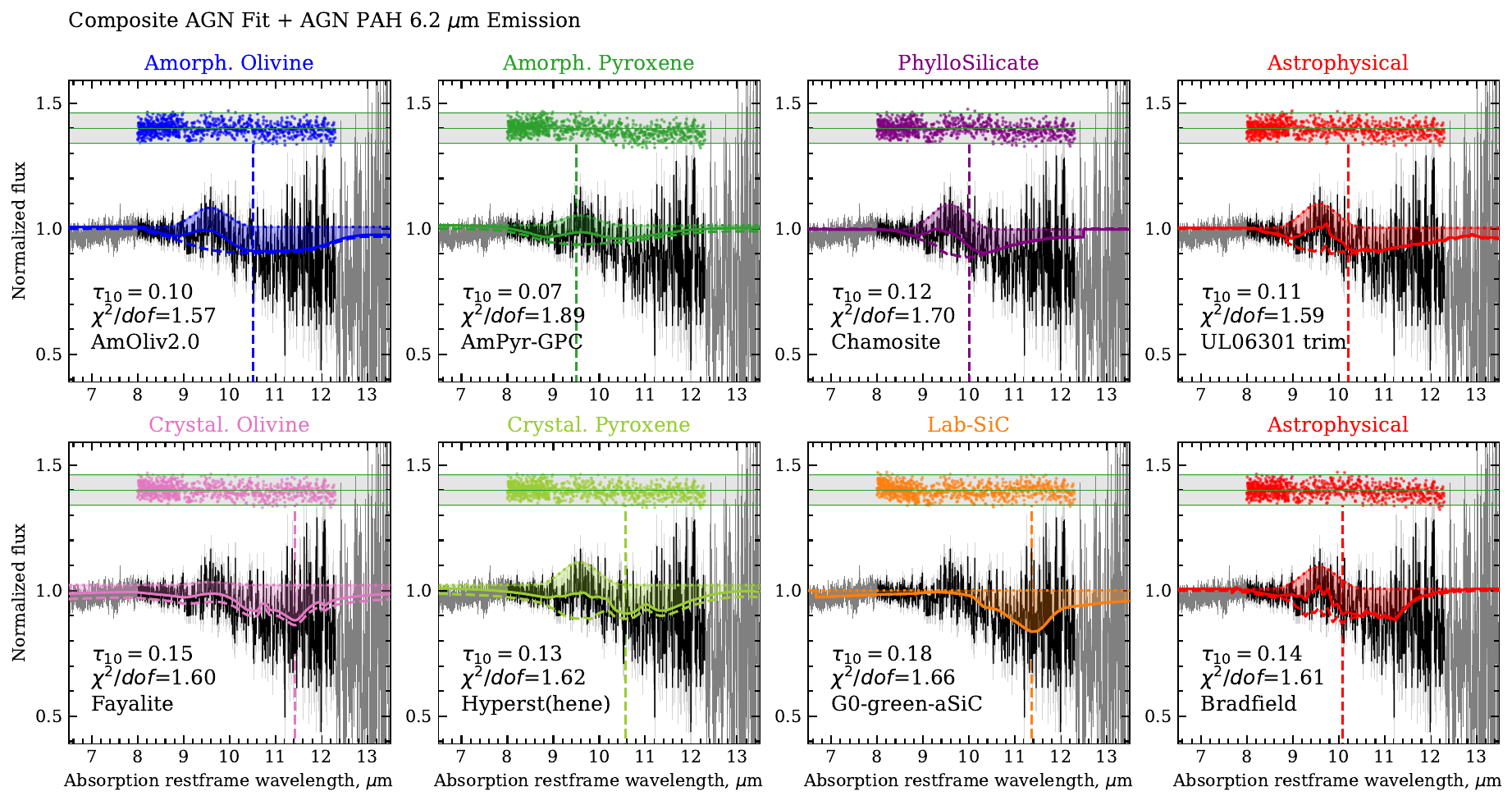}
        \caption{\rm 
        Fit to the silicate absorption in the spectrum of J0901$+$2044, normalized using the constructed AGN composite spectrum (top panels), and to the composite AGN spectrum with variable AGN PAH~6.2$\mu$m emission strength (bottom panels). The description of the panels are the same as in Fig.~\ref{fig:fit-silicate-ao0235-fit1}.
        }
        \label{fig:fit-silicate-j0901-em}
\end{center}
\end{figure*}

\begin{figure*}
\begin{center} 
\includegraphics[width=0.95\textwidth]{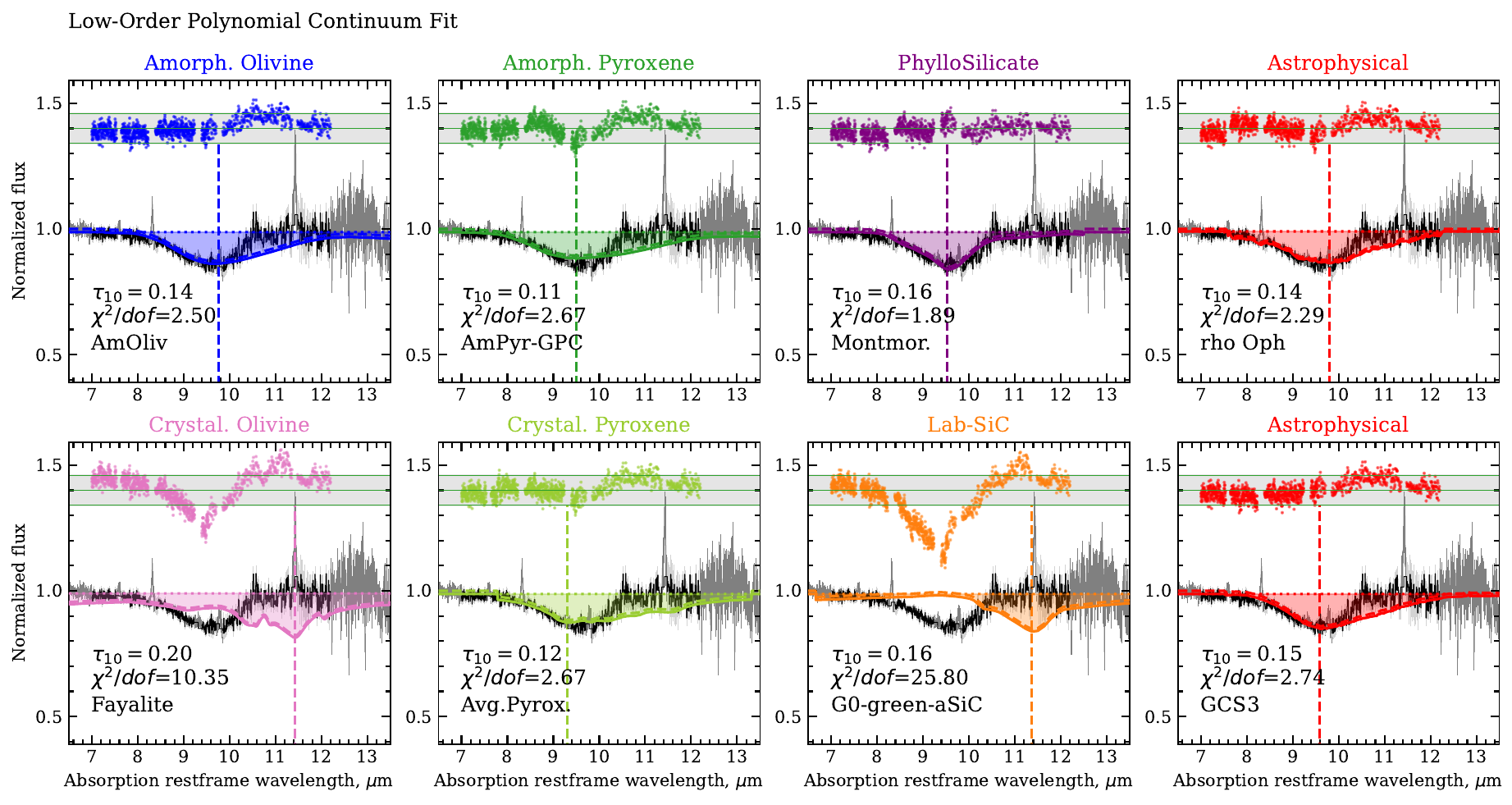}
        \includegraphics[width=0.95\textwidth]{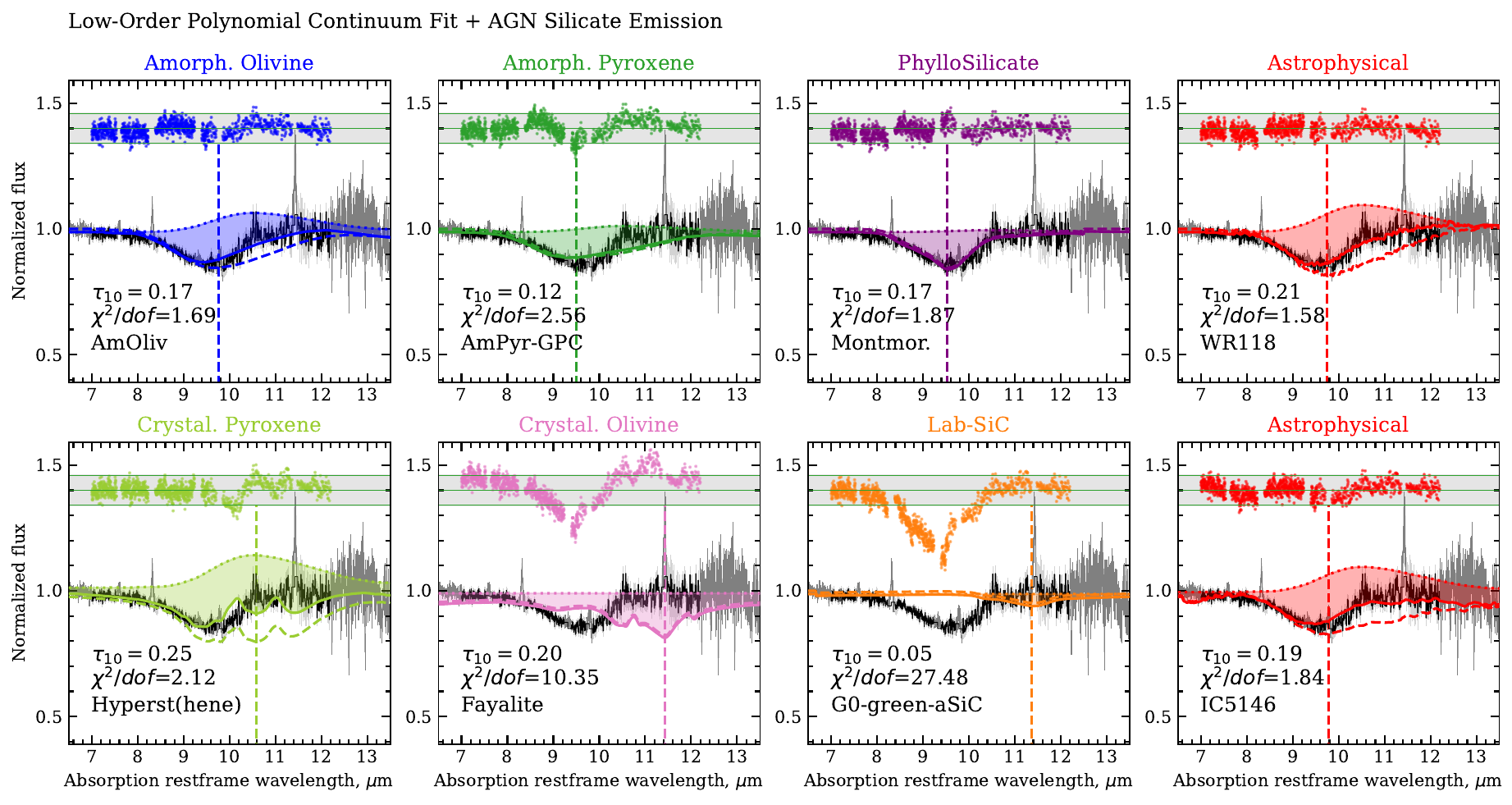}

        \caption{\rm Fit to the silicate absorption in the spectrum of J1007$+$2853, normalized to a low-order polynomial continuum fit, shown without (top panels) and with (bottom panels) the AGN silicate emission component included in the fitting procedure. The description of the panels are the same as in Fig.~\ref{fig:fit-silicate-ao0235-fit1}. 
        }
        \label{fig:fit-silicate-j1007-em}
\end{center}
\end{figure*}

\begin{figure*}
\begin{center}      
\includegraphics[width=0.95\textwidth]{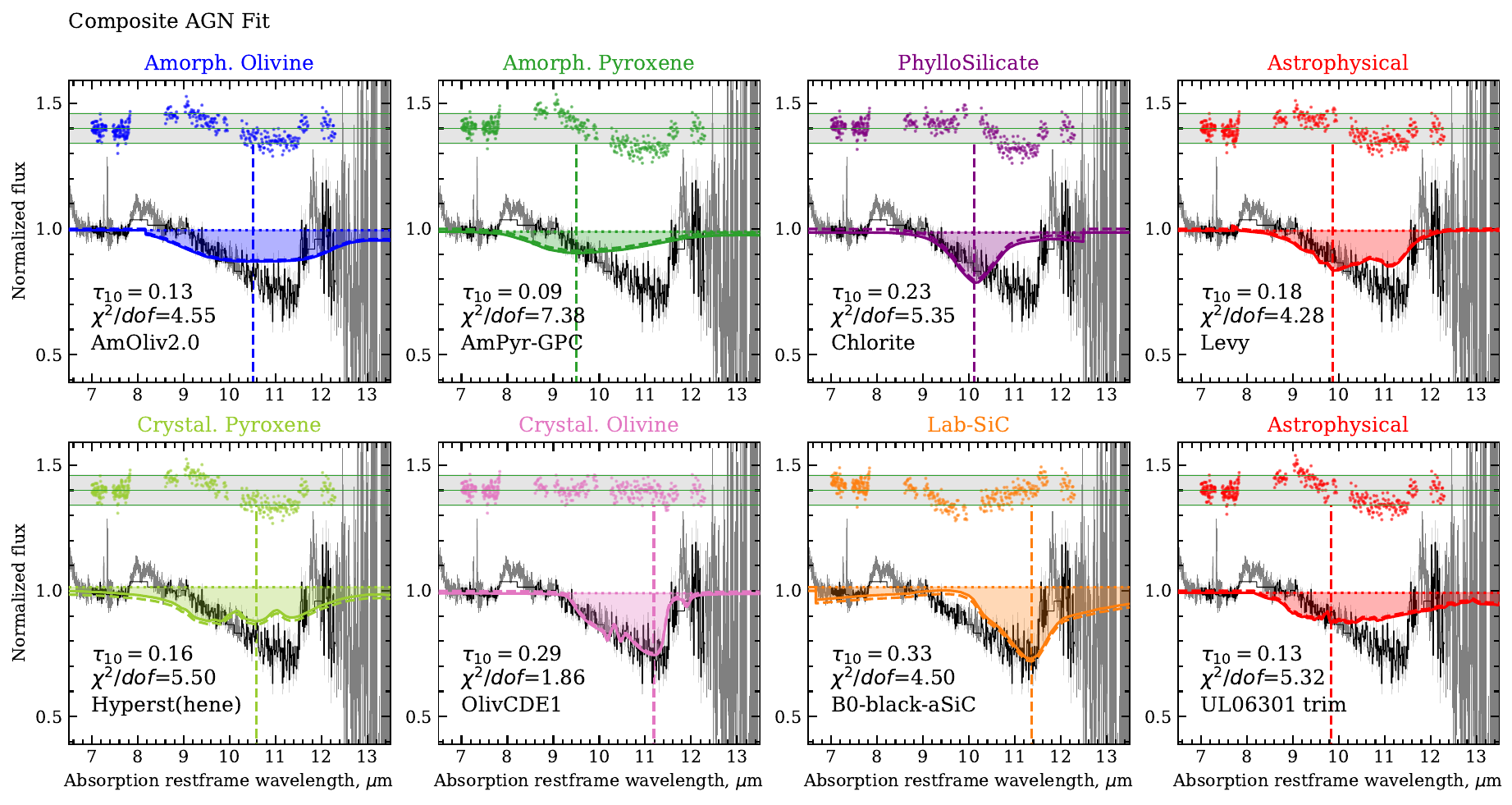}
\includegraphics[width=0.95\textwidth]{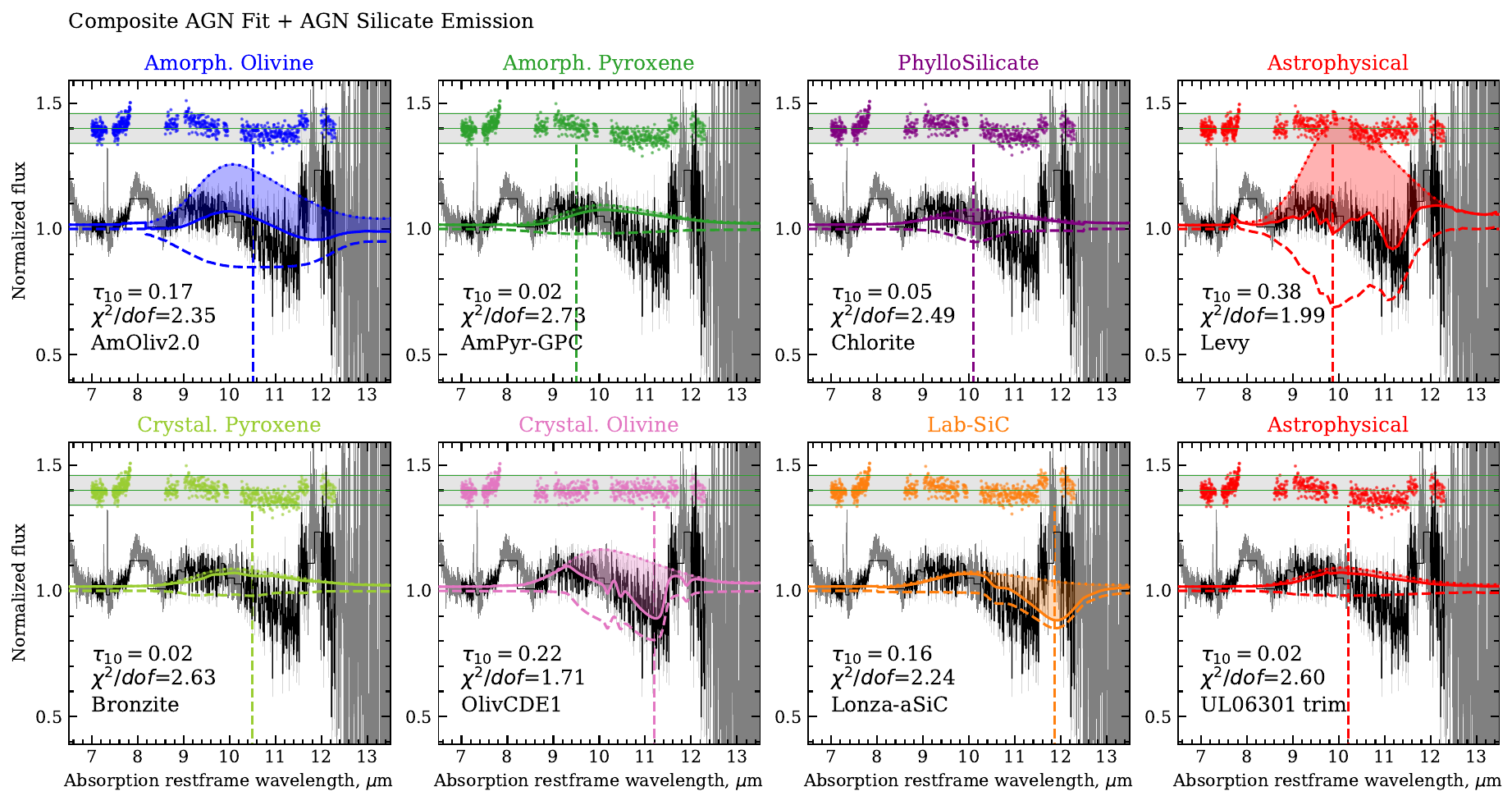}
        \caption{\rm Fit to the silicate absorption in the spectrum of J1017$+$4749, normalized to the constructed AGN composite spectrum (top panels), and to the composite AGN spectrum with variable AGN silicate-emission strength (bottom panels). The description of the panels are the same as in Fig.~\ref{fig:fit-silicate-ao0235-fit1}.}
        \label{fig:fit-silicate-j1017}
\end{center}
\end{figure*}

We present fits to the 10~$\mu$m silicate feature in MIRI MRS quasars spectra from our sample. Figs.~\ref{fig:fit-silicate-ao0235-fit1}-\ref{fig:fit-silicate-j1017} show the best fits for the laboratory-based (amorphous olivine, amorphous pyroxene, phyllo-silicate, crystalline olivine, crystalline pyroxene, and SiC) and observational templates for the astrophysical sources. Each panel provides the reduced $\chi^2$ value, the optical depth value $\tau_{10}$, and the profile name. The template names are described in \cite{Aller2012}. For the astrophysical templates shown in the right panels, we present two fits for different sources that demonstrate better consistency with the observed absorption. Additionally, we present results for profile fitting in spectra obtained with different continuum models (see Paper~I for details): for AO0235+164, we consider 'Fit1' and 'Fit2'; for J0900+0214 - an AGN composite; for J0901+2044 - an AGN composite and a continuum component plus PAH~6.2~$\mu$m emission line; for J1007+2853 and for J1017+4749 - an AGN composite and a continuum component plus 10~$\mu$m silicate emission line.

\end{document}